\documentclass[12pt,a4paper]{article} % use larger type; default would be 10pt

\usepackage[utf8]{inputenc} % set input encoding 

\usepackage[a4paper,margin=1in]{geometry} % to change the page dimensions
\usepackage{graphicx} % support the \includegraphics command and options
\usepackage{xcolor} %Herb added
\usepackage[parfill]{parskip} % Activate to begin paragraphs with an empty line rather than an indent

\usepackage{amsmath} % Herb added for tfrac
\usepackage{amssymb} % Herb added for tfrac
\usepackage{braket} % Herb added for Dirac brackets
\usepackage{booktabs} % for much better looking tables
\usepackage{array} % for better arrays (eg matrices) in maths
\usepackage{paralist} % very flexible & customisable lists (eg. enumerate/itemize, etc.)
\usepackage{verbatim} % adds environment for commenting out blocks of text & for better verbatim
\usepackage{empheq} %Herb added for emphasizing equations
\usepackage{epstopdf} %Herb added
\usepackage{multicol} %Herb added
\usepackage{qtree} %Herb added, for drawing tree diagrams
\usepackage{float}

\usepackage{sectsty}
\allsectionsfont{\sffamily\mdseries\upshape} % This matches ConTeXt defaults

\usepackage[nottoc,notlof,notlot]{tocbibind} % Put the bibliography in the ToC
\usepackage[titles]{tocloft} % Alter the style of the Table of Contents

\numberwithin{equation}{section} %Herbadded

\title{Symplectic transformations of a beam matrix \\ with real Pauli and Dirac matrices}
\author{Herbert E. M\"uller \\ \normalsize http://herbert-mueller.info/}
\date{} 
\DeclareMathOperator{\Tr}{Tr}
\definecolor{myred}{rgb}{0.8,0,0}
\definecolor{mygreen}{rgb}{0,0.6,0}
\definecolor{myblue}{rgb}{0,0,0.8}

\begin{document}
\maketitle

\begin{flushleft}

\begin{abstract}
A basic problem in linear particle optics is to find a symplectic transformation that brings a symmetric matrix $\mathbf{\Sigma}$ (the beam or bunch matrix) to a special diagonal form, called normal form. The conventional way to do this involves an eigenvalue-decomposition of a matrix related to $\mathbf{\Sigma}$, and may be applied to the case of 1, 2 or 3 particle degrees of freedom.  For 2 degrees of freedom, a different normalization method involving "real Dirac matrices" has recently been proposed \cite{Baumgarten1}, \cite{Baumgarten2}. In the present article, the mathematics of real Dirac matrices is presented differently. Another normalization recipe is given, and more general decoupling problems are solved. A 3D visual representation of $\mathbf{\Sigma}$ is provided. The corresponding normalization method for 1 degree of freedom involving "real Pauli matrices" is also given. 
\end{abstract}

\newpage 

\tableofcontents

\newpage 

\section{Introduction}

In accelerator physics, particle beams are usually chopped into bunches of particles. The single particle has 3 degrees of freedom (DFs), and occupies a point in 6-dimensional phase space. The distribution of bunch particles in phase space can be roughly described by a symmetric $6 \times 6$ matrix $\mathbf{\Sigma}$ of $2$nd moments, called bunch matrix. The evolution of the bunch matrix along a beamline can be described with a symplectic $6 \times 6$ matrix $\mathbf{R}$, called transport matrix. This article concerns the basic problem of finding a transport matrix $\mathbf{N}$ that will bring the bunch matrix to a special diagonal form, called normal form. The conventional way to do this involves an eigenvalue-decomposition of the bunch matrix times another $6 \times 6$ symmetric matrix $\boldsymbol{\gamma}$, called the symplectic form \cite{Wolski}. This normalization method is described in \emph{section 2}.

Often, only the two transversal particle DFs are of interest. The bunch matrix then reduces to order 4, and is called a beam matrix. The normalization problem becomes accordingly simpler. For this case, Ch. Baumgarten has recently proposed a different normalization method involving "real Dirac matrices" \cite{Baumgarten1}, \cite{Baumgarten2}. In \emph{section 5}, Baumgarten's method is presented in a mathematically different way: a 2-index numbering of the real Dirac matrices is used; more general elementary transformations are defined; the beam matrix instead of a force matrix is transformed; more general decoupling problems are solved; and an alternative normalization recipe is proposed. Furthermore, general determinant and inversion formulas for square matrices of order 4 are given, and a visual 3D representation of the beam matrix is provided.

If only one particle DF is considered, the beam matrix further reduces to order 2. The normalization method corresponding to Baumgarten's method then involves "real Pauli matrices". This method is hardly ever used, since the normalizing transformation $\mathbf{N}$ can be directly expressed in terms of the components of the $2 \times 2$ beam matrix $\mathbf{\Sigma}$ \cite{Hinterberger}, \cite{Wille}. However, getting familiar with it is a good preparation for the more difficult case of 2 DFs. Normalizing a beam matrix for 1 DF with real Pauli matrices is therefore presented in \emph{section 4}.

Both the real Pauli and Dirac matrices are real representations of Clifford-Algebras \cite{LaMi}, \cite{FoF}, \cite{Mueller}. The details are explained in \emph{section 3}. 

Introducing real Pauli or Dirac matrices into particle optics means, expressing the transformation and beam matrices in a non-standard basis. Since matrix elements in the standard basis come with definite units (usually  mm$^a$mrad$^b$, with $a,b=\ldots -1,0,1 \ldots$), the matrix components in the new basis have mixed units, and have no obvious physical significance. It's a case of adding oranges and apples. This means that the use of real Pauli or Dirac matrices allows mathematical tricks, but doesn't add to physical insight. Another short-coming of real Pauli or Dirac matrices is, that they cannot be used to describe phase vectors, or more generally (phase) tensors of degree other than 2. Finally, there is no matrix algebra of order 6 corresponding to the real Pauli or Dirac matrices, since real irreducible representations of Clifford algebras have order $2, 4, 8 \ldots$. In other words, there is no obvious non-standard matrix basis for 3-DF-particle optics. 

The figures in this article were created with the software package GNU Octave (version 3.6.4 for the operating system Windows) \cite{Octave}. I thank the Octave development community for making this software freely available, and for their excellent work.

\newpage

\section{Linear optics for particles with three degrees of freedom}

\subsection{The particle bunch}

A particle bunch can be described by the distribution of its particles in phase space. A single particle has 3 degrees of freedom, and the particle phase space is 6-dimensional. The phase space coordinate system is chosen co-moving with the bunch, and the coordinates and momenta are denoted by $\mathbf{X}^\intercal=(x, x', y, y', l, \delta)$ \cite{Hinterberger}, \cite{Wille}. 

The main information about the bunch is contained in the lowest moments of the particle distribution in phase space: the vector $\mathbf{\bar{X}}$ of first moments, describing the bunch position, and the matrix $\mathbf{\Sigma}=\overline{(\mathbf{X}-\bar{\mathbf{X}})(\mathbf{X}^\intercal-\bar{\mathbf{X}}^\intercal)}$ of second moments, describing the bunch shape.

\subsection{Symplectic transport}

In this section the following extended versions of the three \emph{real Pauli matrices} $\sigma_3$, $\sigma_1$ and $i\sigma_2$ are used: \\
\footnotesize
\begin{equation}
\boldsymbol\alpha=
\left( \begin{array}{rrrrrr}
1 & 0 & 0 & 0 & 0 & 0\\
0 & -1 & 0 & 0 & 0 & 0\\
0 & 0 & 1 & 0 & 0 & 0\\
0 & 0 & 0 & -1 & 0 & 0\\
0 & 0 & 0 & 0 & 1& 0 \\
0 & 0 & 0 & 0 & 0 & -1\\
\end{array} \right)
\quad
\boldsymbol\beta=
\left( \begin{array}{rrrrrr}
0 & 1 & 0 & 0 & 0 & 0\\
1 & 0 & 0 & 0 & 0 & 0\\
0 & 0 & 0 & 1 & 0 & 0\\
0 & 0 & 1 & 0 & 0 & 0\\
0 & 0 & 0 & 0 & 0 & 1\\
0 & 0 & 0 & 0 & 1 & 0\\
\end{array} \right)
\quad
\boldsymbol\gamma=
\left( \begin{array}{rrrrrr}
0 & 1 & 0 & 0 & 0 & 0\\
-1 & 0 & 0 & 0 & 0 & 0\\
0 & 0 & 0 &  1 & 0 & 0\\
0 & 0 & -1 & 0 & 0 & 0\\
0 & 0 & 0 & 0 & 0 &  1\\
0 & 0 & 0 & 0 & -1 & 0\\
\end{array} \right)
\end{equation} 
\normalsize

We have 
\begin{equation}
\boldsymbol\alpha\boldsymbol\beta=\boldsymbol\gamma \qquad \qquad
\boldsymbol\alpha^2=\boldsymbol\beta^2=\mathbf{1}  \qquad \qquad 
\boldsymbol\gamma^2=-\mathbf{1}
\end{equation}
$\boldsymbol\alpha$ and $\boldsymbol\beta$ are \emph{bireal units}, $\boldsymbol\gamma$ is a \emph{complex unit}, see section 3. $\boldsymbol\gamma$ is the \emph{symplectic form} on the phase space. Now let us return to linear particle optics.

As the bunch runs down the beamline, its particles describe trajectories $\mathbf{X}$ in phase space, and the bunch mean position $\mathbf{\bar{X}}$ and 2nd moments $\mathbf{\Sigma}$ change. The evolution of all these quantities is described with a $6\times 6$ \emph{transport matrix} $\mathbf{R}$:
\begin{equation}
\mathbf{X}=\mathbf{R}\mathbf{X}(0) \qquad \qquad
\mathbf{\bar{X}}=\mathbf{R}\mathbf{\bar{X}}(0) \qquad \qquad
\mathbf{\Sigma}=\mathbf{R}\mathbf{\Sigma}(0)\mathbf{R^\intercal}
\end{equation}

The particle motion is such that the phase space volume enclosed by many neighbouring particles is preserved, and the transport matrix $\mathbf{R}$ is symplectic: 
\begin{equation}
\mathbf{R}\boldsymbol\gamma \mathbf{R}^\intercal = \boldsymbol\gamma \qquad \text{or} \qquad \mathbf{R}^\intercal\boldsymbol\gamma \mathbf{R} = \boldsymbol\gamma \qquad \text{or} \qquad \mathbf{R^{-1}} =-\boldsymbol\gamma \mathbf{R}^\intercal \boldsymbol\gamma
\label{e2.02}
\end{equation}

Now let $\mathbf{r}$ be the logarithm of the transformation $\mathbf{R}$: 
\begin{equation}
\mathbf{r} = \ln\mathbf{R} \qquad  \qquad \mathbf{R} = \exp \mathbf{r}  
\end{equation}

It can be shown that the above condition on $\mathbf{R}$ translates to
\begin{equation}
\mathbf{r}=\boldsymbol\gamma\mathbf{r}^\intercal \boldsymbol\gamma \qquad 
\text{or} \qquad \boldsymbol\gamma\mathbf{r}=-\mathbf{r}^\intercal\boldsymbol\gamma \qquad \text{or} \qquad \mathbf{r}\boldsymbol\gamma=-\boldsymbol\gamma\mathbf{r}^\intercal  \label{e2.01}
\end{equation}

I will call matrices with this kind of symmetry \emph{$\gamma$-symmetric}. The second and third eqn. \eqref{e2.01} say that a $\gamma$-symmetric matrix left- or right-multiplied with $\boldsymbol\gamma$ becomes symmetric. Conversely, a symmetric matrix left- or right-multiplied with $\boldsymbol\gamma$ becomes $\gamma$-symmetric. Therefore, to construct a symplectic transformation $\mathbf{R}$, take any symmetric $6\times 6$ matrix, left- or right-multiply it with $\boldsymbol\gamma$, and exponentiate. 

\subsection{Invariants of the bunch matrix}

Given a $6\times 6$-matrix of 2nd bunch moments $\mathbf{\Sigma}$, three numbers can be calculated that are invariant under symplectic transport. These are the bunch emittances $\epsilon_1, \epsilon_2, \epsilon_3$. They are a measure of the bunch size in phase space. 

To find the emittances, note that  $\mathbf{\Sigma}\boldsymbol\gamma$ transforms according to  
\begin{equation}
\mathbf{\mathbf{\Sigma}\boldsymbol\gamma}=\mathbf{R}\mathbf{\Sigma}(0)\boldsymbol\gamma\mathbf{R}^{-1}
\end{equation}

This similarity transformation leaves the characteristic polynomial of $\mathbf{\Sigma}\boldsymbol\gamma$ invariant. The latter is an even function of its argument, as the following calculation shows: 
\begin{equation*}
\det\left(\mathbf{\Sigma}\boldsymbol\gamma-\lambda \mathbf{1}\right) \overset{\cdot(-\boldsymbol\gamma)}{=}
\det\left(\mathbf{\Sigma}+\lambda\boldsymbol\gamma\right)
\overset{\intercal}{=}
\det\left(\mathbf{\Sigma}-\lambda\boldsymbol\gamma\right)
\overset{\cdot\boldsymbol\gamma}{=}
\det\left(\mathbf{\Sigma}\boldsymbol\gamma+\lambda \mathbf{1}\right)
\end{equation*}

The squared emittances now appear as coefficients in the characteristic polynomial:
\begin{equation}
\det\left(\mathbf{\Sigma}\boldsymbol\gamma-\lambda \mathbf{1}\right)=\left(\lambda^2+\epsilon_{I}^{2} \right)
\left(\lambda^2+\epsilon_{II}^{2} \right)
\left(\lambda^2+\epsilon_{III}^{2} \right)
\end{equation}

The eigenvalues of $\mathbf{\Sigma}\boldsymbol\gamma$ are therefore $\lambda=\pm i\epsilon_{I}, \pm i\epsilon_{II}, \pm i\epsilon_{III}$. 

\subsection{Normalizing the bunch matrix}

\cite{Wolski} We are looking for a symplectic transformation $\mathbf{N}$ that puts the bunch matrix $\mathbf{\Sigma}$ into normal form: 
\begin{equation}
\mathbf{\Sigma}=\mathbf{N}\tilde{\mathbf{\Sigma}}\mathbf{N}^\intercal
\qquad \qquad
\tilde{\mathbf{\Sigma}}=\left( \begin{array}{cccccc}
\epsilon_I & 0 & 0 & 0 & 0 & 0\\
0 & \epsilon_I & 0 & 0 & 0 & 0\\
0 & 0 & \epsilon_{II} & 0 & 0 & 0\\
0 & 0 & 0 & \epsilon_{II} & 0 & 0\\
0 & 0 & 0 & 0 & \epsilon_{III} & 0\\
0 & 0 & 0 & 0 & 0 & \epsilon_{III}\\
\end{array} \right)
\label{e2.03}
\end{equation} 

To this end, right-multiply eqn. \eqref{e2.03} with $\boldsymbol\gamma$ and use eqn. \eqref{e2.02}. The result is 
\begin{equation}
\mathbf{\Sigma}\boldsymbol\gamma=\mathbf{N}\tilde{\mathbf{\Sigma}}\boldsymbol\gamma\mathbf{N}^{-1}
\qquad \qquad
\tilde{\mathbf{\Sigma}}\boldsymbol\gamma=\left( \begin{array}{cccccc}
0 & \epsilon_I & 0 & 0 & 0 & 0\\
-\epsilon_I & 0 & 0 & 0 & 0 & 0\\
0 & 0 & 0 & \epsilon_{II} & 0 & 0\\
0 & 0 & -\epsilon_{II} & 0 & 0 & 0\\
0 & 0 & 0 & 0 & 0 & \epsilon_{III}\\
0 & 0 & 0 & 0 & -\epsilon_{III} & 0\\
\end{array} \right)
\label{e2.04}
\end{equation} 

This looks similar to the eigenvalue-decomposition
\begin{equation}
\mathbf{\Sigma}\boldsymbol\gamma=\mathbf{E}\mathbf{\Lambda}\mathbf{E}^{-1}
\qquad
\mathbf{\Lambda}=\left( \begin{array}{cccccc}
-i\epsilon_I & 0 & 0 & 0 & 0 & 0\\
0 & i\epsilon_I & 0 & 0 & 0 & 0\\
0 & 0 & -i\epsilon_{II} & 0 & 0 & 0\\
0 & 0 & 0 & i\epsilon_{II} & 0 & 0\\
0 & 0 & 0 & 0 & -i\epsilon_{III} & 0\\
0 & 0 & 0 & 0 & 0 & i\epsilon_{III}\\
\end{array} \right)
\label{e2.05}
\end{equation} 
where we can normalize the complex matrix $\mathbf{E}$ to symplectic.

The RHS's of eqn.s \eqref{e2.04} and \eqref{e2.05} can be made to agree by means of an intermediate symplectic transformation $\mathbf{Q}$: 
\begin{equation}
\mathbf{N}=\mathbf{E}\mathbf{Q} \qquad \qquad 
\tilde{\mathbf{\Sigma}}\boldsymbol\gamma=\mathbf{Q}^{-1}\mathbf{\Lambda}\mathbf{Q}
\end{equation} 

A calculation shows that the following $\mathbf{Q}$ does the trick:
\small
\begin{equation}
\mathbf{Q}=\frac{\mathbf{1}+i\boldsymbol\beta}{\sqrt{2}}=\frac{1}{\sqrt{2}}
\left( \begin{array}{cccccc}
1 & i & 0 & 0 & 0 & 0\\
i & 1 & 0 & 0 & 0 & 0\\
0 & 0 & 1 & i & 0 & 0\\
0 & 0 & i & 1 & 0 & 0\\
0 & 0 & 0 & 0 & 1 & i\\
0 & 0 & 0 & 0 & i & 1\\
\end{array} \right) 
\end{equation} 
\normalsize

For brevity of notation, let us define the matrices $\mathbf{1}^{(r)}, \boldsymbol\alpha^{(r)}, \boldsymbol\beta^{(r)}, \boldsymbol\gamma^{(r)}$, with $r=I, II$ or $III$. They are the restrictions of these  $6\times 6$ matrices to the $r^\mathrm{th}$ $2\times 2$ block, e. g. 
\small
\begin{equation*}
\mathbf{1}^{II}=
\left( \begin{array}{rrrrrr}
0 & 0 & 0 & 0 & 0 & 0\\
0 & 0 & 0 & 0 & 0 & 0\\
0 & 0 & 1 & 0 & 0 & 0\\
0 & 0 & 0 & 1 & 0 & 0\\
0 & 0 & 0 & 0 & 0& 0 \\
0 & 0 & 0 & 0 & 0 & 0\\
\end{array} \right)
\quad
\boldsymbol\alpha^{III}=
\left( \begin{array}{rrrrrr}
0 & 0 & 0 & 0 & 0 & 0\\
0 & 0 & 0 & 0 & 0 & 0\\
0 & 0 & 0 & 0 & 0 & 0\\
0 & 0 & 0 & 0 & 0 & 0\\
0 & 0 & 0 & 0 & 1 & 0\\
0 & 0 & 0 & 0 & 0 & -1\\
\end{array} \right)
\qquad \text{etc.}
\end{equation*} 
\normalsize

\fbox{
\begin{minipage}{0.98\linewidth}
To obtain the normal decomposition $\mathbf{\Sigma}= \mathbf{N}\tilde{\mathbf{\Sigma}}\mathbf{N}^\intercal$ of a given bunch matrix $\mathbf{\Sigma}$, proceed as follows:

\begin{enumerate}
\item Calculate the eigenvalue-decomposition $\mathbf{\Sigma}\boldsymbol\gamma=\mathbf{E}\mathbf{\Lambda}\mathbf{E}^{-1}$. 
\item Order the eigenvectors in $\mathbf{E}$ such that
$\mathbf{\Lambda}=-i\boldsymbol\alpha^{I}\epsilon_{I}-i\boldsymbol\alpha^{II}\epsilon_{II}-i\boldsymbol\alpha^{III}\epsilon_{III}$.
\item Normalize the eigenvectors in $\mathbf{E}$ to symplectic: $\mathbf{E}^\intercal\boldsymbol\gamma \mathbf{E}=\boldsymbol\gamma$. 
\item Set $\mathbf{N}=\mathbf{E}(\mathbf{1}+i\boldsymbol\beta)/\sqrt{2}$, and 
$\tilde{\mathbf{\Sigma}}=\mathbf{1}^{I}\epsilon_{I}+\mathbf{1}^{II}\epsilon_{II}+\mathbf{1}^{III}\epsilon_{III}$. 
\end{enumerate}
\end{minipage}
}

In step 3, the real transpose, not the conjugate transpose, must be used. \\
In step 4, $\mathbf{N}$ should come out real.

In appendix A I give a short Octave \cite{Octave} test program (it should run on Matlab, too) that will apply these steps to a random beam matrix.

\newpage

\subsection{Invariance group of the bunch matrix}

Which transformations $\mathbf{I}$ leave the bunch matrix $\mathbf{\Sigma}$ invariant: $\mathbf{I}\mathbf{\Sigma}\mathbf{I}^\intercal= \mathbf{\Sigma}$ ? 

\fbox{
\begin{minipage}{0.98\linewidth}
\vspace{5pt}
If the bunch matrix is normalized, the invariance transformations are phase rotations in the 3 subspaces $(x,x')$, $(y,y')$, $(z,z')$: 
\begin{align}
&\mathbf{I}\left(\psi^{I},\psi^{II},\psi^{III}\right)=\exp\left(\boldsymbol\gamma^{I}\psi^{I}+\boldsymbol\gamma^{II}\psi^{II}+\boldsymbol\gamma^{III}\psi^{III}\right)  \nonumber\\ 
&\text{with } \psi^{I}, \psi^{II}, \psi^{III} \text{ arbitrary} \nonumber
\end{align}
If the bunch matrix is not normalized (which is the general case), let $\mathbf{N}$ be a normalizing transformation. The group of invariance transformations is then given by 
\begin{align}
&\mathbf{I}\left(\psi^{I},\psi^{II},\psi^{III}\right)=\mathbf{N}\exp\left(\boldsymbol\gamma^{I}\psi^{I}+\boldsymbol\gamma^{II}\psi^{II}+\boldsymbol\gamma^{III}\psi^{III}\right)\mathbf{N}^{-1}
\nonumber \\ &\text{with } \psi^{I}, \psi^{II}, \psi^{III} \text{ arbitrary} \nonumber 
\end{align}
\vspace{-10pt}
\end{minipage}
}

\newpage
\section{Matrices and Clifford algebras}

\subsection{Definitions}

The real $d \times d$ matrices form a \emph{non-commutative ring}: you can add and multiply the matrices. Nowadays mathematicians prefer to talk of an  \emph{associative algebra}: the additional property is scalar multiplication, which is the same as multiplication with a multiple of the unit matrix. 

A $d \times d$ matrix $\mathbf{Z}=\left( Z_{mn} \right)$ can be written in vector notation as 
\begin{equation}
\mathbf{Z}=\mathbf{E}^{(1,1)}Z_{11}+\mathbf{E}^{(1,2)}Z_{12}+ \ldots + \mathbf{E}^{(d,d-1)}Z_{d,d-1}+\mathbf{E}^{(d,d)}Z_{d,d}
\end{equation}
Here, the basis of the $d \times d$ matrix algebra is $E^{(k,l)}_{m,n}=\delta_{k,m} \delta_{l,n}$. 

According to the problem at hand, there may be a better choice of the basis, one that simplifies calculation. Particularly attractive are basis matrices that square to $\pm \mathbf{1}$. The general matrix then has the form
\begin{equation}
\mathbf{Z}=\mathbf{1}Z_0+\boldsymbol\beta_1Z_1+ \ldots + \boldsymbol\beta_KZ_K+\boldsymbol\gamma_{1}Z_{K+1}+\ldots + \boldsymbol\gamma_L Z_{K+L}  \label{e3.01}
\end{equation}

\begin{itemize}
 \item $\mathbf{1}$ is the multiplicative identity, or real unit. \\Its representation is the unit matrix. 
 \item $\boldsymbol\beta$ is a bireal unit, i. e. it squares to $+1$: $\boldsymbol\beta^2=\mathbf{1}$. \\Its real representation is a symmetric, traceless matrix.
 \item $\boldsymbol\gamma$ is a complex unit, i. e. it squares to $-1$: $\boldsymbol\gamma^2=-\mathbf{1}$. \\Its real representation is an anti-symmetric (and therefore traceless) matrix.
\end{itemize}

The real representation of real, bireal and complex units is orthogonal, i. e. the transpose of a representative matrix is its inverse. 

Now let
\begin{equation}
\mathbf{Z}^\intercal=\mathbf{1}Z_0+\boldsymbol\beta_1Z_1+\ldots+ \boldsymbol\beta_MZ_M -\boldsymbol\gamma_{1}Z_{M+1}\ldots - \boldsymbol\gamma_N Z_{M+N}
\end{equation}
the transpose of $\mathbf{Z}$. 

Also, let 
\begin{equation}
\Re \mathbf{Z}=Z_0
\end{equation}
the real part of $\mathbf{Z}$. 

If $\mathbf{Z}$ is a $d \times d$ matrix, then 
\begin{equation}
\Re \mathbf{Z}=\tfrac{1}{d}\Tr \mathbf{Z}
\end{equation}
since all imaginary (bireal and complex) units have trace 0.

It is easy to verify that 
\begin{equation}
\left( \mathbf{A} | \mathbf{B} \right) \equiv \Re\left(\mathbf{A}^\intercal \mathbf{B}\right)
\end{equation}
is a scalar product. What's more, all the units form an orthogonal basis with respect to this scalar product. Therefore, the remaining coefficients of $\mathbf{Z}$ are
\begin{equation}
Z_k= 
\Re\left(\boldsymbol\beta_k^\intercal \mathbf{Z} \right) \quad \left(1 \leq k \leq K \right) \qquad
Z_{K+l}= 
\Re\left(\boldsymbol\gamma_l^\intercal \mathbf{Z} \right) \quad \left(1 \leq l \leq L \right) 
\end{equation}

Let us distinguish between an abstract algebra and its matrix representation. An element of an abstract algebra also looks like eqn. \eqref{e3.01}. The addition of two algebra elements given in this form is straightforward; multiplication is specified with a multiplication table of the units $\mathbf{1}, \boldsymbol\beta_k, \boldsymbol\gamma_l$. The matrix representation of the units together with matrix multiplication must reproduce this table. 

\subsection{Example: the complex numbers}

For example, take the algebra of complex numbers $\mathbf{Z}=\mathbf{1}Z_0+\boldsymbol\gamma Z_1$, with $Z_0$ and $Z_1$ real. \\

\parbox[t][4em][s]{0.5\linewidth}{
The multiplication table is 
\vfill
\bgroup
\def\arraystretch{1.2}
\begin{tabular}{| l | r r |}
\hline
$\cdot$  & $\mathbf{1}$  & $\boldsymbol\gamma$   \\
\hline
$\mathbf{1}$  & $\mathbf{1}$  & $\boldsymbol\gamma$   \\
$\boldsymbol\gamma$  & $\boldsymbol\gamma$  & $-\mathbf{1}$  \\
\hline
\end{tabular}
\egroup
}
\parbox[t][4.75em][s]{0.4\linewidth}{
To save space, let's truncate this to 
\vfill
\bgroup
\def\arraystretch{1.2}
\begin{tabular}{| r r | }
\hline
$\mathbf{1}$  & $\boldsymbol\gamma$   \\
$\boldsymbol\gamma$  & $-\mathbf{1}$  \\
\hline
\end{tabular}
\egroup
}

\vspace{5pt}
The multiplication of two complex numbers is the familiar 
\begin{equation*}
\mathbf{A}\cdot\mathbf{B}=(\mathbf{1}A_0+\boldsymbol\gamma A_1)(\mathbf{1}B_0+\boldsymbol\gamma B_1)=\mathbf{1}(A_0 B_0-A_1 B_1)+\boldsymbol\gamma(A_0 B_1+A_1 B_0)
\end{equation*}

The real representation of $\mathbf{Z}$ is a $2 \times 2$ matrix, e. g.
\begin{equation}
\mathbf{Z}=\mathbf{1}Z_0+\boldsymbol\gamma Z_1 \cong
\left( \begin{array}{rr}
Z_0 & Z_1 \\
-Z_1 & Z_0
\end{array}\right)
\end{equation}

Matrix multiplication now reproduces the abstract multiplication law. 

\subsection{Clifford algebras}

Here is a list of the simplest associative algebras.

\begin{itemize}
\item The only algebra with 1 unit is the real numbers.
\item There are two algebras with 2 units: the bireal (or split-complex) numbers $\mathbf{Z}=\mathbf{1}X+\boldsymbol\beta Y$, and the complex numbers $\mathbf{Z}=\mathbf{1}X+\boldsymbol\gamma Y$. 
\item There are no algebras with 3 units. 
\item Algebras with 4 units are the Cockle (or split-) quaternions $\mathbf{Z}=\mathbf{1}Z_0+\boldsymbol\beta_1 Z_1+  \boldsymbol\beta_2 Z_2+\boldsymbol\gamma Z_3$, the tessarines $\mathbf{Z}=\mathbf{1}Z_1+\boldsymbol\beta Z_2+ \boldsymbol\gamma_{1}Z_3+ \boldsymbol\gamma_2 Z_4$, and the Hamilton quaternions: $\mathbf{Z}=\mathbf{1}Z_0+\boldsymbol\gamma_{1}Z_1+\boldsymbol\gamma_{2}Z_2+ \boldsymbol\gamma_3 Z_3$.  
\end{itemize}

How can we construct associative algebras in a systematic way? There are basically two ways to do this: by forming the tensor product of two algebras, and with a scheme due to W. K. Clifford. I will present the latter, since it automatically classifies the algebra. 

\newpage
\cite{LaMi}, \cite{FoF} Clifford algebras over a $\mathbb{R}$ are denoted by $Cl_{p,q}(\mathbb{R})$. The units and multiplication table are obtained as follows.

\begin{minipage}[c]{0.6 \linewidth}
1. Start with the real unit $\mathbf{1}$ and the $N=p+q$ generator units $\boldsymbol\beta_1 \ldots \boldsymbol\beta_p, \boldsymbol\gamma_1 \ldots \boldsymbol\gamma_q$. The symmetric part of the multiplication table of these units is given to the right. Any two different generators anti-commute!
\end{minipage}
\hspace{0.03 \linewidth}
\begin{minipage}[c]{0.35 \linewidth}
\bgroup
\def\arraystretch{1.2}
\begin{tabular}{| l | c c c |}
\hline
$\tfrac{1}{2}\{\,,\}$  & $\mathbf{1}$  & $\boldsymbol\beta_l$  &  $\boldsymbol\gamma_l$  \\
\hline
$\mathbf{1}$ & $\mathbf{1}$ & $\boldsymbol\beta_l$  & $\boldsymbol\gamma_n$  \\
$\boldsymbol\beta_k$  & $\boldsymbol\beta_k$ & $\delta_{kl}\mathbf{1}$ & $\mathbf{0}$ \\
$\boldsymbol\gamma_k$  & $\boldsymbol\gamma_k$ & $\mathbf{0}$ & $-\delta_{mn}\mathbf{1}$ \\
\hline
\end{tabular}
\egroup
\end{minipage}

2. Define $N(N-1)/2$ new units as product of two different generators: $\boldsymbol\gamma_{q+1} \equiv \boldsymbol\beta_1\boldsymbol\beta_2=-\boldsymbol\beta_2\boldsymbol\beta_1, 
\boldsymbol\gamma_{q+2} \equiv \boldsymbol\gamma_1\boldsymbol\gamma_2=-\boldsymbol\gamma_2\boldsymbol\gamma_1, \ldots, \boldsymbol\beta_{p+1} \equiv \boldsymbol\beta_1\boldsymbol\gamma_1=-\boldsymbol\gamma_1\boldsymbol\beta_1,\ldots$

3. Define $N(N-1)(N-2)/6$ new units as product of three different generators.

$\ldots$

$N$. Define the last new unit $\boldsymbol\omega=\boldsymbol\beta_1 \ldots \boldsymbol\beta_p  \boldsymbol\gamma_1 \ldots \boldsymbol\gamma_q$.

The full multiplication table of \emph{all} the $2^N$ units now follows from the symmetric multiplication table of the $N$ generator units. 

\subsection{Matrix representations}

Here is a table of irreducible matrix representations of Clifford algebras $Cl_{p,q}$ with $p, q \leq 4$.

\begin{table}[h!]
\begin{tabular}{ l | c | c | c | c | c  }
$Cl_{p\downarrow,q \rightarrow}$ & $0$ & $1$ & $2$ & $3$ & $4$  \\
\hline
$0$ & $\mathbb{R}$  & $\mathbb{C}$ & $\mathbb{H}$ & $\mathbb{H} \oplus \mathbb{H}$ & $\mathbb{H}(2)$ \\
\hline
$1$ & $\mathbb{R} \oplus \mathbb{R}$ & $\mathbb{R}(2)$ & $\mathbb{C}(2)$ & $\mathbb{H}(2)$ & $\color{gray}\mathbb{H}(2) \oplus \mathbb{H}(2) $  \\
\hline
$2$ & $\mathbb{R}(2)$ & $\mathbb{R}(2) \oplus \mathbb{R}(2)$ & $\mathbb{R}(4)$ & $\color{gray}\mathbb{C}(4)$ & $\color{gray}\mathbb{H}(4)$ \\
\hline
$3$ & $\mathbb{C}(2)$ & $\mathbb{R}(4)$ & $\color{gray}\mathbb{R}(4) \oplus \mathbb{R}(4)$ & $\color{gray}\mathbb{R}(8)$ & $\color{gray}\mathbb{C}(8)$ \\
\hline
$4$ & $\mathbb{H}(2)$  & $\color{gray}\mathbb{C}(4)$ & $\color{gray}\mathbb{R}(8)$ &\color{gray} $\mathbb{R}(8) \oplus \mathbb{R}(8)$ & $\color{gray}\mathbb{R}(16)$ \\
\end{tabular}
\end{table}

It turns out that every Clifford algebra is isomorphic to 
\begin{itemize}
\item a full $d \times d$ real matrix algebra $\mathbb{R}(d)$, or
\item a full $d \times d$ complex matrix algebra $\mathbb{C}(d)$, or
\item a full $d \times d$ quaternionic matrix algebra $\mathbb{H}(d)$, or 
\item to the direct sum of two such algebras. 
\end{itemize}

It also appears that some Clifford algebras with different $p, q$, but the same overall number of generators $p+q$, are isomorphic, i. e. their multiplication tables can be brought to coincide after re-numbering the units (sign-changes may be necessary, too).

We are interested in the algebra of the real $2 \times 2$ matrices $\mathbb{R}(2)$, and of the real $4 \times 4$ matrices $\mathbb{R}(4)$. Both algebras can be constructed with Clifford's scheme: 
\begin{equation*}
\mathbb{R}(2) \cong Cl_{2,0}(\mathbb{R}) \cong  Cl_{1,1}(\mathbb{R}) \qquad \qquad \mathbb{R}(4) \cong Cl_{3,1}(\mathbb{R}) \cong Cl_{2,2}(\mathbb{R})
\end{equation*}

The multiplication table and representation of these two algebras is given in sections 4 and 5. For the other Clifford algebras see ref. \cite{Mueller}. 

\newpage
\section{One particle degree of freedom: linear optics with real Pauli matrices}

\subsection{Cockle quaternions}

The real $2 \times 2$ matrices  $\mathbb{R}(2)$ are a representation of the Clifford-Algebras $Cl_{2,0}(\mathbb{R})$ and $Cl_{1,1}(\mathbb{R})$. This is the algebra of the Cockle quaternions (or split-quaternions).

Units of $Cl_{2,0}(\mathbb{R})$: $\mathbf{1}, \color{myred}\boldsymbol\beta_1, \boldsymbol\beta_2, \color{black}\boldsymbol\gamma=\boldsymbol\beta_1\boldsymbol\beta_2$

Units of $Cl_{1,1}(\mathbb{R})$: $\mathbf{1}, \color{myred}\boldsymbol\beta_1, \boldsymbol\gamma, \color{black}\boldsymbol\beta_2=\boldsymbol\beta_1\boldsymbol\gamma$

\subsubsection*{Multiplication table}

\bgroup
\def\arraystretch{1.2}
\begin{tabular}{ | c c c c | }
\hline
$\mathbf{1}$ & $\boldsymbol\beta_1$  & $\boldsymbol\beta_2$ & $\boldsymbol\gamma$  \\
$\boldsymbol\beta_1$  & $\mathbf{1}$ & $\color{myblue}\boldsymbol\gamma$ & $\color{myblue}\boldsymbol\beta_2$ \\
$\boldsymbol\beta_2$ & $\color{myblue}-\boldsymbol\gamma$  & $\mathbf{1}$ & $\color{myblue}-\boldsymbol\beta_1$  \\
$\boldsymbol\gamma$ & $\color{myblue}-\boldsymbol\beta_2$  & $\color{myblue}\boldsymbol\beta_1$ & $-\mathbf{1}$  \\
\hline
\end{tabular}
\egroup
\hspace{50pt}\color{myblue}Blue: \color{black}anti-commuting products.

\subsubsection*{General Cockle quaternion and its real representation}

\begin{equation}
\mathbf{Z}=\mathbf{1}Z_0+\boldsymbol\beta_1Z_1+\boldsymbol\beta_2Z_2+\boldsymbol\gamma Z_3 \cong
\left( \begin{array}{rr}
Z_0+Z_1 & Z_2+Z_3\\
Z_2-Z_3 & Z_0-Z_1
\end{array}\right)
\end{equation} 

\subsubsection*{Remarks} 

\begin{itemize}
\item There are 1 real ($\mathbf{1}$), 2 bireal ($\boldsymbol\beta_1$, $\boldsymbol\beta_2$) and 1 complex ($\boldsymbol\gamma$) units. \\They are represented by the "real Pauli matrices" $\sigma_0$, $\sigma_3$, $\sigma_1$, $i\sigma_2$.
\item The general symmetric $2 \times 2$ matrix is $\enspace 
\mathbf{\Sigma}=\mathbf{1}\Sigma_0+\boldsymbol\beta_1\Sigma_1+\boldsymbol\beta_2\Sigma_2$.
\item The general skew-symmetric $2 \times 2$ matrix is 
$\enspace \mathbf{A}=\boldsymbol\gamma A_3 \enspace $.
\item The general diagonal matrix is $\enspace \mathbf{D}=\mathbf{1}D_0+\boldsymbol\beta_1 D_1 \enspace $. 
\item $\boldsymbol\gamma$ is the symplectic form.
\end{itemize}

\subsubsection*{Determinant and Eigenvalues} 
\begin{equation}
\det\mathbf{Z}=\mathbf{Z}\bar{\mathbf{Z}}=(Z_0)^2-Z_1^2-Z_2^2+Z_3^2 \quad  \qquad\lambda_\pm=Z_0 \pm \sqrt{Z_1^2+Z_2^2-Z_3^2}
\end{equation}

\subsubsection*{Inverse matrix}

\begin{equation}
\mathbf{Z}^{-1}=\dfrac{\bar{\mathbf{Z}}}{\mathbf{Z}\bar{\mathbf{Z}}}=\dfrac{\mathbf{1}Z_0-\boldsymbol\beta_1 Z_1-\boldsymbol\beta_2 Z_2-\boldsymbol\gamma Z_3}{Z_0^2-Z_1^2-Z_2^2+Z_3^2}
\end{equation}

\subsubsection*{Vector notation}

In the following it will be helpful to combine the part $\boldsymbol\beta_1Z_1+\boldsymbol\beta_2Z_2$ of $\mathbf{Z}$ to a vector: 
\begin{equation}
\mathbf{Z}=\mathbf{1}Z_0+\vec{\boldsymbol\beta}\cdot\vec{Z}+\boldsymbol\gamma Z_3 \qquad \text{where} \qquad
\vec{\boldsymbol\beta}=
\left( \begin{array}{rr}
\boldsymbol\beta_1 \\ \boldsymbol\beta_2
\end{array}\right)
\qquad \text{and} \qquad 
\vec{Z}=
\left( \begin{array}{rr}
Z_1 \\ Z_2
\end{array}\right)
\end{equation}

\subsection{The beam matrix}

The general beam matrix is
\begin{equation}
\mathbf{\Sigma}=\mathbf{1}\Sigma_0+\vec{\boldsymbol\beta}\cdot\vec{\Sigma} \cong
\left( \begin{array}{r r}
\Sigma_0+\Sigma_1 & \Sigma_2\\
\Sigma_2 & \Sigma_0-\Sigma_1
\end{array}\right)
\end{equation}

The beam emittance is  
\begin{equation}
\epsilon=\sqrt{\det\left(\mathbf{\Sigma}\right)}
=\sqrt{\Sigma_0^2-\vec{\Sigma}^2}
\end{equation}

A beam matrix in normal form has vanishing $\vec{\Sigma}$, and is
\begin{equation}
\tilde{\mathbf{\Sigma}}=\mathbf{1}\epsilon \cong
\left( \begin{array}{c c}
\epsilon & 0\\
0 & \epsilon
\end{array}\right)
\end{equation}

\subsection{Symplectic transformations}

The symplectic condition $\mathbf{R}\boldsymbol\gamma\mathbf{R}^\intercal=\boldsymbol\gamma$ translates to $\det\mathbf{R}=(R_0)^2-R_1^2-R_2^2+R_3^2=1$. 

This formula is not really helpful for constructing symplectic transformations. However, it is easy to write down the logarithm of a symplectic transformation:
\begin{equation}
\ln\mathbf{R}\equiv\mathbf{r}=\vec{\boldsymbol\beta}\cdot\vec{r}+\boldsymbol\gamma r_3
\end{equation}
This is a $\boldsymbol\gamma$-symmetric matrix, see eqn. \eqref{e2.01}. 

\subsection{Elementary symplectic transformations}

The RHS of eqn. \eqref{e2.01} is a linear combination of the "units" $\vec{\boldsymbol\beta}\cdot\vec{e}$ (squares to $+1$, since $\vec{e}$ is a unit vector) and $\boldsymbol\gamma$ (squares to $-1$). These units generate the "elementary transformations":
\begin{align}
&\text{Boosts: } &&\mathbf{R}_\beta &&=\exp \left( \vec{\boldsymbol\beta}\cdot\vec{e} \chi \right) &&=\mathbf{1}\cosh\chi+ \vec{\boldsymbol\beta}\cdot\vec{e}\sinh\chi \qquad \\
&\text{Rotations: } &&\mathbf{R}_\gamma &&=\exp \left( \boldsymbol\gamma \psi \right) &&=\mathbf{1}\cos\psi+ \boldsymbol\gamma\sin\psi \qquad 
\end{align}

\subsubsection*{Boosting $\mathbf{\Sigma}$}

To describe the action of a boost along the unit vector $\vec{e}=(e_1,e_2)$, let 
\begin{equation}
\vec{\Sigma}=\vec{e} \Sigma_\parallel + \vec{n} \Sigma_\perp
\end{equation}
where
\begin{equation}
\vec{e}= \left( \begin{array}{c} e_1 \\ e_2 \end{array}\right) \qquad  
\vec{n}= \left( \begin{array}{c} -e_2 \\ e_1 \end{array}\right) \qquad 
\Sigma_\parallel=\vec{e}\cdot\vec{\Sigma} \qquad \Sigma_\perp=\vec{n}\cdot\vec{\Sigma}=\vec{e}\wedge\vec{\Sigma}
\end{equation}

The action of a Boost on $\mathbf{\Sigma}$ is :
\begin{align*}
\mathbf{\Sigma}'&=\mathbf{R}_\beta\mathbf{\Sigma}\mathbf{R}_\beta^\intercal \\
&=\exp \left( \vec{\boldsymbol\beta}\cdot\vec{e} \chi \right) \left( 1\Sigma_0+\vec{\boldsymbol\beta}\cdot\vec{\Sigma} \right) \exp \left( \vec{\boldsymbol\beta}\cdot\vec{e} \chi \right) \\
&=1\left( \cosh(2\chi) \Sigma_0 + \sinh(2\chi) \Sigma_\parallel \right)+ \vec{\boldsymbol\beta}\cdot \left[ \vec{e} \left( \cosh(2\chi) \Sigma_\parallel +   \sinh(2\chi) \Sigma_0 \right)+\vec{n} \Sigma_\perp \right]
\end{align*}

\begin{figure}[ht]
\begin{minipage}[c]{0.5 \linewidth}
\centering
\includegraphics[trim = 0pt 0pt 10pt 0pt, width=1\textwidth]{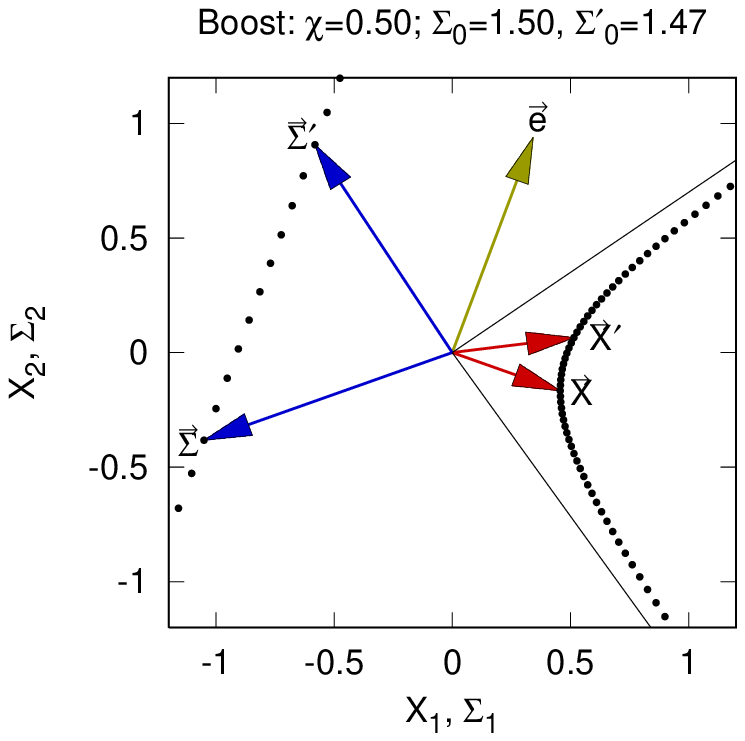}
\end{minipage}
\hspace{0.03 \linewidth}
\begin{minipage}[c]{0.45 \linewidth}
The components of the transformed beam matrix are 
\begin{align}
\Sigma'_0 &= \cosh(2\chi) \Sigma_0 + \sinh(2\chi) \Sigma_\parallel \nonumber \\
\Sigma'_\parallel &=  \cosh(2\chi) \Sigma_\parallel + \sinh(2\chi) \Sigma_0  \nonumber \\ \Sigma'_\perp &= \Sigma_\perp
\end{align}

\vspace{10pt}
\caption{ }
$\mathbf{R}_\beta=\exp  \vec{\boldsymbol\beta}\cdot\vec{e} \chi$ boosts a phase vector $\vec{X}$ on a hyperbola with $\vec{e}$ as the conjugate axis, and boosts the vector part of a beam matrix $\mathbf{\Sigma}=1\Sigma_0+\vec{\boldsymbol\beta}\cdot\vec{\Sigma}$ in the direction of $\vec{e}$.
\end{minipage}
\end{figure}

The boost invariants are  
\begin{equation}
\Sigma_0 ^2-\vec{\Sigma}^2 \qquad \text{and} \qquad \Sigma_\perp
\end{equation}

Boosts parallel to $\vec{\Sigma}$, i. e. of the form 
\begin{equation}
\mathbf{R}=\exp \left( \vec{\boldsymbol\beta}\cdot\vec{e} \chi \right)
\qquad \qquad \vec{e}=\frac{\vec{\Sigma}}{|\vec{\Sigma}|}
\end{equation}
leave the direction of $\vec{\Sigma}$ invariant. If the boost angle is given by 
\begin{equation}
\tanh (2\chi)=-\frac{|\vec{\Sigma}|}{\Sigma_0}=-\frac{\sqrt{\Sigma_1^2+\Sigma_2^2}}{\Sigma_0}
\end{equation} 
then $\vec{\Sigma}'$ vanishes.

\subsubsection*{Rotating $\mathbf{\Sigma}$}

The action of a rotation on $\mathbf{\Sigma}$ is (see also figure 2):
\begin{align*}
\mathbf{\Sigma}'&=\mathbf{R}_\gamma\mathbf{\Sigma}\mathbf{R}_\gamma^\intercal \\
&=\exp \left( \boldsymbol\gamma \psi \right) \left(\mathbf{1} \Sigma_0 +  \vec{\boldsymbol\beta}\cdot\vec{\Sigma} \right) \exp \left( -\boldsymbol\gamma \psi \right) \\
&=\mathbf{1}\Sigma_0 + \vec{\boldsymbol\beta}\cdot \left[ \vec{\Sigma} \cos(2\psi) + \boldsymbol\epsilon \vec{\Sigma}  \sin(2\psi)\right] 
\end{align*}
Here, $\boldsymbol\epsilon$ is the skew-symmetric tensor (it looks like $\boldsymbol\gamma$), and $\boldsymbol\epsilon \vec{\Sigma}=(\Sigma_2,-\Sigma_1)^\intercal$. 

\begin{figure}[ht]
\begin{minipage}[c]{0.5 \linewidth}
\centering
\includegraphics[trim = 0pt 0pt 10pt 0pt, width=1\textwidth]{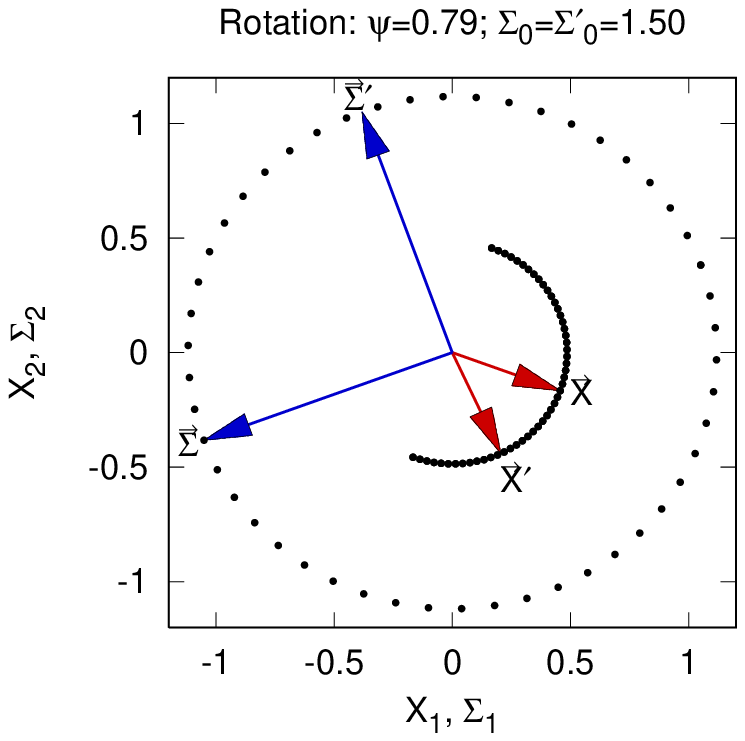}
\end{minipage}
\hspace{0.03 \linewidth}
\begin{minipage}[c]{0.45 \linewidth}
The components of the transformed beam matrix are 
\begin{align}
\Sigma'_0 &= \Sigma_0 \\
\vec{\Sigma}'&=\vec{\Sigma} \cos(2\psi) + \boldsymbol\epsilon \vec{\Sigma}  \sin(2\psi) \nonumber 
\end{align}

\vspace{10pt}
\caption{ }

$\mathbf{R}_\gamma=\exp  \boldsymbol\gamma\psi$ rotates a phase vector $\vec{X}$ clockwise by the angle $\psi$, and rotates the vector part of a beam matrix $\mathbf{\Sigma}=\mathbf{1}\Sigma_0+\vec{\boldsymbol\beta}\cdot\vec{\Sigma}$ clockwise by the angle $2\psi$, while leaving the scalar part invariant.
\end{minipage}
\end{figure}

The rotational invariants are 
\begin{equation}
\Sigma_0 \qquad \text{and} \qquad \vec{\Sigma}^2
\end{equation}

\subsection{Diagonalizing the beam matrix}

The beam matrix is diagonal when $\Sigma_{12}=\Sigma_{21}=0$ (standard matrix basis), or when $\Sigma_{2}= 0$ (real Pauli matrix basis). The two phase space coordinates $x$ and $x'$ are then uncorrelated, or decoupled. 

\fbox{
\begin{minipage}{0.98\linewidth}
\vspace{5pt}
To diagonalize a beam matrix $\mathbf{\Sigma}$, decompose $\mathbf{\Sigma}=\mathbf{1}\Sigma_0+\vec{\boldsymbol\beta}\cdot\vec{\Sigma}$. Rotate $\mathbf{\Sigma}$ with 
\begin{equation*}
\mathbf{R}_\gamma=\exp \left( \boldsymbol\gamma \psi_1 \right) \qquad \qquad \tan (2\psi_1) =\frac{\Sigma_2}{\Sigma_1}
\end{equation*}
The new beam matrix   is diagonal: $\mathbf{\Sigma}'=\mathbf{R}_\gamma\mathbf{\Sigma}\mathbf{R}_\gamma^\intercal=\mathbf{1}\Sigma_0'+\boldsymbol\beta_1\Sigma_1'$. 

\vspace{5pt}
The diagonal form is preserved under the group of boosts 
$\mathbf{R}_\beta=\exp \left( \boldsymbol\beta_1 \chi \right)$ (free parameter $\chi$).
\end{minipage}
}

\subsection{Normalizing the beam matrix}

The beam matrix is normal when $\Sigma_{12}=\Sigma_{21}= 0$ and $\Sigma_{11}=\Sigma_{22}=\epsilon$ (standard matrix basis), or when $\Sigma_{1}=\Sigma_{2}= 0$ (real Pauli matrix basis). 

\fbox{
\begin{minipage}{0.98\linewidth}
\vspace{5pt}
To normalize a beam matrix $\mathbf{\Sigma}$, first diagonalize it to $\mathbf{\Sigma}'=\mathbf{1}\Sigma_0'+\boldsymbol\beta_1\Sigma_1'$, see section 4.5. \\
Then boost $\mathbf{\Sigma}'$ with
\begin{equation*}
\mathbf{R}_\beta=\exp \left( \boldsymbol\beta_1 \chi \right) \qquad \qquad \tanh (2\chi)=-\frac{\Sigma_1'}{\Sigma_0'}
\end{equation*}
The new beam matrix is in normal form: $\mathbf{\Sigma}''=\mathbf{R}_\beta\mathbf{\Sigma}'\mathbf{R}_\beta^\intercal=\mathbf{1}\epsilon$.

\vspace{5pt}
The normal form is preserved under the group of rotations 
$\mathbf{R}_\gamma=\exp\left(\boldsymbol\gamma\psi \right)$ \\(free parameter $\chi$).
\vspace{5pt}
\end{minipage}
}

The normalization can also be done without the intermediate diagonalization, as follows. 

\fbox{
\begin{minipage}{0.98\linewidth}
\vspace{5pt}
To normalize a beam matrix $\mathbf{\Sigma}$, decompose $\mathbf{\Sigma}=\mathbf{1}\Sigma_0+\vec{\boldsymbol\beta}\cdot\vec{\Sigma}$. Boost $\mathbf{\Sigma}$ with
\begin{equation*}
\mathbf{R}_\beta=\exp \left( \vec{\boldsymbol\beta} \cdot \vec{e} \chi \right) \qquad \qquad \vec{e}=\frac{\vec{\Sigma}}{|\vec{\Sigma}|} \qquad \qquad \tanh (2\chi)=-\frac{|\vec{\Sigma}|}{\Sigma_0}
\end{equation*}
The new beam matrix is in normal form: $\mathbf{\Sigma}'=\mathbf{R}_\beta\mathbf{\Sigma}\mathbf{R}_\beta=\mathbf{1}\epsilon$.
\vspace{5pt}
\end{minipage}
}

\subsection{Invariance group of the beam matrix}

Which transformations $\mathbf{I}$ leave the beam matrix $\mathbf{\Sigma}$ invariant: $\mathbf{I}\mathbf{\Sigma}\mathbf{I}^\intercal= \mathbf{\Sigma}$ ? 

\fbox{
\begin{minipage}{0.98\linewidth}
\vspace{5pt}
Let $\mathbf{N}$ be a normalizing transformation to a beam matrix $\mathbf{\Sigma}$, i. e.  $\mathbf{\Sigma}=\mathbf{N}\mathbf{1}\epsilon\mathbf{N}^\intercal$; the beam matrix is invariant under the group of transformations
$\mathbf{I}\left(\psi\right)=\mathbf{N}\exp\left(\boldsymbol\gamma\psi\right)\mathbf{N}^{-1}$ (free parameter $\psi$).
\vspace{5pt}
\end{minipage}
}

\newpage
\section{Two particle degrees of freedom: linear optics with real Dirac matrices}

\subsection{The Clifford algebra $Cl_{3,1}(\mathbb{R})$}
%\subsection{The Clifford algebra Cl(3,1) = Cl(2,2)}

The real $4 \times 4$ matrices  $\mathbb{R}(2)$ are a representation of the Clifford-Algebra $Cl_{3,1}(\mathbb{R})\cong$ $Cl_{2,2}(\mathbb{R})$, see section 3.4. The 16 units of this algebra can be grouped into 4 "scalars": $\mathbf{1}$, $\boldsymbol\gamma^1$, $\boldsymbol\gamma^2$, $\boldsymbol\gamma^3$, and 4 "3-vectors": $\boldsymbol\zeta_k \rightarrow \vec{\boldsymbol\zeta}$, $\boldsymbol\beta^1_k \rightarrow \vec{\boldsymbol\beta}\mathstrut^1$, $\boldsymbol\beta^2_k \rightarrow \vec{\boldsymbol\beta}\mathstrut^2$, $\boldsymbol\beta^3_k \rightarrow \vec{\boldsymbol\beta}\mathstrut^3$, with $k = 1,2,3$.

Units of $Cl_{3,1}(\mathbb{R})$: 

\small
$\mathbf{1} \quad \enspace \color{myred}\boldsymbol\beta^2_k,  \boldsymbol\gamma^1 \quad \enspace \color{black} \boldsymbol\beta^3_k=\boldsymbol\beta^2_k\boldsymbol\gamma^1,  \boldsymbol\zeta_k=\tfrac{1}{2}\epsilon_{klm}\boldsymbol\beta^2_l\boldsymbol\beta^2_m \qquad  \boldsymbol\beta^1_k=
\boldsymbol\zeta_k\boldsymbol\gamma^1 \quad \enspace \boldsymbol\gamma^2=-\boldsymbol\beta^2_1\boldsymbol\beta^2_2\boldsymbol\beta^2_3 \quad \enspace \boldsymbol\gamma^3=\boldsymbol\gamma^2\boldsymbol\gamma^1$

\normalsize
Units of $Cl_{2,2}(\mathbb{R})$: 

\small
$\mathbf{1} \quad \enspace \color{myred}\boldsymbol\beta^1_2, \boldsymbol\beta^1_3, \boldsymbol\gamma^2, \boldsymbol\gamma^3 \quad \enspace \color{black}\boldsymbol\beta^2_2=\boldsymbol\beta^1_2\boldsymbol\gamma^3, \boldsymbol\beta^2_3=\boldsymbol\beta^1_3\boldsymbol\gamma^3, \boldsymbol\beta^3_2=-\boldsymbol\beta^1_2\boldsymbol\gamma^2,   \boldsymbol\beta^3_3=-\boldsymbol\beta^1_3\boldsymbol\gamma^3, \boldsymbol\zeta_1=\boldsymbol\beta^1_2\boldsymbol\beta^1_3,  \boldsymbol\gamma^1=\boldsymbol\gamma^2\boldsymbol\gamma^3$

$\boldsymbol\beta^2_1= \boldsymbol\beta^1_2\boldsymbol\beta^1_3\boldsymbol\gamma^2,
\boldsymbol\beta^3_1=\boldsymbol\beta^1_2\boldsymbol\beta^1_3\boldsymbol\gamma^3, \boldsymbol\zeta_2=\boldsymbol\beta^1_2\boldsymbol\gamma^2\boldsymbol\gamma^3, \boldsymbol\zeta_3=\boldsymbol\beta^1_3\boldsymbol\gamma^2\boldsymbol\gamma^3 \quad \enspace \boldsymbol\beta^1_1=-
\boldsymbol\beta^1_2\boldsymbol\beta^1_3\boldsymbol\gamma^2\boldsymbol\gamma^3$
\normalsize

\subsubsection*{Multiplication table}

\bgroup
\def\arraystretch{1.2}
\begin{tabular}{  | r r r r |}
\hline
$\mathbf{1}$ & $\boldsymbol\gamma^n$ & $\boldsymbol\zeta_l$ & $\boldsymbol\beta^n_l$\\
$\boldsymbol\gamma^m$ & $-\delta_{mn}\mathbf{1}\color{myblue}-\epsilon_{mno}\boldsymbol\gamma^o$ &  $\boldsymbol\beta^m_l$ & 
$-\delta_{mn}\boldsymbol\zeta_l\color{myblue}-\epsilon_{mno}\boldsymbol\beta^o_l$  \\
$\boldsymbol\zeta_k$ &  $\boldsymbol\beta^n_k$ & $-\delta_{kl}\mathbf{1}\color{myblue}-\epsilon_{kli}\boldsymbol\zeta_i$ & $-\delta_{kl}\boldsymbol\gamma_n\color{myblue}-\epsilon_{kli}\boldsymbol\beta^n_i$ \\
$\boldsymbol\beta^m_k$ & $-\delta_{mn}\boldsymbol\zeta_k\color{myblue}-\epsilon_{mno}\boldsymbol\beta^o_k$ & $-\delta_{kl}\boldsymbol\gamma_m\color{myblue}-\epsilon_{kli}\boldsymbol\beta^m_i$ & $\delta_{kl}\delta_{mn}\mathbf{1}\color{myblue}+\delta_{kl}\epsilon_{mno}\boldsymbol\gamma_o$ \\
& & & $\ldots\color{myblue}+\epsilon_{kli}\delta_{mn}\boldsymbol\zeta_i \color{black} +\epsilon_{kli}\epsilon_{mno}\boldsymbol\beta^o_i$ \\
\hline
\end{tabular}
\egroup

\color{myblue}Blue\color{black}: anti-commuting products.

\subsubsection*{General $Cl_{3,1}$-sedenion and its representation} 

Sedenion means "number with 16 components". These components are the coefficients that come with the units. Like the units, the coefficients can be grouped into 4 "scalars": $Z_0^0$, $Z_0^1$, $Z_0^2$, $Z_0^3$, and 4 "3-vectors": $Z_k^0 \rightarrow \vec{Z}\mathstrut^0$, $Z_k^1\rightarrow \vec{Z}\mathstrut^1$, $Z_k^2\rightarrow \vec{Z}\mathstrut^2$, $Z_k^3\rightarrow \vec{Z}\mathstrut^3$, with $k = 1,2,3$. In products of units and/or coefficients with indices, summation over equal indices is required.
\begin{equation}
\mathbf{Z}=\mathbf{1}Z_0^0 + \boldsymbol\zeta_k Z^0_k + \boldsymbol\gamma^m Z^m_0 + \boldsymbol\beta\mathstrut^m_k Z^m_k \equiv \mathbf{1}Z_0^0 + \vec{\boldsymbol\zeta}\cdot\vec{Z^0} + \boldsymbol\gamma^m Z^m_0 + \vec{\boldsymbol\beta}\mathstrut^m\!\cdot\vec{Z}^m \cong
\end{equation}
\small
\[ \left(
\begin{array}{ r r r r}
Z^0_0+Z^1_1+Z^2_2+Z^3_3 & -Z^0_1+Z_0^1-Z^2_3+Z^3_2 & -Z^0_2+Z^1_3+Z^2_0-Z^3_1 & -Z^0_3-Z^1_2+Z^2_1+Z^3_0 \\
Z^0_1-Z^1_0-Z^2_3+Z^3_2 & Z^0_0+Z^1_1-Z^2_2-Z^3_3 & Z^0_3+Z^1_2+Z^2_1+Z^3_0 & -Z^0_2+Z^1_3-Z^2_0+Z^3_1 \\
Z^0_2+Z^1_3-Z^2_0-Z^3_1 & -Z^0_3+Z^1_2+Z^2_1-Z^3_0 & Z^0_0-Z^1_1+Z^2_2-Z^3_3 & Z^0_1+Z^1_0+Z^2_3+Z^3_2 \\
Z^0_3-Z^1_2+Z^2_1-Z^3_0 & Z^0_2+Z^1_3+Z^2_0+Z^3_1 & -Z^0_1-Z^1_0+Z^2_3+Z^3_2 & Z^0_0-Z^1_1-Z^2_2+Z^3_3 \\
\end{array}
\right)\]
\normalsize

See also appendix B.1. 

\subsubsection*{Remarks} 

\begin{itemize}
\item There are 1 real ($\mathbf{1}$), 9 bireal ($\boldsymbol\beta^m_k$) and 6 complex ($\boldsymbol\zeta_k$, $\boldsymbol\gamma^m$) units ($k,m = 1,2,3$). They are represented by "real Dirac matrices".
\item The general symmetric $4 \times 4$ matrix is $\enspace 
\mathbf{\Sigma}=\mathbf{1}\Sigma_0^0 + \vec{\boldsymbol\beta}\mathstrut^1 \!\cdot \vec{\Sigma}^1 + \vec{\boldsymbol\beta}\mathstrut^2 \!\cdot \vec{\Sigma}^2 + \vec{\boldsymbol\beta}\mathstrut^3 \!\cdot \vec{\Sigma}^3 \enspace $.
\item The general skew-symmetric $4 \times 4$ matrix is 
$\enspace \mathbf{A}=\boldsymbol\zeta_k A^0_k + \boldsymbol\gamma^m A^m_0 \enspace $.
\item The general diagonal matrix is $\enspace \mathbf{D}=\mathbf{1}D_0^0+\boldsymbol\beta^1_1 D^1_1+\boldsymbol\beta^2_2D^2_2+\boldsymbol\beta^3_3D^3_3 \enspace$.
\item $\boldsymbol\gamma^1$ is the symplectic form. 
\item The 16 components $Z_\kappa^\lambda$ of the number $\mathbf{Z}$ are arranged as a $4 \times 4$ matrix, not as a vector. Don't confuse this matrix with the representative matrix $\left(Z_{\mu\nu}\right)$! 

Representative matrix of $\mathbf{Z}$:  \hspace{49pt} Component matrix of $\mathbf{Z}$:

$\left(Z_{\mu \nu}\right)=
\left(
\begin{array}{ r r r r}
Z_{11} & Z_{12} & Z_{13} & Z_{14} \\
Z_{21} & Z_{22} & Z_{23} & Z_{24} \\
Z_{31} & Z_{32} & Z_{33} & Z_{34} \\
Z_{41} & Z_{42} & Z_{43} & Z_{44} \\
\end{array}
\right)$ \hspace{30pt} 
$\left(Z_\kappa^\lambda\right)=
\left(
\begin{array}{ r r r r}
Z^0_0 & Z_0^1 & Z_0^2 & Z_0^3 \\
Z_1^0 & Z_1^1 & Z_1^2 & Z_1^3 \\
Z_2^0 & Z_2^1 & Z_2^2 & Z_2^3 \\
Z_3^0 & Z_3^1 & Z_3^2 & Z_3^3 \\
\end{array}
\right)$

\end{itemize}

\subsubsection*{Determinants}

The anti-symmetric $4 \times 4$ matrix $\enspace \mathbf{A}=\boldsymbol\zeta_k A^0_k + \boldsymbol\gamma^m A^m_0 \enspace$ has the determinant
\begin{equation}
\det{\mathbf{A}}=\left[(A^1_0)^2+(A^2_0)^2+(A^3_0)^2-(A_1^0)^2-(A_2^0)^2-(A_3^0)^2\right]^2
\end{equation}

Let $\qquad \: \, \mathbf{A}=\mathbf{\Sigma}\boldsymbol\gamma^1\mathbf{\Sigma}=\boldsymbol\zeta_k A^0_k + \boldsymbol\gamma^m A^m_0 \enspace$

where $\quad  \: \! A^1_0=\left(\Sigma^0_0\right)^2+ \vec{\Sigma}^1\!\cdot\vec{\Sigma}^1-\vec{\Sigma}^2\!\cdot\vec{\Sigma}^2-\vec{\Sigma}^3\!\cdot\vec{\Sigma}^3\quad A^2_0=2\vec{\Sigma}^2\!\cdot\vec{\Sigma}^1 \quad
A^3_0=2\vec{\Sigma}^3\!\cdot\vec{\Sigma}^1$

and $\qquad \vec{A}^0=2\left(\vec{\Sigma}^2\!\wedge\vec{\Sigma}^3 -\Sigma^0_0\vec{\Sigma}^1\right)$

The symmetric $4 \times 4$ matrix $\enspace \mathbf{\Sigma}=\mathbf{1}\Sigma_0^0 + \vec{\boldsymbol\beta}\mathstrut^m \!\cdot \vec{\Sigma}^m \enspace$ has the determinant
\begin{equation}
\det{\mathbf{\Sigma}}=\sqrt{\det\left(\mathbf{\Sigma}\boldsymbol\gamma^1\mathbf{\Sigma}\right)} = \sqrt{\det\mathbf{A}}= (A^1_0)^2+(A^2_0)^2+(A^3_0)^2-\vec{A}^0\!\cdot\vec{A}^0  \label{e5.02}
\end{equation}

Let $\qquad \: \, \mathbf{A}=\mathbf{Z}\boldsymbol\gamma^1\mathbf{Z}^\intercal=\boldsymbol\zeta_k A^0_k + \boldsymbol\gamma^m A^m_0 \enspace$

where $\quad  \: \! A^1_0=Z^0_\mu Z^0_\mu+Z^1_\mu Z^1_\mu-Z^2_\mu Z^2_\mu-Z^3_\mu Z^3_\mu$

$\qquad \qquad A^2_0=2\left(Z^2_\mu Z^1_\mu-Z^0_\mu Z^3_\mu\right) \quad
A^3_0=2\left(Z^0_\mu Z^2_\mu+Z^3_\mu Z^1_\mu\right)$

and $\qquad \vec{A}^0=2\left(Z^1_0\vec{Z}^0-Z^0_0\vec{Z}^1 + \vec{Z}^0\!\wedge\vec{Z}^1 + Z^3_0\vec{Z}^2-Z^2_0\vec{Z}^3 + \vec{Z}^2\!\wedge\vec{Z}^3\right)$

The general $4\times 4$ matrix $\enspace \mathbf{1}Z_0^0 + \vec{\boldsymbol\zeta}\cdot\vec{Z^0} + \boldsymbol\gamma^m Z^m_0 + \vec{\boldsymbol\beta}\mathstrut^m\!\cdot\vec{Z}^m \enspace$ has the determinant
\begin{equation}
\det\mathbf{Z}=\sqrt{\det\left(\mathbf{Z}\boldsymbol\gamma^1\mathbf{Z}^\intercal\right)} = \sqrt{\det\mathbf{A}}= (A^1_0)^2+(A^2_0)^2+(A^3_0)^2 - \vec{A}^0\!\cdot\vec{A}^0
\end{equation}

(Choosing $\boldsymbol\gamma^1$ instead of $\boldsymbol\gamma^2, \boldsymbol\gamma^3, \boldsymbol\zeta_1, \boldsymbol\zeta_2$ or $\boldsymbol\zeta_3$ for constructing a skew-symmetric matrix $\mathbf{A}$ out of $\mathbf{\Sigma}$ or $\mathbf{Z}$ is arbitrary. But only  $\mathbf{A}=\mathbf{\Sigma}\boldsymbol\gamma^1\mathbf{\Sigma}$ transforms in the same way as $\mathbf{\Sigma}$, to wit $\mathbf{A}'= \mathbf{R}\mathbf{A}\mathbf{R}^\intercal$.)

\newpage
\subsubsection*{Inverse matrix} 

The anti-symmetric $4 \times 4$ matrix $\quad \mathbf{A}=\boldsymbol\zeta_k A^0_k + \boldsymbol\gamma^m A^m_0 \quad$ has the inverse
\begin{equation}
\mathbf{A}^{-1}=\dfrac{ -\boldsymbol\gamma^m A^m_0 + \boldsymbol\zeta_k A^0_k }{ (A^1_0)^2+(A^2_0)^2+(A^3_0)^2-(A_1^0)^2-(A_2^0)^2-(A_3^0)^2 }
\end{equation}
The symmetric $4 \times 4$ matrix $\quad \mathbf{\Sigma}=\mathbf{1}\Sigma_0^0 + \vec{\boldsymbol\beta}\mathstrut^m \!\cdot \vec{\Sigma}^m \quad$ has the inverse
\begin{equation}
\mathbf{\Sigma}^{-1}=\boldsymbol\gamma^1\mathbf{\Sigma}\mathbf{A}^{-1}\qquad \quad  \text{where} \quad \qquad \mathbf{A}=\mathbf{\Sigma}\boldsymbol\gamma^1\mathbf{\Sigma}
\end{equation}
For the components of $\mathbf{A}$ in terms of $\mathbf{\Sigma}$ see the paragraph \emph{Determinants}.

The general $4\times 4$ matrix $\quad \mathbf{1}Z_0^0 + \vec{\boldsymbol\zeta}\cdot\vec{Z^0} + \boldsymbol\gamma^m Z^m_0 + \vec{\boldsymbol\beta}\mathstrut^m\!\cdot\vec{Z}^m \quad$ has the inverse
\begin{equation}
\mathbf{Z}^{-1}=\boldsymbol\gamma^1\mathbf{Z}^\intercal\mathbf{A}^{-1} \qquad \quad \! \text{where} \qquad \quad \mathbf{A}=\mathbf{Z}\boldsymbol\gamma^1\mathbf{Z}^\intercal
\end{equation}
For the components of $\mathbf{A}$ in terms of $\mathbf{Z}$ see the paragraph \emph{Determinants}.

\subsection{The beam matrix}

The general beam matrix is 
\begin{equation}
\mathbf{\Sigma}=\mathbf{1}\Sigma_0^0 + \vec{\boldsymbol\beta}\mathstrut^1 \!\cdot \vec{\Sigma}^1 + \vec{\boldsymbol\beta}\mathstrut^2 \!\cdot \vec{\Sigma}^2 + \vec{\boldsymbol\beta}\mathstrut^3 \!\cdot \vec{\Sigma}^3 
\cong
\end{equation}
\small
\begin{equation*}
\left(
\begin{array}{ r r r r}
\Sigma_0^0+\Sigma^1_1+\Sigma^2_2+\Sigma^3_3 & -\Sigma^2_3+\Sigma^3_2 & -\Sigma^3_1+\Sigma^1_3 & -\Sigma^1_2+\Sigma^2_1 \\
-\Sigma^2_3+\Sigma^3_2 & \Sigma_0^0+\Sigma^1_1-\Sigma^2_2-\Sigma^3_3 & \Sigma^1_2+\Sigma^2_1 & \Sigma^3_1+\Sigma^1_3 \\
-\Sigma^3_1+\Sigma^1_3 & \Sigma^1_2+\Sigma^2_1 & \Sigma_0^0-\Sigma^1_1+\Sigma^2_2-\Sigma^3_3 & \Sigma^2_3+\Sigma^3_2 \\
-\Sigma^1_2+\Sigma^2_1 & \Sigma^3_1+\Sigma^1_3 & \Sigma^2_3+\Sigma^3_2 & \Sigma_0^0-\Sigma^1_1-\Sigma^2_2+\Sigma^3_3 \\
\end{array}
\right)
\end{equation*}
\normalsize

The two emittances of $\mathbf{\Sigma}$ appear in the characteristic polynomial of $\mathbf{\Sigma}\boldsymbol\gamma^1$, see section 2.3:
\begin{equation}
\det\left(\mathbf{\Sigma}\boldsymbol\gamma^1-\lambda \mathbf{1}\right)=
\left(\lambda^2+\epsilon_{I}^{2} \right)
\left(\lambda^2+\epsilon_{II}^{2} \right)
\end{equation}
On the other hand we have 
\begin{equation}
\det\left(\mathbf{\Sigma}\boldsymbol\gamma^1-\lambda \mathbf{1}\right)=
\sqrt{\det\left(\mathbf{A}+\lambda^2 \boldsymbol\gamma^1 \right)}
=\left(A^1_0+\lambda^2\right)^2+\left(A^2_0\right)^2+\left(A^3_0\right)^2 -\vec{A}^0\!\cdot\vec{A}^0
\end{equation}
with $\mathbf{A}=\mathbf{\Sigma}\boldsymbol\gamma^1\mathbf{\Sigma}$ given in the paragraph \emph{Determinants}. The comparison gives
\begin{equation}
\epsilon_{I,II}^2=A^1_0 \pm \sqrt{\vec{A}^0\!\cdot\vec{A}^0
-\left(A^2_0\right)^2-\left(A^3_0\right)^2}
\end{equation}

Like the emittances, the quantities $A^1_0$ and $\vec{A}^0\!\cdot\vec{A}^0
-\left(A^2_0\right)^2-\left(A^3_0\right)^2$ are transformation invariants.

\newpage
\subsection{3D visualization  and numerical example}

Here is a stereo view of a random beam coefficient matrix: you should look at the left figure with your left eye, and at the right figure with your right eye. The beam matrix consists of a scalar $\color{gray}\Sigma^0_0$, shown as a gray bar along the upright 1-axis, and three vectors $\color{myred}\vec{\Sigma}^1, \color{mygreen}\vec{\Sigma}^2, \color{myblue}\vec{\Sigma}^3$, shown in primary colors. The scalar products of the three beam vectors are given in the figure heading: (R,G) means "red vector dot green vector" etc. The several decoupling transformations in the later sections will all be applied to this initial beam. The visualization is often helpful for finding a decoupling procedure.

\begin{figure}[H]
\begin{minipage}{0.7 \linewidth}
\centering
\includegraphics[trim = 0pt 0pt 0pt 0pt, width=1\textwidth]{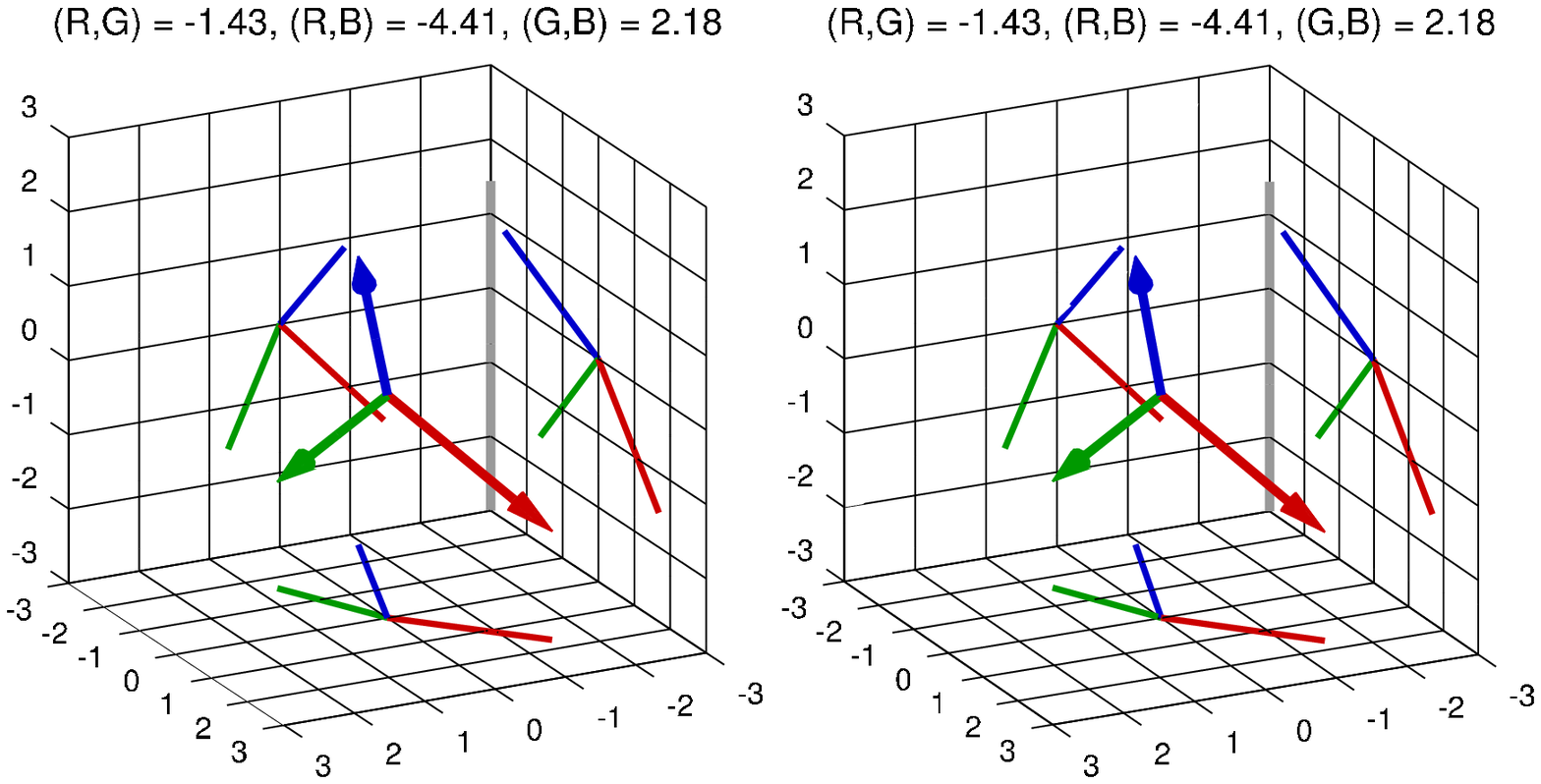}
\end{minipage}
\hspace{0.02 \linewidth}
\begin{minipage}{0.26 \linewidth}
\caption{ }

Random initial beam with coefficients $\Sigma_\kappa^\mu=$
\small
$\left(\begin{array}{cccc}
4.4 & 0 & 0 & 0\\
0 & \bar{1}.5 & \bar{1}.6 & 0.9\\
0 & \bar{1}.5 & 0.7 & \bar{0}.9 \\
0 & 1.7 & \bar{1}.6 & \bar{2}.6
\end{array}\right)$
\normalsize
\phantom{patati}\\
1-axis upright \\
2-axis to your left\\
3-axis to your right
\end{minipage}
\end{figure}

\subsection{Symplectic transformations}

The symplectic condition $\mathbf{A}\equiv\mathbf{R}\boldsymbol\gamma^1\mathbf{R}^\intercal=\boldsymbol\gamma$ translates to 

$A^1_0=R^0_\mu R^0_\mu+R^1_\mu R^1_\mu-R^2_\mu R^2_\mu-R^3_\mu R^3_\mu=\det\mathbf{R}=1$

$A^2_0=2\left(R^2_\mu R^1_\mu-R^0_\mu R^3_\mu\right)=0 \quad
A^3_0=2\left(R^0_\mu R^2_\mu+R^3_\mu R^1_\mu\right)=0$

$\vec{A}^0=2\left(R^1_0\vec{R}^0-R^0_0\vec{R}^1 + \vec{R}^0\!\wedge\vec{R}^1 + R^3_0\vec{R}^2-R^2_0\vec{R}^3 + \vec{R}^2\!\wedge\vec{R}^3\right)=0$

It can be shown that the conditions $A^2_0=A^3_0=0$ split up into

$R^0_\mu R^2_\mu=R^0_\mu R^3_\mu=R^1_\mu R^2_\mu =R^1_\mu R^3_\mu=0$,

and that the conditions $\vec{A}^0=0$ splits up into

$R^1_0\vec{R}^0-R^0_0\vec{R}^1 + \vec{R}^2\!\wedge\vec{R}^3 = R^3_0\vec{R}^2-R^2_0\vec{R}^3 + \vec{R}^0\!\wedge\vec{R}^1=0$

These formulas are not very helpful for constructing symplectic transformations. How- ever, it is easy to write down the logarithm of a symplectic transformation:
\begin{equation}
\ln\mathbf{R}\equiv\mathbf{r}=\vec{\boldsymbol\zeta}\cdot\vec{r}\,^0+\boldsymbol\gamma^1 r^1_0+\vec{\boldsymbol\beta}\mathstrut^2\!\cdot\vec{r}\,^2+\vec{\boldsymbol\beta}\mathstrut^3\!\cdot\vec{r}\,^{3} \label{e5.01}
\end{equation}

This is a $\boldsymbol\gamma$-symmetric matrix, see eqn. \eqref{e2.01}. 

\subsection{Elementary symplectic transformations}

The RHS of eqn. \eqref{e5.01} is a linear combination of the four units $\vec{\boldsymbol\zeta}\cdot\vec{e}\mathstrut^{\:0}$ (squares to $-1$), $\boldsymbol\gamma$ (squares to $-1$), $\vec{\boldsymbol\beta}\mathstrut^2\!\cdot\vec{e}\mathstrut^{\:2}$ (squares to $+1$) $\vec{\boldsymbol\beta}\mathstrut^3\!\cdot\vec{e}\mathstrut^{\:3}$ (squares to $+1$) , where $\vec{e}\mathstrut^{\:\lambda}$ is a unit vector. These units generate the "elementary transformations":
\begin{align}
&\boldsymbol\zeta\text{-Rotations: } &&\mathbf{R}_\zeta &&=\exp \left( \vec{\boldsymbol\zeta}\cdot\vec{e} \psi \right) &&=\mathbf{1}\cos\psi+ \vec{\boldsymbol\zeta}\cdot\vec{e}\sin\psi \qquad \\
&\boldsymbol\gamma\text{-Rotations: } &&\mathbf{R}_{\gamma} &&=\exp \left( \boldsymbol\gamma^1 \phi \right) &&=\mathbf{1}\cos\phi+ \boldsymbol\gamma^1\sin\phi \qquad \\
&\boldsymbol\beta^2\text{-Boosts: } &&\mathbf{R}_{\beta 2} &&=\exp \left( \vec{\boldsymbol\beta}\mathstrut^2\!\cdot\vec{e} \chi \right) &&=\mathbf{1}\cosh\chi+ \vec{\boldsymbol\beta}\mathstrut^2\!\cdot\vec{e}\sinh\chi \qquad \\
&\boldsymbol\beta^3\text{-Boosts: } &&\mathbf{R}_{\beta 3} &&=\exp \left( \vec{\boldsymbol\beta}\mathstrut^3\!\cdot\vec{e} \chi \right) &&=\mathbf{1}\cosh\chi+ \vec{\boldsymbol\beta}\mathstrut^3\!\cdot\vec{e}\sinh\chi \qquad
\end{align}

Below I give the transformed beam matrix $\mathbf{\Sigma}'= \mathbf{R}\mathbf{\Sigma}\mathbf{R}^\intercal$ under the above elementary transformations. The calculation is laborious, details are given in Appendix B.2.

Earlier we have found that the invariants of $\mathbf{\Sigma}$ under the 10 parameter group of symplectic transformations are the two emittances $\epsilon_{I}$, $\epsilon_{II}$. Under the 1-parameter subgroup of elementary elementary transformations, $\mathbf{\Sigma}$ has many additional invariants. These will be listed together with the transformation laws. 

\subsubsection*{$\boldsymbol\zeta$-Rotation}

The action of a $\boldsymbol\zeta$-rotation on $\mathbf{\Sigma}$ is:
\begin{align}
\mathbf{\Sigma}'=&\;\mathbf{R}_{\zeta}\mathbf{\Sigma}\mathbf{R}_{\zeta}^\intercal \nonumber\\
=&\,\exp \left( \vec{\boldsymbol\zeta}\cdot\vec{e} \psi \right) \left( \Sigma^0_0+\vec{\boldsymbol\beta}\mathstrut^m\!\cdot \vec{\Sigma}^m \right) \exp \left( -\vec{\boldsymbol\zeta}\cdot\vec{e} \psi \right) \nonumber\\
=&\;\mathbf{1}  \Sigma^0_0+ \nonumber\\
&\;\vec{\boldsymbol\beta}\mathstrut^1\!\cdot \left[ \vec{e} \Sigma^1_\parallel+\cos \left( 2\psi \right) \vec{\Sigma}^1_\perp-\sin \left( 2\psi \right) \vec{e} \wedge \vec{\Sigma}^1  \right]+\nonumber\\
&\;\vec{\boldsymbol\beta}\mathstrut^2\!\cdot \left[ \vec{e} \Sigma^2_\parallel+\cos \left( 2\psi \right) \vec{\Sigma}^2_\perp-\sin \left( 2\psi \right) \vec{e} \wedge \vec{\Sigma}^2  \right]+\nonumber\\
&\;\vec{\boldsymbol\beta}\mathstrut^3\!\cdot \left[ \vec{e} \Sigma^3_\parallel+\cos \left( 2\psi \right) \vec{\Sigma}^3_\perp-\sin \left( 2\psi \right) \vec{e} \wedge \vec{\Sigma}^3  \right] 
\end{align}

\begin{figure}[H]
\begin{minipage}{0.7 \linewidth}
\centering
\includegraphics[trim = 0pt 0pt 0pt 0pt, width=1\textwidth]{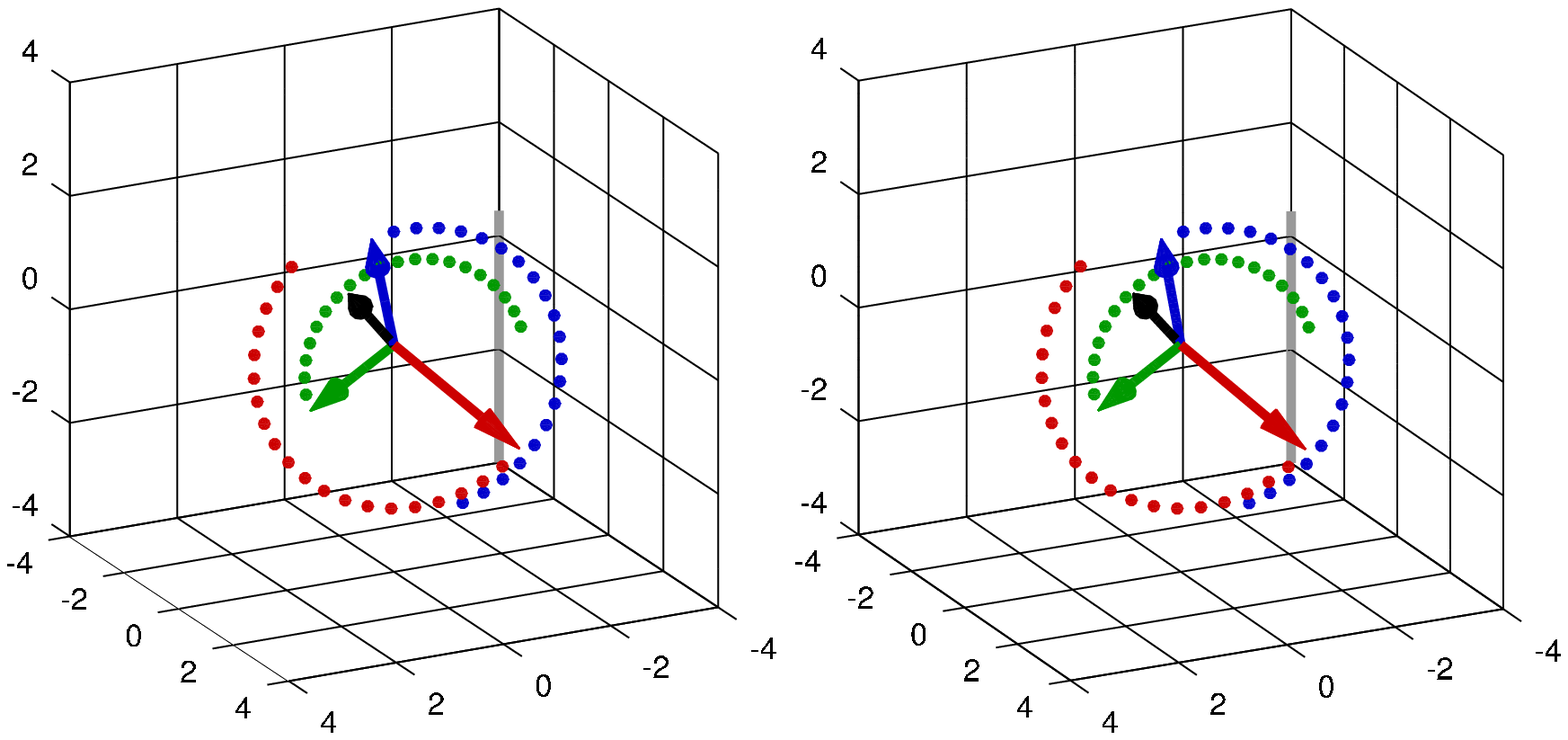}
\end{minipage}
\hspace{0.02 \linewidth}
\begin{minipage}{0.26 \linewidth}
\caption{ }

A $\boldsymbol\zeta$-transformation leaves $\color{gray}\Sigma^0_0$ unchanged and rotates $\color{myred}\vec{\Sigma}^1$, $\color{mygreen}\vec{\Sigma}^2$ and $\color{myblue}\vec{\Sigma}^3$ cw around $\vec{e}$ by $2\psi$. \\ ($\vec{e}=(1,1,1)^\intercal/\sqrt{3}$)
\end{minipage}
\end{figure}

The transformation invariants are 
\begin{equation}
\Sigma_0^0 \qquad \text{all scalar products} \enspace \vec{\Sigma}^l\!\cdot\vec{\Sigma}^m \enspace \text{where} \enspace l,m=1,2,3 \qquad \Sigma^1_\parallel \quad \Sigma^2_\parallel \quad \Sigma^3_\parallel
\end{equation} 

\subsubsection*{$\boldsymbol\gamma$-Rotation}

The action of a $\boldsymbol\gamma^1$-rotation on $\mathbf{\Sigma}$ is:
\begin{align}
\mathbf{\Sigma}'=&\;\mathbf{R}_{\gamma}\mathbf{\Sigma}\mathbf{R}_{\gamma }^\intercal \nonumber\\
=&\,\exp \left( \boldsymbol\gamma^1 \phi \right) \left(1 \Sigma^0_0+\vec{\boldsymbol\beta}\mathstrut^m\!\cdot \vec{\Sigma}^m \right) \exp \left( -\boldsymbol\gamma^1 \phi \right) \nonumber\\
=&\;\mathbf{1} \Sigma^0_0+ \vec{\boldsymbol\beta}\mathstrut^1 \!\cdot \vec{\Sigma}^1 + \nonumber\\ 
&\;\vec{\boldsymbol\beta}\mathstrut^2\!\cdot \left[ \cos \left( 2\phi \right) \vec{\Sigma}^2+\sin \left( 2\phi \right) \vec{\Sigma}^3  \right] +\nonumber\\
&\;\vec{\boldsymbol\beta}\mathstrut^3\!\cdot \left[ -\sin \left( 2\phi \right) \vec{\Sigma}^2+\cos \left( 2\phi \right) \vec{\Sigma}^3  \right]
\end{align}

\begin{figure}[H]
\begin{minipage}{0.7 \linewidth}
\centering
\includegraphics[trim = 0pt 0pt 0pt 0pt, width=1\textwidth]{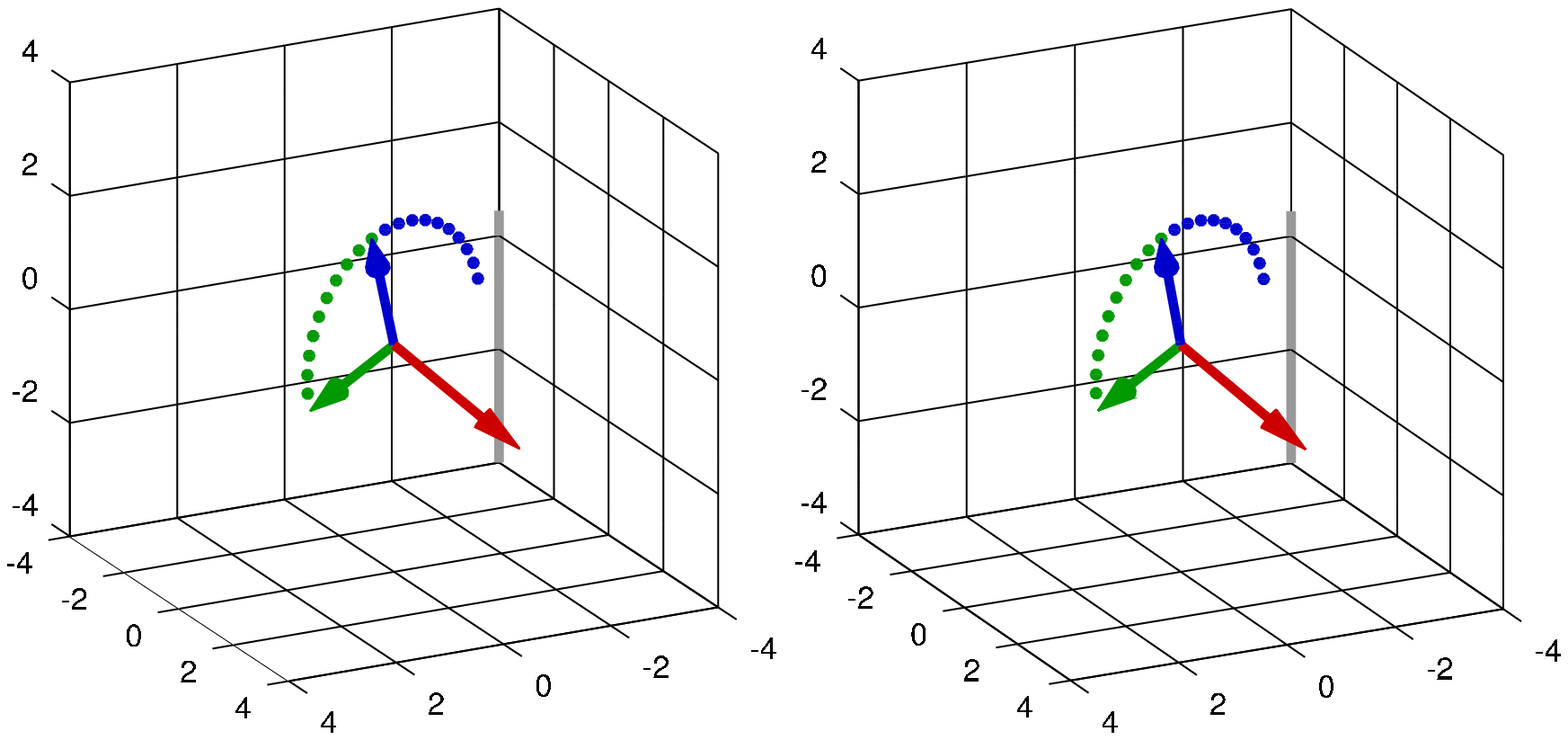}
\end{minipage}
\hspace{0.02 \linewidth}
\begin{minipage}{0.26 \linewidth}
\caption{ }

A $\boldsymbol\gamma$-transformation leaves $\color{gray}\Sigma_0^0$ and $\color{myred}\vec{\Sigma}^1$ un-changed and rotates $\color{mygreen}\vec{\Sigma}^2$ \color{black}\& $\color{myblue}\vec{\Sigma}^3$ along the in-variant ellipse spanned by these two vectors. 
\end{minipage}
\end{figure}

The transformation invariants are 
\begin{equation}
\Sigma_0^0 \qquad \vec{\Sigma}^1 \qquad \vec{\Sigma}^2\!\wedge\vec{\Sigma}^3
\qquad (\vec{\Sigma}^2)^2+(\vec{\Sigma}^3)^2 \qquad (\vec{\Sigma}^2\!\cdot\vec{\Sigma}^1)^2+(\vec{\Sigma}^3\!\cdot\vec{\Sigma}^1)^2 \end{equation} 

To rotate $\vec{\Sigma}'^2$ and $\vec{\Sigma}'^3$ into main axes of the invariant ellipse ($\vec{\Sigma}'^2\!\cdot\vec{\Sigma}'^3=0$), choose the rotation angle according to
\begin{equation}
\tan (4\psi)=\frac{2 \vec{\Sigma}^2 \!\cdot \vec{\Sigma}^3}{(\vec{\Sigma}^2)^2 - (\vec{\Sigma}^3)^2}
\end{equation} 

\subsubsection*{$\boldsymbol\beta^2$-Boost}

The action of a $\boldsymbol\beta^2$-boost on $\mathbf{\Sigma}$  is:
\begin{align}
\mathbf{\Sigma}'=&\; \mathbf{R}_{\beta 2}\mathbf{\Sigma}\mathbf{R}_{\beta 2}^\intercal \nonumber\\
=&\,\exp \left( \vec{\boldsymbol\beta}\mathstrut^2\!\cdot\vec{e} \chi \right) \left( \mathbf{1} \Sigma^0_0+\vec{\boldsymbol\beta}\mathstrut^m\!\cdot \vec{\Sigma}^m \right) \exp \left( \vec{\boldsymbol\beta}\mathstrut^2\!\cdot\vec{e} \chi \right) \nonumber\\
=&\;\mathbf{1}\left[ \cosh \left( 2\chi \right) \Sigma^0_0+\sinh \left( 2\chi \right) \Sigma^2_\parallel \right] + \nonumber\\
&\;\vec{\boldsymbol\beta}\mathstrut^1\!\cdot \left[ \vec{e}\Sigma^1_\parallel + \sinh(2\chi) \vec{e} \wedge \vec{\Sigma}^3 + \cosh(2\chi) \vec{\Sigma}^1_\perp \right] +\nonumber \\
&\;\vec{\boldsymbol\beta}\mathstrut^2\!\cdot \left[\vec{e}  \left(\sinh \left( 2\chi \right) \Sigma^0_0 + \cosh \left( 2\chi \right) \Sigma^2_\parallel\right) + \vec{\Sigma}^2_\perp \right] +\nonumber\\
&\;\vec{\boldsymbol\beta}\mathstrut^3\!\cdot \left[ \vec{e}\Sigma^3_\parallel+\cosh \left( 2\chi \right) \vec{\Sigma}^3_\perp-\sinh \left( 2\chi \right) \vec{e} \wedge \vec{\Sigma}^1  \right]
\end{align}
where
$\Sigma^m_\parallel=\vec{e} \cdot \vec{\Sigma}^m$ and $\vec{\Sigma}^m_\perp=\vec{\Sigma}^m-\vec{e}\Sigma^m_\parallel$. 

\begin{figure}[H]
\begin{minipage}{0.7 \linewidth}
\centering
\includegraphics[trim = 0pt 0pt 0pt 0pt, width=1\textwidth]{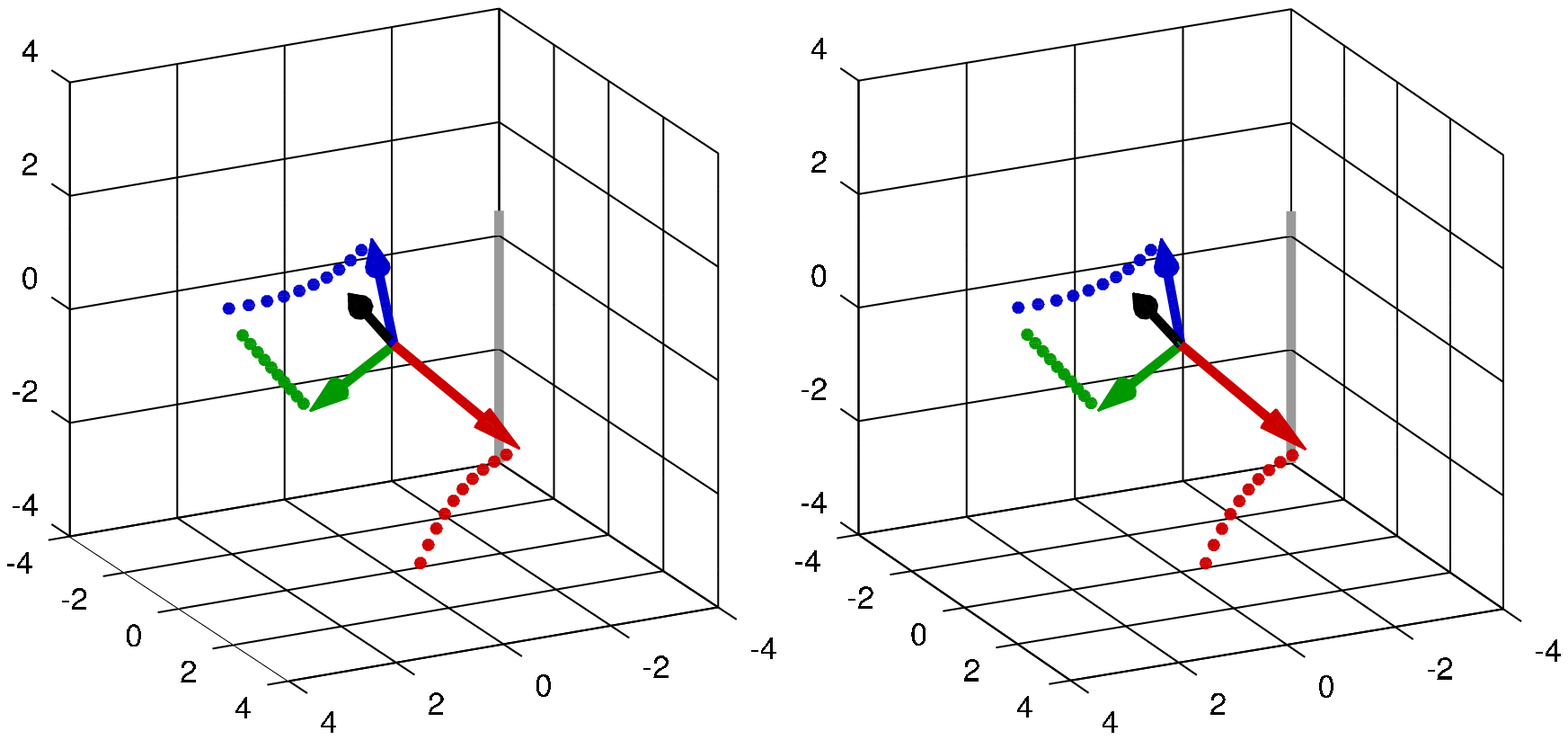}
\end{minipage}
\hspace{0.02 \linewidth}
\begin{minipage}{0.26 \linewidth}
\caption{ }

A $\boldsymbol\beta^2$-transformation mixes $\color{gray}\Sigma^0_0$ and $\color{mygreen}\vec{\Sigma}^2_\parallel$, and mixes $\color{myred}\vec{\Sigma}^1_\perp$ and $\color{myblue}\vec{\Sigma}^3_\perp$. \\ ($\vec{e}=(1,1,1)^\intercal/\sqrt{3}$) 
\end{minipage}
\end{figure}

The transformation invariants are 
\begin{equation}
\Sigma^1_\parallel \qquad \vec{\Sigma}^2_\perp \qquad \Sigma^3_\parallel \qquad  \vec{\Sigma}^3\!\cdot\vec{\Sigma}^1 
\end{equation}
\begin{equation*}
(\Sigma^0_0)^2-(\vec{\Sigma}^2)^2 \qquad (\vec{\Sigma}^1)^2-(\vec{\Sigma}^3)^2 \qquad \left(\vec{\Sigma}^2\!\wedge\vec{\Sigma}^3-\Sigma^0_0\vec{\Sigma}^1\right)^2  -\left(\vec{\Sigma}^2\!\cdot\vec{\Sigma}^1\right)^2 
\end{equation*} 

A $\boldsymbol\beta^2$-boost also allows to supress $\vec{\Sigma}^2$. This is achieved by setting 
\begin{equation}
\vec{e}=\frac{\vec{\Sigma^2} }{|\vec{\Sigma}^2|} \qquad
\tanh (2\chi)=-\frac{|\vec{\Sigma}^2|}{\Sigma^0_0 }
\end{equation} 

A $\boldsymbol\beta^2$-boost allows to adjust $\vec{\Sigma}^2 \!\cdot \vec{\Sigma}^1$ while keeping $\vec{\Sigma}^3 \!\cdot \vec{\Sigma}^1$ constant. 

To make $\vec{\Sigma}'^2 \!\cdot \vec{\Sigma}'^1 = 0$, set
\begin{equation}
\vec{e}=\frac{\vec{\Sigma}^2\!\wedge\vec{\Sigma}^3 -\Sigma^0_0\vec{\Sigma}^1} {|\vec{\Sigma}^2\!\wedge\vec{\Sigma}^3 -\Sigma^0_0\vec{\Sigma}^1|} 
\qquad \tanh (2\chi)=\frac{\vec{\Sigma}^2\!\cdot\vec{\Sigma}^1} {|\vec{\Sigma}^2\!\wedge\vec{\Sigma}^3 -\Sigma^0_0\vec{\Sigma}^1|}
\end{equation} 

A $\boldsymbol\beta^2$-boost allows to adjust $\vec{\Sigma}^2 \!\cdot \vec{\Sigma}^3$ while keeping $\vec{\Sigma}^3 \!\cdot \vec{\Sigma}^1$ constant. 

To make $\vec{\Sigma}'^2 \!\cdot \vec{\Sigma}'^3 = 0$, set
\begin{equation}
\vec{e}=\frac{\vec{\Sigma}^1\!\wedge\vec{\Sigma}^2 -\Sigma^0_0\vec{\Sigma}^3} {|\vec{\Sigma}^1\!\wedge\vec{\Sigma}^2 -\Sigma^0_0\vec{\Sigma}^3|} 
\qquad \tanh (2\chi)=\frac{\vec{\Sigma}^2\!\cdot\vec{\Sigma}^3} {|\vec{\Sigma}^1\!\wedge\vec{\Sigma}^2 -\Sigma^0_0\vec{\Sigma}^3|}
\end{equation} 

\subsubsection*{$\boldsymbol\beta^3$-Boost}

The action of a $\boldsymbol\beta^3$-boost on $\mathbf{\Sigma}$ is:
\begin{align}
\mathbf{\Sigma}'=&\; \mathbf{R}_{\beta3}\mathbf{\Sigma}\mathbf{R}_{\beta3}^\intercal \nonumber\\
=&\,\exp \left( \vec{\boldsymbol\beta}\mathstrut^3\!\cdot\vec{e} \chi \right) \left( \mathbf{2} \Sigma^0_0+\vec{\boldsymbol\beta}\mathstrut^m\!\cdot \vec{\Sigma}^m \right) \exp \left( \vec{\boldsymbol\beta}\mathstrut^3\!\cdot\vec{e} \chi \right) \nonumber\\
=&\;\mathbf{1}\left[ \cosh \left( 2\chi \right) \Sigma^0_0+\sinh \left( 2\chi \right) \Sigma^3_\parallel \right] + \nonumber\\
&\;\vec{\boldsymbol\beta}\mathstrut^1\!\cdot \left[ \vec{e}\Sigma^1_\parallel - \sinh(2\chi) \vec{e} \wedge \vec{\Sigma}^2 + \cosh(2\chi) \vec{\Sigma}^1_\perp \right] +\nonumber\\
&\;\vec{\boldsymbol\beta}\mathstrut^2\!\cdot \left[ \vec{e}\Sigma^2_\parallel+\cosh \left( 2\chi \right) \vec{\Sigma}^2_\perp+\sinh \left( 2\chi \right) \vec{e} \wedge \vec{\Sigma}^1  \right]+\nonumber\\
&\;\vec{\boldsymbol\beta}\mathstrut^3\!\cdot \left[\vec{e}  \left(\sinh \left( 2\chi \right) \Sigma^0_0 + \cosh \left( 2\chi \right) \Sigma^3_\parallel\right) + \vec{\Sigma}^3_\perp \right] 
\end{align}

\begin{figure}[H]
\begin{minipage}{0.7 \linewidth}
\centering
\includegraphics[trim = 0pt 0pt 0pt 0pt, width=1\textwidth]{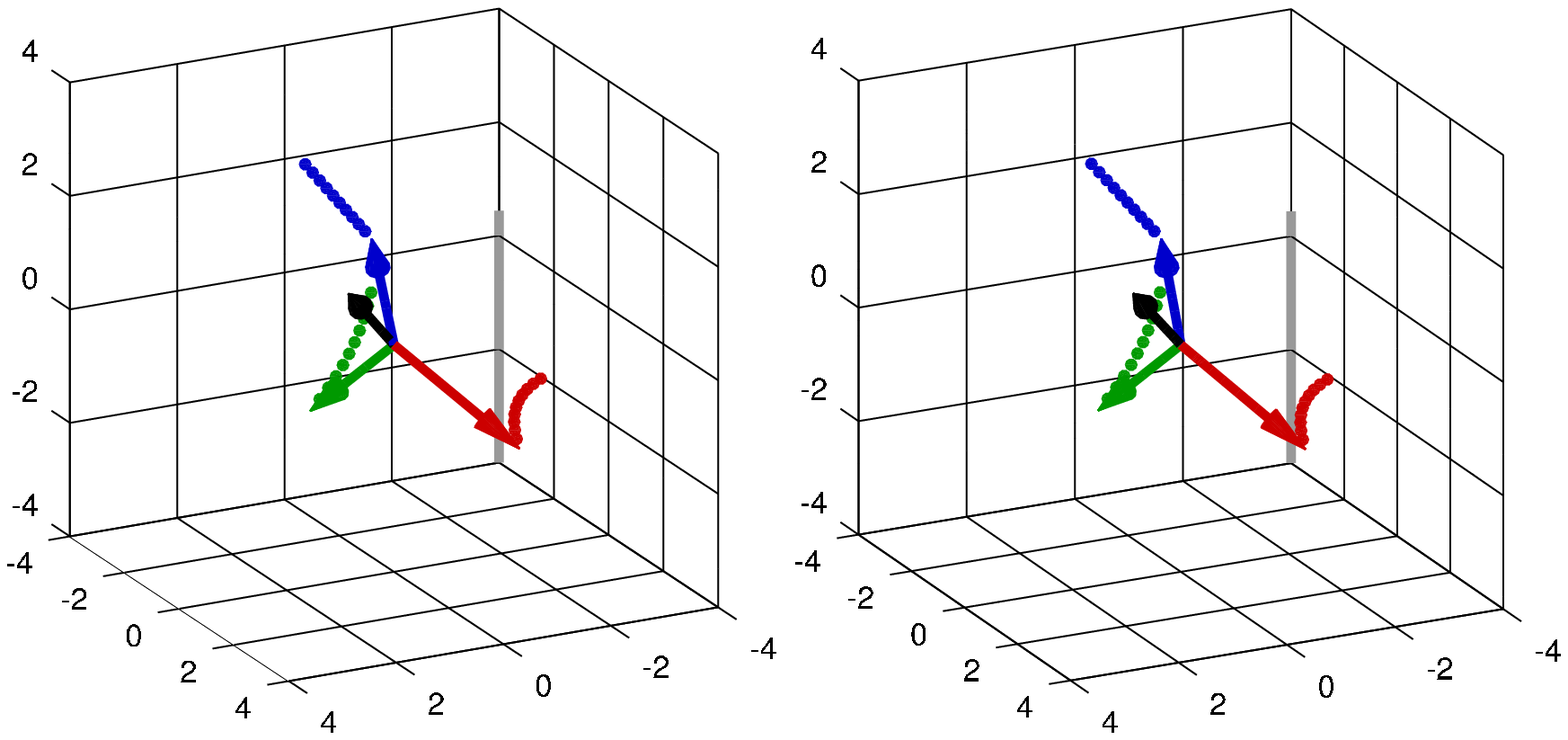}
\end{minipage}
\hspace{0.02 \linewidth}
\begin{minipage}{0.26 \linewidth}
\caption{ }

A $\boldsymbol\beta^3$-transformation mixes $\color{gray}\Sigma^0_0$ and $\color{myblue}\vec{\Sigma}^3_\parallel$, and mixes $\color{myred}\vec{\Sigma}^1_\perp$ and $\color{mygreen}\vec{\Sigma}^2_\perp$. \\ ($\vec{e}=(1,1,1)^\intercal/\sqrt{3}$) 
\end{minipage}
\end{figure}

The transformation invariants are: 
\begin{equation}
\Sigma^1_\parallel \qquad \Sigma^2_\parallel \qquad \vec{\Sigma}^3_\perp \qquad
\vec{\Sigma}^2\!\cdot\vec{\Sigma}^1 
\end{equation}
\begin{equation*}
(\Sigma^0_0)^2-(\vec{\Sigma}^3)^2 \qquad (\vec{\Sigma}^1)^2-(\vec{\Sigma}^2)^2 \qquad
\left(\vec{\Sigma}^2\!\wedge\vec{\Sigma}^3-\Sigma^0_0\vec{\Sigma}^1\right)^2-\left(\vec{\Sigma}^3\!\cdot\vec{\Sigma}^1\right)^2 
\end{equation*} 

A $\boldsymbol\beta^3$-boost also allows to suppress $\vec{\Sigma}^3$. This is achieved by setting 
\begin{equation}
\vec{e}=\frac{\vec{\Sigma}^3 }{|\vec{\Sigma}^3|} \qquad
\tanh (2\chi)=-\frac{|\vec{\Sigma}^3|}{\Sigma^0_0 }
\end{equation} 

A $\boldsymbol\beta^3$-boost allows to adjust $\vec{\Sigma}^3 \!\cdot \vec{\Sigma}^1$ while keeping $\vec{\Sigma}^2 \!\cdot \vec{\Sigma}^1$ constant. 

To make $\vec{\Sigma}'^3 \!\cdot \vec{\Sigma}'^1 = 0$, set
\begin{equation}
\vec{e}=\frac{\vec{\Sigma}^2\!\wedge\vec{\Sigma}^3 -\Sigma^0_0\vec{\Sigma}^1} {|\vec{\Sigma}^2\!\wedge\vec{\Sigma}^3 -\Sigma^0_0\vec{\Sigma}^1|} 
\qquad \tanh (2\chi)=\frac{\vec{\Sigma}^3\!\cdot\vec{\Sigma}^1} {|\vec{\Sigma}^2\!\wedge\vec{\Sigma}^3 -\Sigma^0_0\vec{\Sigma}^1|}
\end{equation} 

A $\boldsymbol\beta^3$-boost allows to adjust $\vec{\Sigma}^3 \!\cdot \vec{\Sigma}^2$ while keeping $\vec{\Sigma}^2 \!\cdot \vec{\Sigma}^1$ constant. 

To make $\vec{\Sigma}'^3 \!\cdot \vec{\Sigma}'^2 = 0$, set
\begin{equation}
\vec{e}=\frac{\vec{\Sigma}^3\!\wedge\vec{\Sigma}^1 -\Sigma^0_0\vec{\Sigma}^2} {|\vec{\Sigma}^3\!\wedge\vec{\Sigma}^1 -\Sigma^0_0\vec{\Sigma}^2|} 
\qquad \tanh (2\chi)=\frac{\vec{\Sigma}^3\!\cdot\vec{\Sigma}^2} {|\vec{\Sigma}^3\!\wedge\vec{\Sigma}^1 -\Sigma^0_0\vec{\Sigma}^2|}
\end{equation}

\newpage
\subsection{Decoupling two pairs of phase space coordinates}

Decoupling two complementary sets $A$ and $\bar{A}$ of phase space coordinates means, making the correlation between $A$ and $\bar{A}$ vanish. We take the 4 phase space coordinates to be $\left\{X_1, X_2, X_3, X_4\right\} = \left\{ x, x', y, y'\right\} (=A\cup\bar{A})$. 

\subsubsection*{Decoupling $x,x'$ from $y, y'$}

This means bringing the representative matrix and component matrix into the form

\vspace{5pt}
$\mathbf{\Sigma}=
\left(
\begin{array}{c c c c}
\Sigma_{11} & \Sigma_{12} & 0 & 0 \\
\Sigma_{21} & \Sigma_{22} & 0 & 0 \\
0 & 0 & \Sigma_{33} & \Sigma_{34} \\
0 & 0 & \Sigma_{43} & \Sigma_{44} \\
\end{array}
\right)\qquad \Rightarrow \qquad (\Sigma_\kappa^\lambda)=
\left(
\begin{array}{r r r r}
\Sigma_0^0 & 0 & 0 & 0 \\
0 & \Sigma_1^1 & 0 & 0\\
0 & 0 & \Sigma_2^2 & \Sigma_2^3  \\
0 & 0 & \Sigma_3^2  & \Sigma_3^3 \\
\end{array}
\right)$

\vspace{5pt}
\fbox{
\begin{minipage}{0.98\linewidth}
\vspace{5pt}

To decouple $x,x'$ from $y, y'$, proceed as follows (steps 1 and 2 can be interchanged):

\vspace{5pt}
\begin{enumerate}
\item 
Decompose $\mathbf{\Sigma}=\mathbf{1}\Sigma^0_0+\vec{\boldsymbol\beta}\mathstrut^m\!\cdot \vec{\Sigma}^m$. \\
Boost $\vec{\Sigma}^2 \!\cdot \vec{\Sigma}^1$ to 0 with $\mathbf{\Sigma'}=\mathbf{R}_{\beta 2}{\mathbf{\Sigma}}\mathbf{R}_{\beta 2}^\intercal$, where 
\begin{equation*}
\mathbf{R}_{\beta 2}=\exp \left( \vec{\boldsymbol\beta}\mathstrut^2 \!\cdot \vec{e} \chi \right) 
\qquad \vec{e}=\frac{\vec{\Sigma}^2\!\wedge\vec{\Sigma}^3 -\Sigma^0_0\vec{\Sigma}^1}{|\vec{\Sigma}^2\!\wedge\vec{\Sigma}^3 -\Sigma^0_0\vec{\Sigma}^1|} 
\qquad \tanh (2\chi)=\frac{\vec{\Sigma}^2\!\cdot\vec{\Sigma}^1} {|\vec{\Sigma}^2\!\wedge\vec{\Sigma}^3 -\Sigma^0_0\vec{\Sigma}^1|}
\end{equation*} 
\item 
Rename $\mathbf{\Sigma'} \rightarrow\mathbf{\Sigma}$ and decompose $\mathbf{\Sigma}=\mathbf{1}\Sigma^0_0+\vec{\boldsymbol\beta}\mathstrut^m\!\cdot \vec{\Sigma}^m$. \\
Boost $\vec{\Sigma}^3 \!\cdot \vec{\Sigma}^1$ to 0 (keeping $\vec{\Sigma}^2 \!\cdot \vec{\Sigma}^1 = 0$) with $\mathbf{\Sigma'}=\mathbf{R}_{\beta 3}{\mathbf{\Sigma}}\mathbf{R}_{\beta 3}^\intercal$, where
\begin{equation*}
\mathbf{R}_{\beta 3}=\exp \left( \vec{\boldsymbol\beta}\mathstrut^3 \!\cdot \vec{e} \chi \right) 
\qquad 
\vec{e}=\frac{\vec{\Sigma}^2\!\wedge\vec{\Sigma}^3-\Sigma^0_0\vec{\Sigma}^1}{|\vec{\Sigma}^2\!\wedge\vec{\Sigma}^3-\Sigma^0_0\vec{\Sigma}^1|} 
\qquad
\tanh (2\chi)=\frac{\vec{\Sigma}^3 \!\cdot \vec{\Sigma}^1} {|\vec{\Sigma}^2\!\wedge\vec{\Sigma}^3-\Sigma^0_0\vec{\Sigma}^1|}
\end{equation*} 
\item 
Rename $\mathbf{\Sigma'} \rightarrow\mathbf{\Sigma}$ and decompose $\mathbf{\Sigma}=\mathbf{1}\Sigma^0_0+\vec{\boldsymbol\beta}\mathstrut^m\!\cdot \vec{\Sigma}^m$. \\
Rotate $\vec{\Sigma}^1$ parallel to the 1-axis  with  
$\mathbf{\Sigma'}=\mathbf{R}_{\zeta}{\mathbf{\Sigma}}\mathbf{R}_{\zeta}^\intercal$, where \begin{equation*}
\mathbf{R}_{\zeta}=\exp \left(\vec{\boldsymbol\zeta}\cdot\vec{e} \psi\right)
\qquad \vec{e}=\frac{(0,\Sigma_3^1,-\Sigma_2^1)^\intercal}{\sqrt{(\Sigma_2^1)^2+(\Sigma_3^1)^2}} 
\qquad  \tan (2\psi)=-\frac{\sqrt{(\Sigma_2^1)^2+(\Sigma_3^1)^2}}{\Sigma_1^1}
\end{equation*}

\end{enumerate}

\vspace{5pt}
The decoupling is preserved under the transformation group 
$\mathbf{R}=\exp\left[\boldsymbol\zeta_1 \psi + \boldsymbol\gamma^1\phi \right.$ $\left. + \boldsymbol\beta_2^2 \chi_1+\boldsymbol\beta_3^2 \chi_2+ \boldsymbol\beta_2^3 \chi_3+\boldsymbol\beta_3^3 \chi_4 \right]$ (free parameters $\psi, \phi, \chi_1, \chi_2, \chi_3, \chi_4$). 
\vspace{5pt}

\end{minipage}
}

\vspace{10pt}
\begin{figure}[H]
\begin{minipage}{0.68 \linewidth}
\centering
\includegraphics[trim = 0pt 0pt 0pt 0pt, width=1\textwidth]{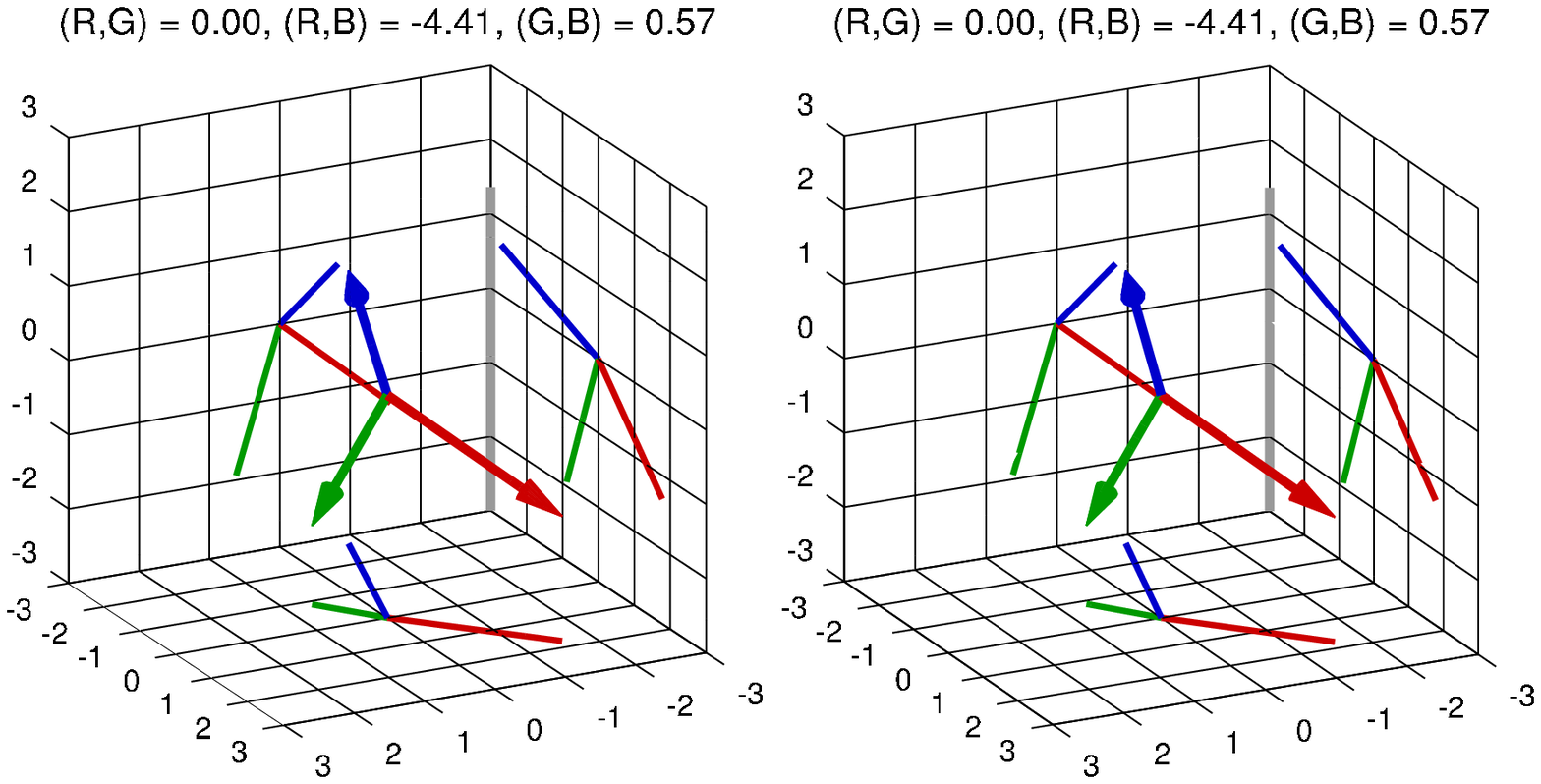}
\end{minipage}
\hspace{0.02 \linewidth}
\begin{minipage}{0.28 \linewidth}
\caption{ }

Dec. $x,x' \;|\; y, y'$, step 1:\\
$\color{myred}\vec{\Sigma}^1\color{black}\!\perp\color{mygreen}\vec{\Sigma}^2$\\

Coefficient matrix $\Sigma_\kappa^\lambda=$

\small
$\left(\begin{array}{cccc}
4.3 & 0 & 0 & 0\\
0 & \bar{1}.3 & \bar{1}.9 & 0.7 \\
0 & \bar{1}.6 & 0.6 & \bar{0}.8 \\
0 & 1.8 & \bar{0}.9 & \bar{2}.7
\end{array}\right)$
\normalsize 
\end{minipage}
\end{figure}

\vspace{10pt}
\begin{figure}[H]
\begin{minipage}{0.68 \linewidth}
\centering
\includegraphics[trim = 0pt 0pt 0pt 0pt, width=1\textwidth]{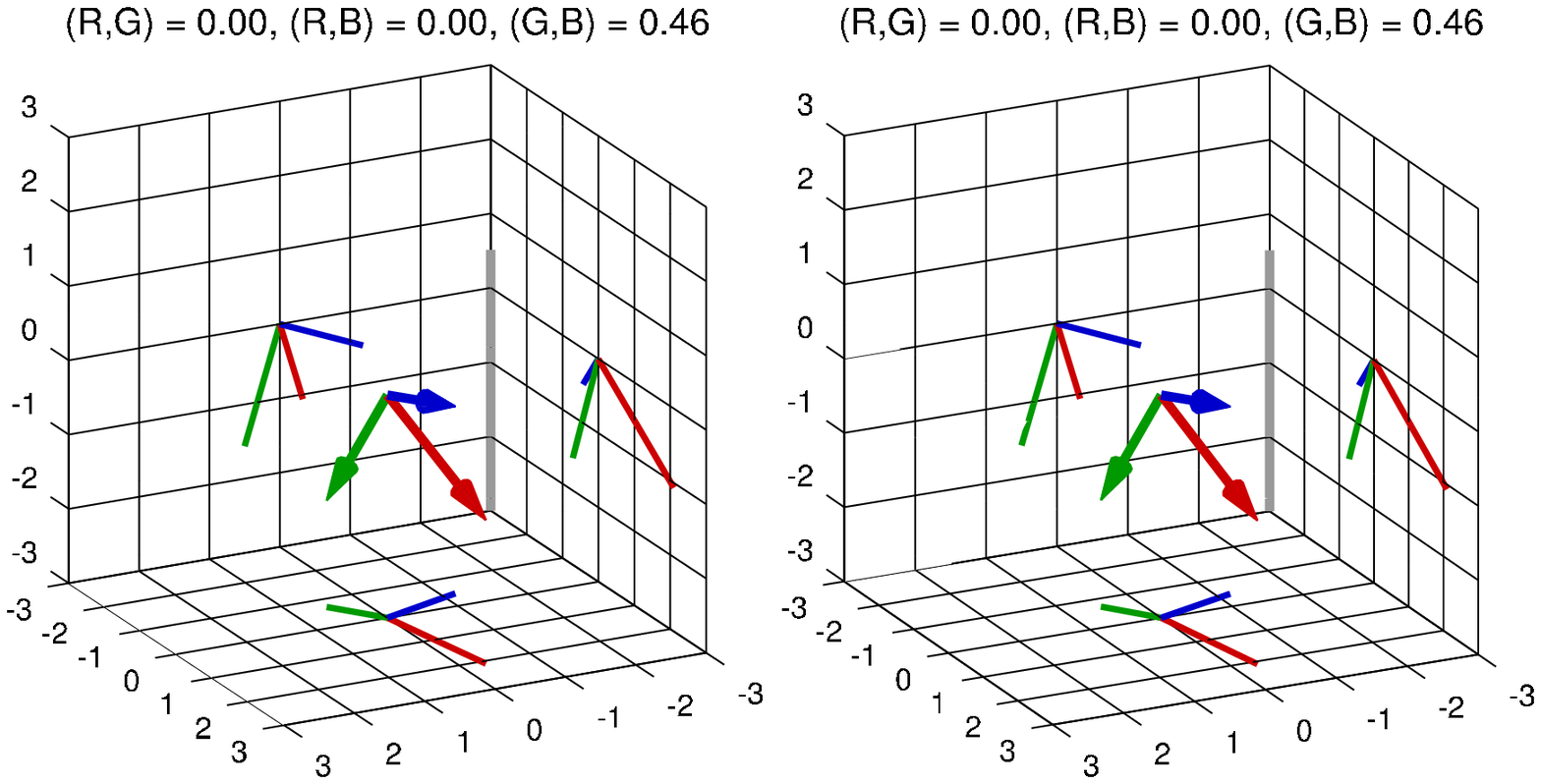}
\end{minipage}
\hspace{0.02 \linewidth}
\begin{minipage}{0.28 \linewidth}
\caption{ }

Dec. $x,x' \;|\; y, y'$, step 2:\\
$\color{myred}\vec{\Sigma}^1\color{black}\!\perp\color{mygreen}\vec{\Sigma}^2$, 
$\color{myred}\vec{\Sigma}^1\color{black}\!\perp\color{myblue}\vec{\Sigma}^3$ \\

Coefficient matrix $\Sigma_\kappa^\lambda=$

\small
$\left(\begin{array}{cccc}
3.5 & 0 & 0 & 0\\
0 & \bar{1}.1 & \bar{1}.6 & \bar{0}.5 \\
0 & \bar{0}.3 & 0.5 & \bar{1}.2 \\
0 & 2.1 & \bar{0}.7 & \bar{0}.4
\end{array}\right)$
\normalsize
\end{minipage}
\end{figure}

\vspace{10pt}
\begin{figure}[H]
\begin{minipage}{0.68 \linewidth}
\centering
\includegraphics[trim = 0pt 0pt 0pt 0pt, width=1\textwidth]{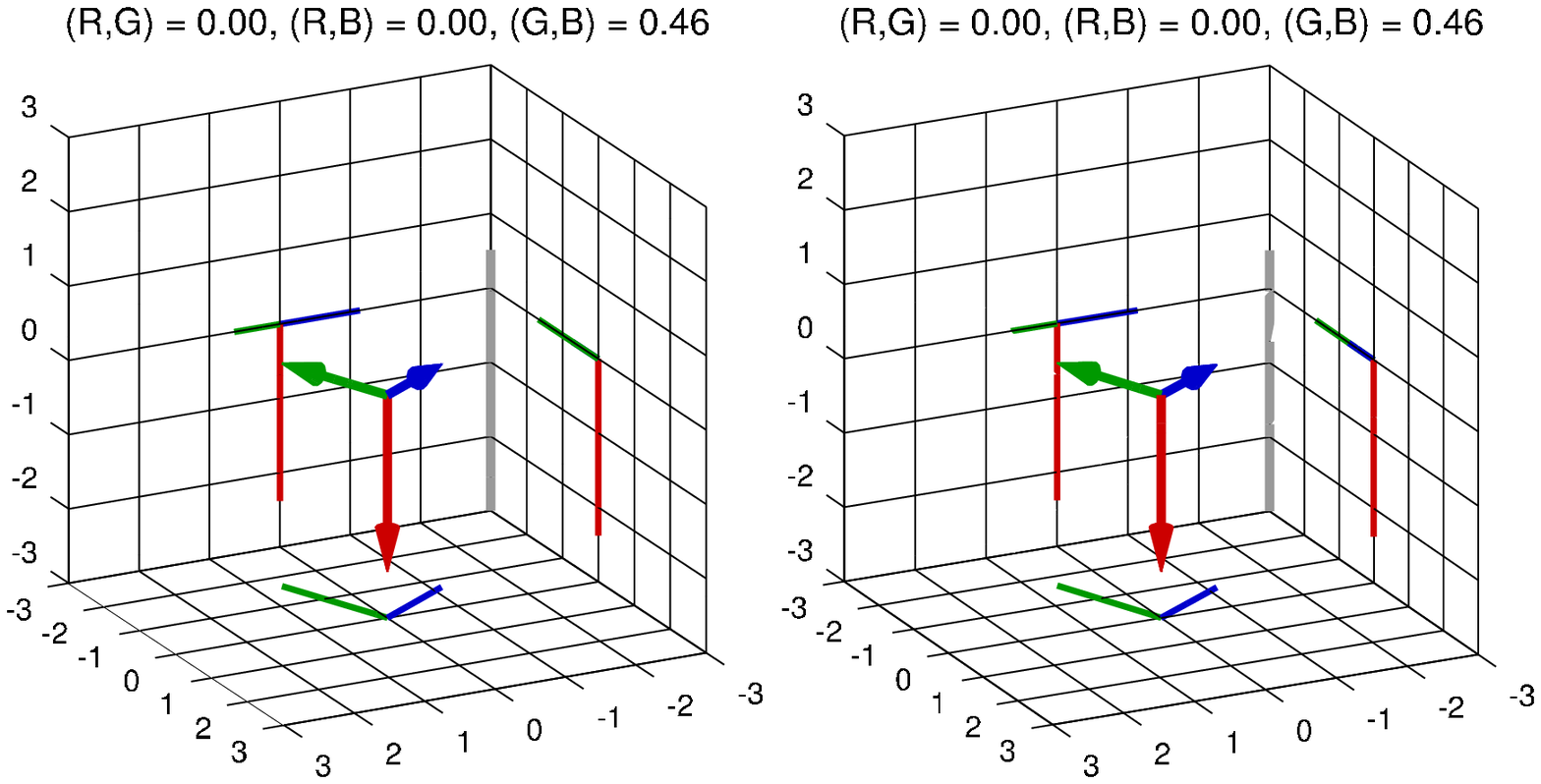}
\end{minipage}
\hspace{0.02 \linewidth}
\begin{minipage}{0.28 \linewidth}
\caption{ }
Dec. $x,x' \;|\; y, y'$, step 3:\\ 
$\color{myred}\vec{\Sigma}^1\color{black}\!\perp\color{mygreen}\vec{\Sigma}^2$,
$\color{myred}\vec{\Sigma}^1\color{black}\!\perp\color{myblue}\vec{\Sigma}^3$\\ $\color{myred}\vec{\Sigma}^1\color{black}\!\parallel$ 1-axis \\

Coefficient matrix $\Sigma_\kappa^\lambda=$

\small
$\left(\begin{array}{cccc}
3.5 & 0 & 0 & 0\\
0 & \bar{2}.3 & 0 & 0\\
0 & 0 & 0.6 & \bar{1}.1 \\
0 & 0 & \bar{1}.7 & \bar{0}.7
\end{array}\right)$
\normalsize
\end{minipage}
\end{figure}

\newpage
\subsubsection*{Decoupling $x, y$ from $x', y'$}

This means bringing the representative matrix and component matrix into the form

\vspace{5pt}
$\mathbf{\Sigma}=
\left(
\begin{array}{c c c c}
\Sigma_{11} & 0 & \Sigma_{13} & 0\\
0 & \Sigma_{22} & 0 & \Sigma_{14} \\
\Sigma_{31} & 0 & \Sigma_{33} & 0 \\
0 & \Sigma_{41} & 0 & \Sigma_{44} \\
\end{array}
\right)\qquad \Rightarrow \qquad (\Sigma_\kappa^\lambda)=
\left(
\begin{array}{c c c c}
\Sigma_0^0 & 0 & 0 & 0 \\
0 & \Sigma_1^1 & 0 & \Sigma_1^3 \\
0 & 0 & \Sigma_2^2 & 0 \\
0 & \Sigma_3^1 & 0 & \Sigma_3^3 \\
\end{array}
\right)$

\vspace{5pt}
\fbox{
\begin{minipage}{0.98\linewidth}
\vspace{5pt}

To decouple $x,y$ from $x',y'$, proceed as follows:

\vspace{5pt}
\begin{enumerate}
\item 
Decompose $\mathbf{\Sigma}=\mathbf{1}\Sigma^0_0+\vec{\boldsymbol\beta}\mathstrut^m\!\cdot \vec{\Sigma}^m$. \\
Boost $\vec{\Sigma}^2 \!\cdot \vec{\Sigma}^1$ to 0 with $\mathbf{\Sigma'}=\mathbf{R}_{\beta 2}{\mathbf{\Sigma}}\mathbf{R}_{\beta 2}^\intercal$, where 
\begin{equation*}
\mathbf{R}_{\beta 2}=\exp \left( \vec{\boldsymbol\beta}\mathstrut^2 \!\cdot \vec{e} \chi \right) 
\qquad \vec{e}=\frac{\vec{\Sigma}^2\!\wedge\vec{\Sigma}^3 -\Sigma^0_0\vec{\Sigma}^1}{|\vec{\Sigma}^2\!\wedge\vec{\Sigma}^3 -\Sigma^0_0\vec{\Sigma}^1|} 
\qquad \tanh (2\chi)=\frac{\vec{\Sigma}^2\!\cdot\vec{\Sigma}^1} {|\vec{\Sigma}^2\!\wedge\vec{\Sigma}^3 -\Sigma^0_0\vec{\Sigma}^1|}
\end{equation*} 
\item 
Rename $\mathbf{\Sigma'} \rightarrow\mathbf{\Sigma}$ and decompose $\mathbf{\Sigma}=\mathbf{1}\Sigma^0_0+\vec{\boldsymbol\beta}\mathstrut^m\!\cdot \vec{\Sigma}^m$. \\
Boost $\vec{\Sigma}^3 \!\cdot \vec{\Sigma}^2$ to 0 (keeping $\vec{\Sigma}^2 \!\cdot \vec{\Sigma}^1 = 0$) with $\mathbf{\Sigma'}=\mathbf{R}_{\beta 3}{\mathbf{\Sigma}}\mathbf{R}_{\beta 3}^\intercal$, where
\begin{equation*}
\mathbf{R}_{\beta 3}=\exp \left( \vec{\boldsymbol\beta}\mathstrut^3 \!\cdot \vec{e} \chi \right) 
\qquad 
\vec{e}=\frac{\vec{\Sigma}^3\!\wedge\vec{\Sigma}^1-\Sigma^0_0\vec{\Sigma}^2}{|\vec{\Sigma}^3\!\wedge\vec{\Sigma}^1-\Sigma^0_0\vec{\Sigma}^2|} 
\qquad
\tanh (2\chi)=\frac{\vec{\Sigma}^3 \!\cdot \vec{\Sigma}^2} {|\vec{\Sigma}^3\!\wedge\vec{\Sigma}^1-\Sigma^0_0\vec{\Sigma}^2|}
\end{equation*} 
\item 
Rename $\mathbf{\Sigma'} \rightarrow\mathbf{\Sigma}$ and decompose $\mathbf{\Sigma}=\mathbf{1}\Sigma^0_0+\vec{\boldsymbol\beta}\mathstrut^m\!\cdot \vec{\Sigma}^m$. \\
Rotate $\vec{\Sigma}^2$ parallel to the 2-axis  with  
$\mathbf{\Sigma'}=\mathbf{R}_{\zeta}{\mathbf{\Sigma}}\mathbf{R}_{\zeta}^\intercal$, where 
\begin{equation*}
\mathbf{R}_{\zeta}=\exp \left(\vec{\boldsymbol\zeta}\cdot\vec{e} \psi\right)
\qquad \vec{e}=\frac{(-\Sigma_3^2,0,\Sigma_1^2)^\intercal}{\sqrt{(\Sigma_1^2)^2+(\Sigma_3^2)^2}} 
\qquad  \tan (2\psi)=-\frac{\sqrt{(\Sigma_1^2)^2+(\Sigma_3^2)^2}}{\Sigma_2^2}
\end{equation*}

\end{enumerate}

\vspace{5pt}
The decoupling is preserved under the transformation group 
$\mathbf{R}=\exp\left[\boldsymbol\zeta_2 \psi + \boldsymbol\beta_2^2 \chi_1 \right.$ $\left. + \boldsymbol\beta_1^3 \chi_2+\boldsymbol\beta_3^3 \chi_3 \right]$ (free parameters $\psi, \chi_1, \chi_2, \chi_3$). 
\vspace{5pt}

\end{minipage}
}

\vspace{10pt}
\begin{figure}[H]
\begin{minipage}{0.68 \linewidth}
\centering
\includegraphics[trim = 0pt 0pt 0pt 0pt, width=1\textwidth]{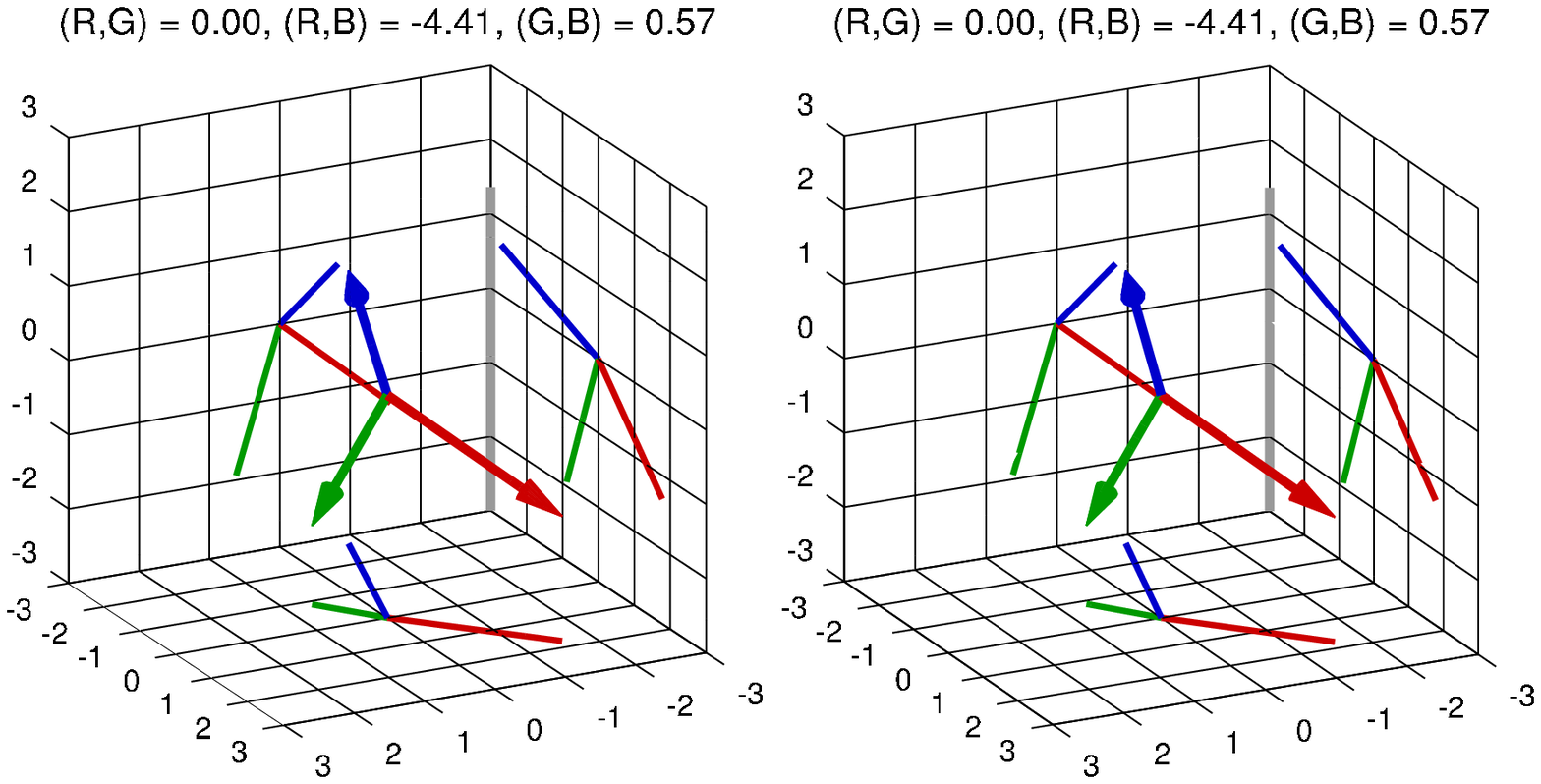}
\end{minipage}
\hspace{0.02 \linewidth}
\begin{minipage}{0.28 \linewidth}
\caption{ }

Dec. $x,x' \;|\; y, y'$, step 1:\\
$\color{myred}\vec{\Sigma}^1\color{black}\!\perp\color{mygreen}\vec{\Sigma}^2$\\

Coefficient matrix $\Sigma_\kappa^\lambda=$

\small
$\left(\begin{array}{cccc}
4.3 & 0 & 0 & 0\\
0 & \bar{1}.3 & \bar{1}.9 & 0.7 \\
0 & \bar{1}.6 & 0.6 & \bar{0}.8 \\
0 & 1.8 & \bar{0}.9 & \bar{2}.7
\end{array}\right)$
\normalsize 
\end{minipage}
\end{figure}

\vspace{10pt}
\begin{figure}[H]
\begin{minipage}{0.68 \linewidth}
\centering
\includegraphics[trim = 0pt 0pt 0pt 0pt, width=1\textwidth]{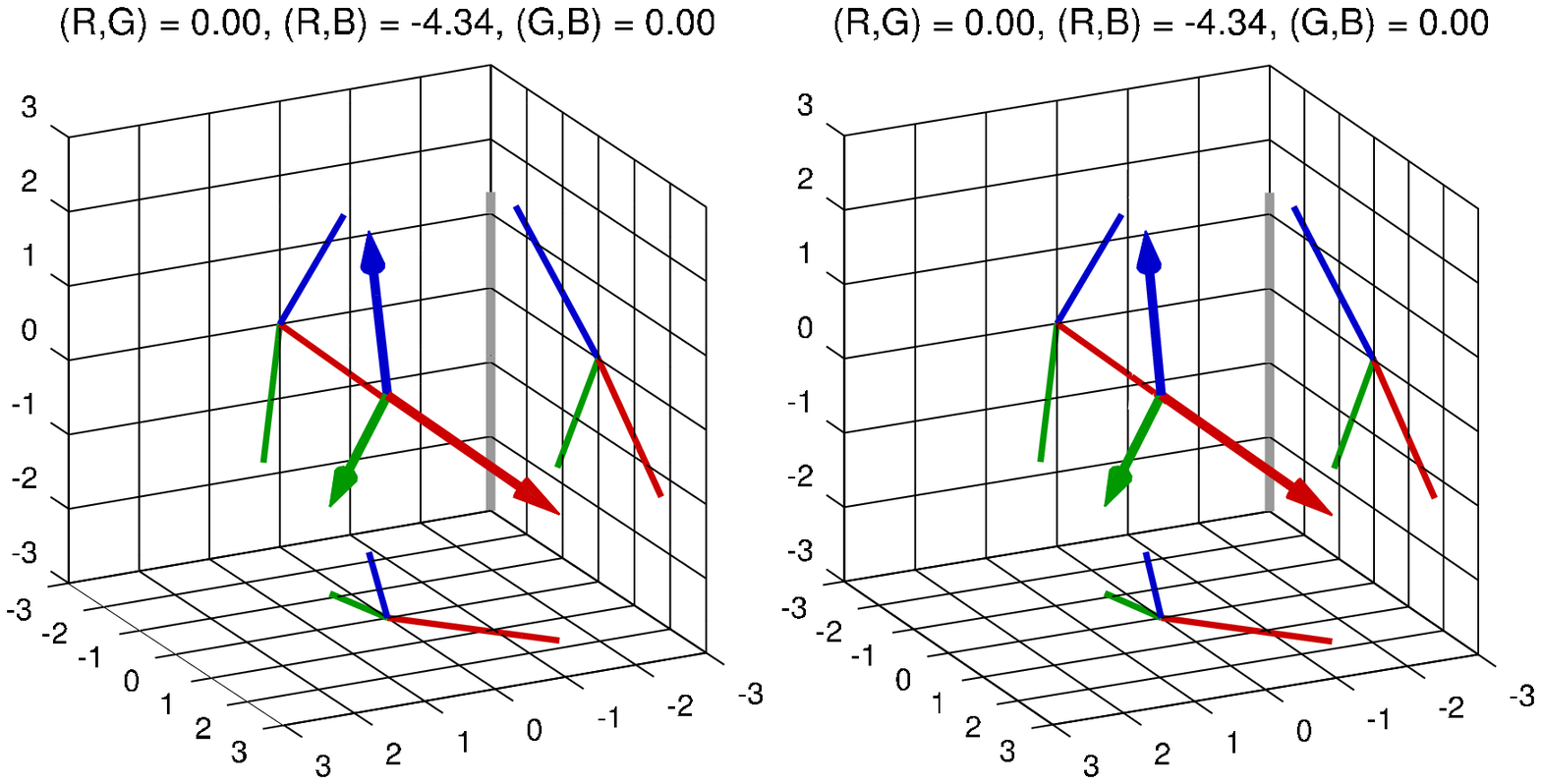}
\end{minipage}
\hspace{0.02 \linewidth}
\begin{minipage}{0.28 \linewidth}
\caption{ }

Dec. $x,x' \;|\; y, y'$, step 2:\\
$\color{myred}\vec{\Sigma}^1\color{black}\!\perp\color{mygreen}\vec{\Sigma}^2$, 
$\color{mygreen}\vec{\Sigma}^2\color{black}\!\perp\color{myblue}\vec{\Sigma}^3$ \\

Coefficient matrix $\Sigma_\kappa^\lambda=$

\small
$\left(\begin{array}{cccc}
4.3 & 0 & 0 & 0\\
0 & \bar{1}.3 & \bar{1}.8 & 1.3 \\
0 & \bar{1}.5 & 0.2 & \bar{0}.9 \\
0 & 1.7 & \bar{1}.1 & \bar{2}.3
\end{array}\right)$
\normalsize
\end{minipage}
\end{figure}

\vspace{10pt}
\begin{figure}[H]
\begin{minipage}{0.68 \linewidth}
\centering
\includegraphics[trim = 0pt 0pt 0pt 0pt, width=1\textwidth]{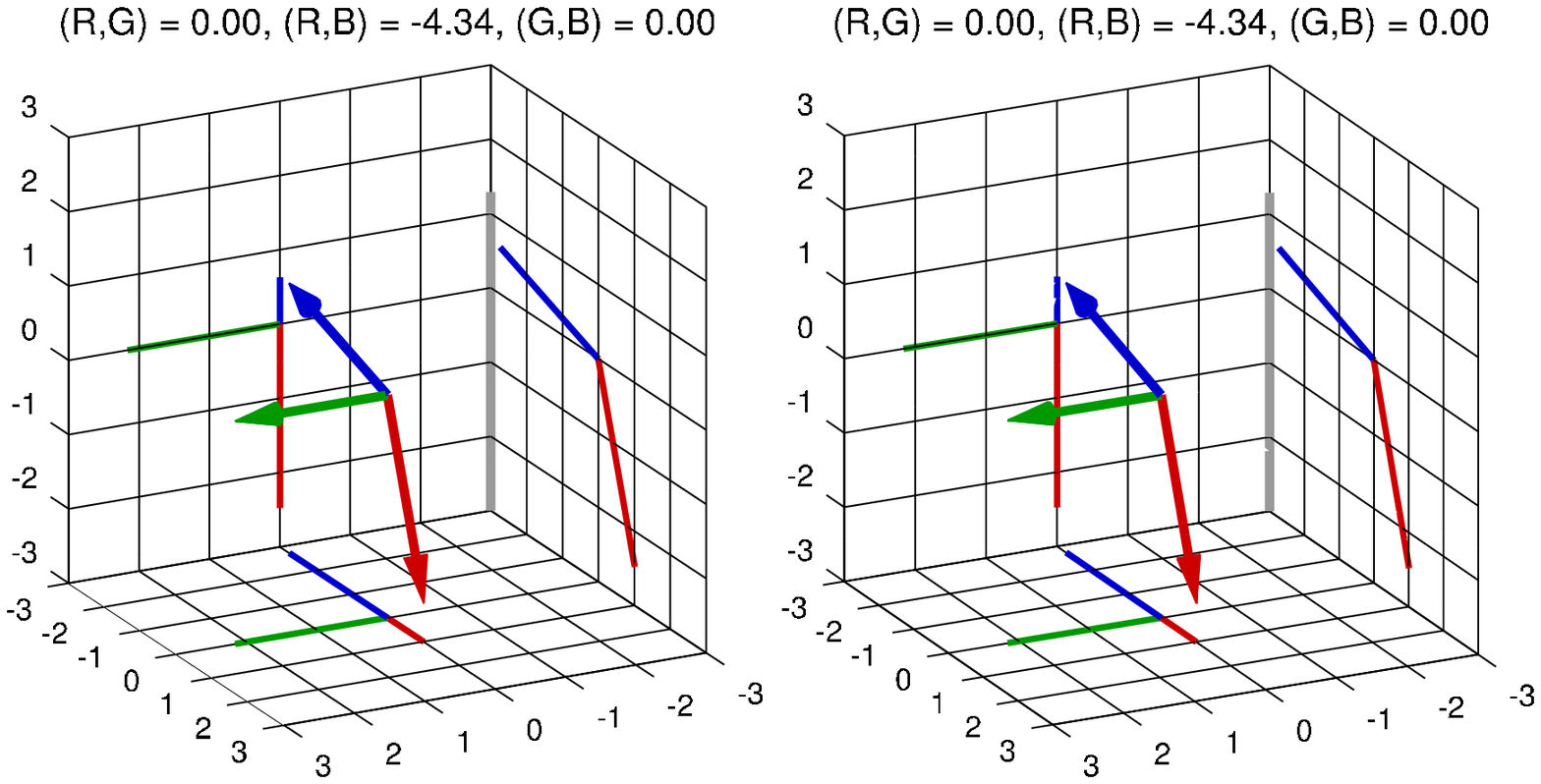}
\end{minipage}
\hspace{0.02 \linewidth}
\begin{minipage}{0.28 \linewidth}
\caption{ }
Dec. $x,x' \;|\; y, y'$, step 3:\\ 
$\color{myred}\vec{\Sigma}^1\color{black}\!\perp\color{mygreen}\vec{\Sigma}^2$,
$\color{mygreen}\vec{\Sigma}^2\color{black}\!\perp\color{myblue}\vec{\Sigma}^3$\\ $\color{mygreen}\vec{\Sigma}^2\color{black}\!\parallel$ 2-axis \\

Coefficient matrix $\Sigma_\kappa^\lambda=$

\small
$\left(\begin{array}{cccc}
4.3 & 0 & 0 & 0\\
0 & \bar{2}.5 & 0 & 0.6\\
0 & 0 & 2.2 & 0 \\
0 & 1.0 & 0 & \bar{2}.7
\end{array}\right)$
\normalsize
\end{minipage}
\end{figure}

\newpage
\subsubsection*{Decoupling $x,y'$ from $x', y$}

This means bringing the representative matrix and component matrix into the form

\vspace{5pt}
$\mathbf{\Sigma}=
\left(
\begin{array}{c c c c}
\Sigma_{11} & 0 & 0 & \Sigma_{14}\\
0 & \Sigma_{22} & \Sigma_{23} & 0 \\
0 & \Sigma_{32}  & \Sigma_{33} & 0 \\
\Sigma_{41} & 0 & 0  & \Sigma_{44} \\
\end{array}
\right)\qquad \Rightarrow \qquad (\Sigma_\kappa^\lambda)=
\left(
\begin{array}{r r r r}
\Sigma_0^0 & 0 & 0 & 0 \\
0 & \Sigma_1^1 & \Sigma_2^3 & 0 \\
0 & \Sigma_3^2 & \Sigma_2^2 & 0 \\
0 & 0 & 0 & \Sigma_3^3 \\
\end{array}
\right)$

\vspace{5pt}
\fbox{
\begin{minipage}{0.98\linewidth}
\vspace{5pt}

To decouple $x,y'$ from $x',y$, proceed as follows:

\vspace{5pt}
\begin{enumerate}
\item 
Decompose $\mathbf{\Sigma}=\mathbf{1}\Sigma^0_0+\vec{\boldsymbol\beta}\mathstrut^m\!\cdot \vec{\Sigma}^m$. \\
Boost $\vec{\Sigma}^1 \!\cdot \vec{\Sigma}^3$ to 0 with $\mathbf{\Sigma'}=\mathbf{R}_{\beta 3}{\mathbf{\Sigma}}\mathbf{R}_{\beta 3}^\intercal$, where 
\begin{equation*}
\mathbf{R}_{\beta 3}=\exp \left( \vec{\boldsymbol\beta}\mathstrut^2 \!\cdot \vec{e} \chi \right) 
\qquad \vec{e}=\frac{\vec{\Sigma}^2\!\wedge\vec{\Sigma}^3 -\Sigma^0_0\vec{\Sigma}^1}{|\vec{\Sigma}^2\!\wedge\vec{\Sigma}^3 -\Sigma^0_0\vec{\Sigma}^1|} 
\qquad \tanh (2\chi)=\frac{\vec{\Sigma}^1\!\cdot\vec{\Sigma}^3} {|\vec{\Sigma}^2\!\wedge\vec{\Sigma}^3 -\Sigma^0_0\vec{\Sigma}^1|}
\end{equation*} 
\item 
Rename $\mathbf{\Sigma'} \rightarrow\mathbf{\Sigma}$ and decompose $\mathbf{\Sigma}=\mathbf{1}\Sigma^0_0+\vec{\boldsymbol\beta}\mathstrut^m\!\cdot \vec{\Sigma}^m$. \\
Boost $\vec{\Sigma}^2 \!\cdot \vec{\Sigma}^3$ to 0 (keeping $\vec{\Sigma}^1 \!\cdot \vec{\Sigma}^3 = 0$) with $\mathbf{\Sigma'}=\mathbf{R}_{\beta 2}{\mathbf{\Sigma}}\mathbf{R}_{\beta 2}^\intercal$, where
\begin{equation*}
\mathbf{R}_{\beta 2}=\exp \left( \vec{\boldsymbol\beta}\mathstrut^2 \!\cdot \vec{e} \chi \right) 
\qquad 
\vec{e}=\frac{\vec{\Sigma}^1\!\wedge\vec{\Sigma}^2-\Sigma^0_0\vec{\Sigma}^3}{|\vec{\Sigma}^1\!\wedge\vec{\Sigma}^2-\Sigma^0_0\vec{\Sigma}^3|} 
\qquad
\tanh (2\chi)=\frac{\vec{\Sigma}^2 \!\cdot \vec{\Sigma}^3} {|\vec{\Sigma}^1\!\wedge\vec{\Sigma}^2-\Sigma^0_0\vec{\Sigma}^3|}
\end{equation*} 
\item 
Rename $\mathbf{\Sigma'} \rightarrow\mathbf{\Sigma}$ and decompose $\mathbf{\Sigma}=\mathbf{1}\Sigma^0_0+\vec{\boldsymbol\beta}\mathstrut^m\!\cdot \vec{\Sigma}^m$. \\
Rotate $\vec{\Sigma}^3$ parallel to the 3-axis  with  
$\mathbf{\Sigma'}=\mathbf{R}_{\zeta}{\mathbf{\Sigma}}\mathbf{R}_{\zeta}^\intercal$, where 
\begin{equation*}
\mathbf{R}_{\zeta}=\exp \left(\vec{\boldsymbol\zeta}\cdot\vec{e} \psi\right)
\qquad \vec{e}=\frac{(\Sigma_2^3,-\Sigma_1^3,0)^\intercal}{\sqrt{(\Sigma_1^3)^2+(\Sigma_2^3)^2}} 
\qquad  \tan (2\psi)=-\frac{\sqrt{(\Sigma_1^3)^2+(\Sigma_2^3)^2}}{\Sigma_3^3}
\end{equation*}

\end{enumerate}

\vspace{5pt}
The decoupling is preserved under the transformation group 
$\mathbf{R}=\exp\left[\boldsymbol\zeta_2 \psi + \boldsymbol\beta_2^2 \chi_1 \right.$ $\left. + \boldsymbol\beta_1^3 \chi_2+\boldsymbol\beta_3^3 \chi_3 \right]$ (free parameters $\psi, \chi_1, \chi_2, \chi_3$). 
\vspace{5pt}

\end{minipage}
}

\vspace{10pt}
\begin{figure}[H]
\begin{minipage}{0.68 \linewidth}
\centering
\includegraphics[trim = 0pt 0pt 0pt 0pt, width=1\textwidth]{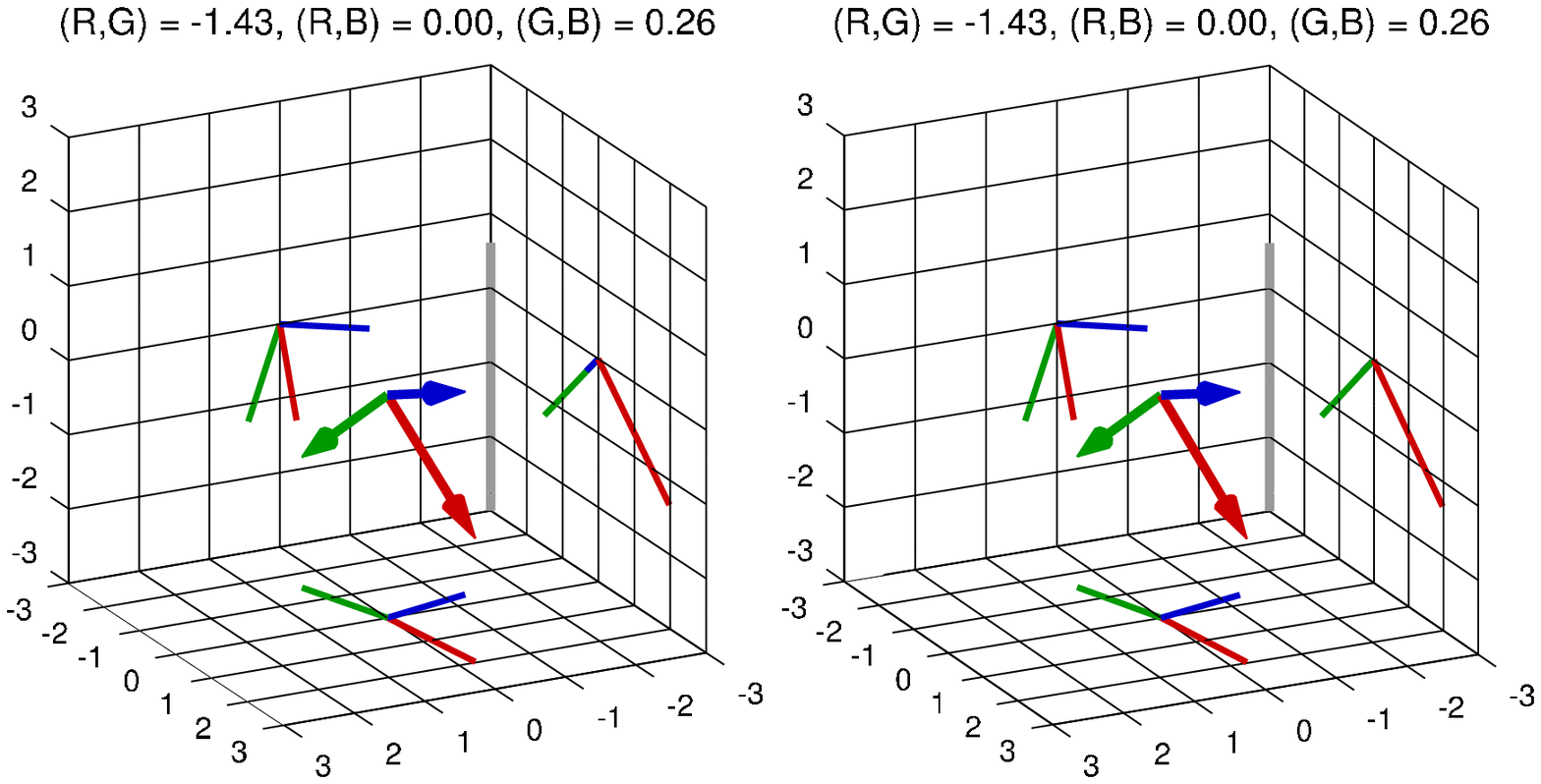}
\end{minipage}
\hspace{0.02 \linewidth}
\begin{minipage}{0.28 \linewidth}
\caption{ }

Dec. $x,x' \;|\; y, y'$, step 1:\\
$\color{myred}\vec{\Sigma}^1\color{black}\!\perp\color{myblue}\vec{\Sigma}^3$\\

Coefficient matrix $\Sigma_\kappa^\lambda=$

\small
$\left(\begin{array}{cccc}
3.6 & 0 & 0 & 0\\
0 & \bar{1}.3 & \bar{1}.2 & \bar{0}.3 \\
0 & \bar{0}.2 & 0.4 & \bar{1}.3 \\
0 & 2.0 & \bar{1}.5 & \bar{0}.3
\end{array}\right)$
\normalsize 
\end{minipage}
\end{figure}

\vspace{10pt}
\begin{figure}[H]
\begin{minipage}{0.68 \linewidth}
\centering
\includegraphics[trim = 0pt 0pt 0pt 0pt, width=1\textwidth]{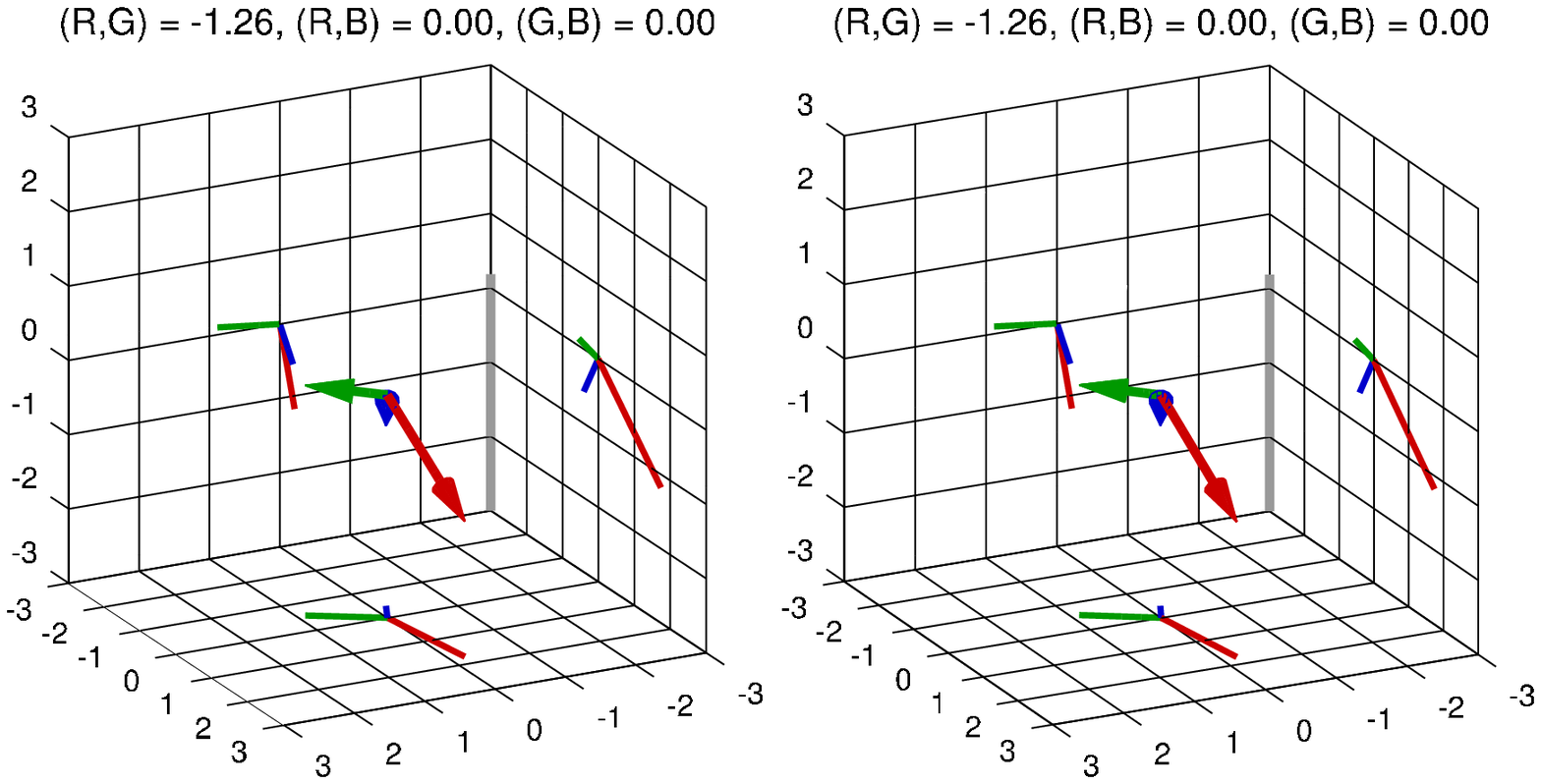}
\end{minipage}
\hspace{0.02 \linewidth}
\begin{minipage}{0.28 \linewidth}
\caption{ }

Dec. $x,x' \;|\; y, y'$, step 2:\\
$\color{myred}\vec{\Sigma}^1\color{black}\!\perp\color{myblue}\vec{\Sigma}^3$, 
$\color{mygreen}\vec{\Sigma}^2\color{black}\!\perp\color{myblue}\vec{\Sigma}^3$ \\

Coefficient matrix $\Sigma_\kappa^\lambda=$

\small
$\left(\begin{array}{cccc}
3.2 & 0 & 0 & 0\\
0 & \bar{1}.2 & 0.1 & \bar{0}.6 \\
0 & \bar{0}.2 & 0.9 & \bar{0}.2 \\
0 & 1.7 & \bar{0}.5 & \bar{0}.4
\end{array}\right)$
\normalsize
\end{minipage}
\end{figure}

\vspace{10pt}
\begin{figure}[H]
\begin{minipage}{0.68 \linewidth}
\centering
\includegraphics[trim = 0pt 0pt 0pt 0pt, width=1\textwidth]{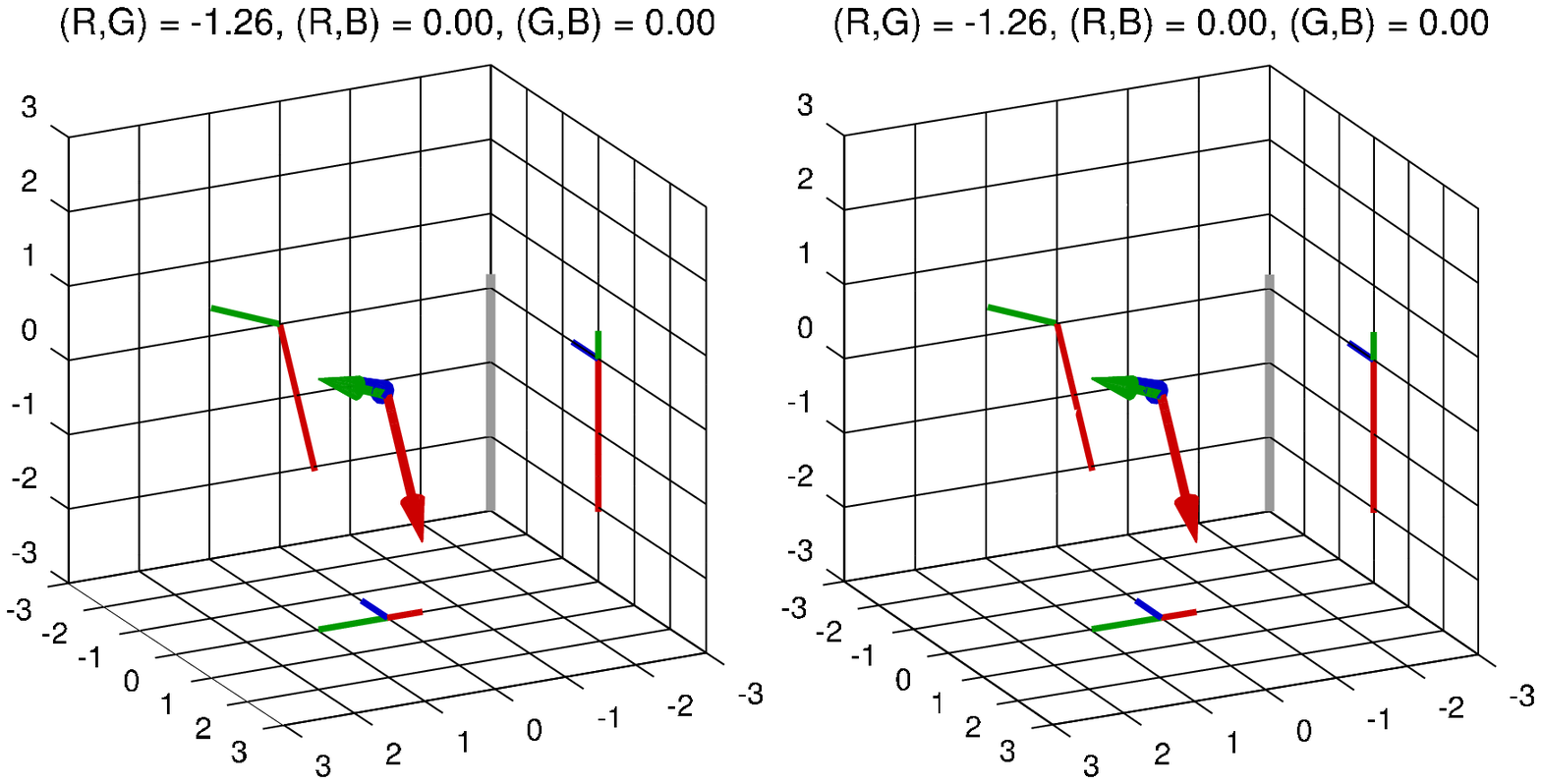}
\end{minipage}
\hspace{0.02 \linewidth}
\begin{minipage}{0.28 \linewidth}
\caption{ }
Dec. $x,x' \;|\; y, y'$, step 3:\\ 
$\color{myred}\vec{\Sigma}^1\color{black}\!\perp\color{myblue}\vec{\Sigma}^3$,
$\color{mygreen}\vec{\Sigma}^2\color{black}\!\perp\color{myblue}\vec{\Sigma}^3$\\ $\color{myblue}\vec{\Sigma}^1\color{black}\!\parallel$ 1-axis \\

Coefficient matrix $\Sigma_\kappa^\lambda=$

\small
$\left(\begin{array}{cccc}
3.2 & 0 & 0 & 0\\
0 & \bar{2}.1 & 0.4 & 0\\
0 & \bar{0}.5 & 1.0 & 0 \\
0 & 0 & 0 & \bar{0}.7
\end{array}\right)$
\normalsize
\end{minipage}
\end{figure}

\vspace{5pt}
\subsection{Diagonalizing the beam matrix}

Diagonalizing the beam matrix means mutually decoupling all 4 phase space coordinates. The beam component matrix is then also diagonal, i. e. it has $\vec{\Sigma}^1$, $\vec{\Sigma}^2$ and $\vec{\Sigma}^3$ aligned to the 1-, 2- and 3-axis respectively. 

\vspace{5pt}
$\mathbf{\Sigma}=
\left(
\begin{array}{c c c c}
\Sigma_{11} & 0 & 0 & 0 \\
0 & \Sigma_{22} & 0 & 0 \\
0 & 0 & \Sigma_{33} & 0 \\
0 & 0 & 0 & \Sigma_{44} \\
\end{array}
\right)\qquad \Rightarrow \qquad (\Sigma_\kappa^\lambda)=
\left(
\begin{array}{r r r r}
\Sigma_0^0 & 0 & 0 & 0 \\
0 & \Sigma_1^1 & 0 & 0\\
0 & 0 & \Sigma_2^2 & 0 \\
0 & 0 & 0 & \Sigma_3^3 \\
\end{array}
\right)$

\vspace{5pt}
\newpage
\fbox{
\begin{minipage}{0.98\linewidth}
\vspace{5pt}

To diagonalize a beam matrix, first transform the beam matrix to block-diagonal form, see steps 1 to 3 in section 5.6. Then continue as follows.

\vspace{5pt}
\begin{enumerate}
\setcounter{enumi}{3}
\item 
Rename $\mathbf{\Sigma'}\rightarrow\mathbf{\Sigma}$ and decompose $\mathbf{\Sigma}=\mathbf{1}\Sigma^0_0+\vec{\boldsymbol\beta}\mathstrut^m\!\cdot \vec{\Sigma}^m$. \\
Rotate $\vec{\Sigma}^1 \!\cdot \vec{\Sigma}^2$ to 0 (keeping $\vec{\Sigma}^2 \!\cdot \vec{\Sigma}^3 = \vec{\Sigma}^1 \!\cdot \vec{\Sigma}^3 = 0$) with $\mathbf{\Sigma}'=\mathbf{R}_{\gamma}\mathbf{\Sigma}\mathbf{R}_{\gamma}^\intercal\rightarrow\mathbf{\Sigma}$, where 
\begin{equation*}
\mathbf{R}_{\gamma}=\exp \left( \boldsymbol\gamma^1 \phi \right) \qquad  \tan (4\phi)=\frac{2 \vec{\Sigma}^2 \!\cdot \vec{\Sigma}^3}{(\vec{\Sigma}^2)^2 - (\vec{\Sigma}^3)^2}
\end{equation*}
\item Rename $\mathbf{\Sigma'}\rightarrow\mathbf{\Sigma}$ and decompose $\mathbf{\Sigma}=\mathbf{1}\Sigma^0_0+\vec{\boldsymbol\beta}\mathstrut^m\!\cdot \vec{\Sigma}^m$. \\
The three vectors $\vec{\Sigma}^1$, $\vec{\Sigma}^2$, $\vec{\Sigma}^3$ are now mutually perpendicular.\\
Align $\vec{\Sigma}^2$ and $\vec{\Sigma}^3$ to the 2- and 3-axis with $\mathbf{\Sigma}'=\mathbf{R}_{\zeta}\mathbf{\Sigma}\mathbf{R}_{\zeta}^\intercal\rightarrow\mathbf{\Sigma}$, where 
\begin{equation*}
\mathbf{R_\zeta}=\exp \left( \boldsymbol\zeta_1 \psi \right) \qquad  \tan (2\psi)=\frac{\Sigma_3^2}{\Sigma_2^2}=-\frac{\Sigma_2^3}{\Sigma_3^3}
\end{equation*}
\end{enumerate}

The diagonal form is preserved under the transformation group
$\mathbf{R}=\exp\left[ \boldsymbol\beta_2^2 \chi_2+\boldsymbol\beta_3^3 \chi_3 \right]$ (free parameters $\chi_2, \chi_3$). 
\vspace{5pt}
\end{minipage}
}

\vspace{10pt}
\begin{figure}[H]
\begin{minipage}{0.68 \linewidth}
\centering
\includegraphics[trim = 0pt 0pt 0pt 0pt, width=1\textwidth]{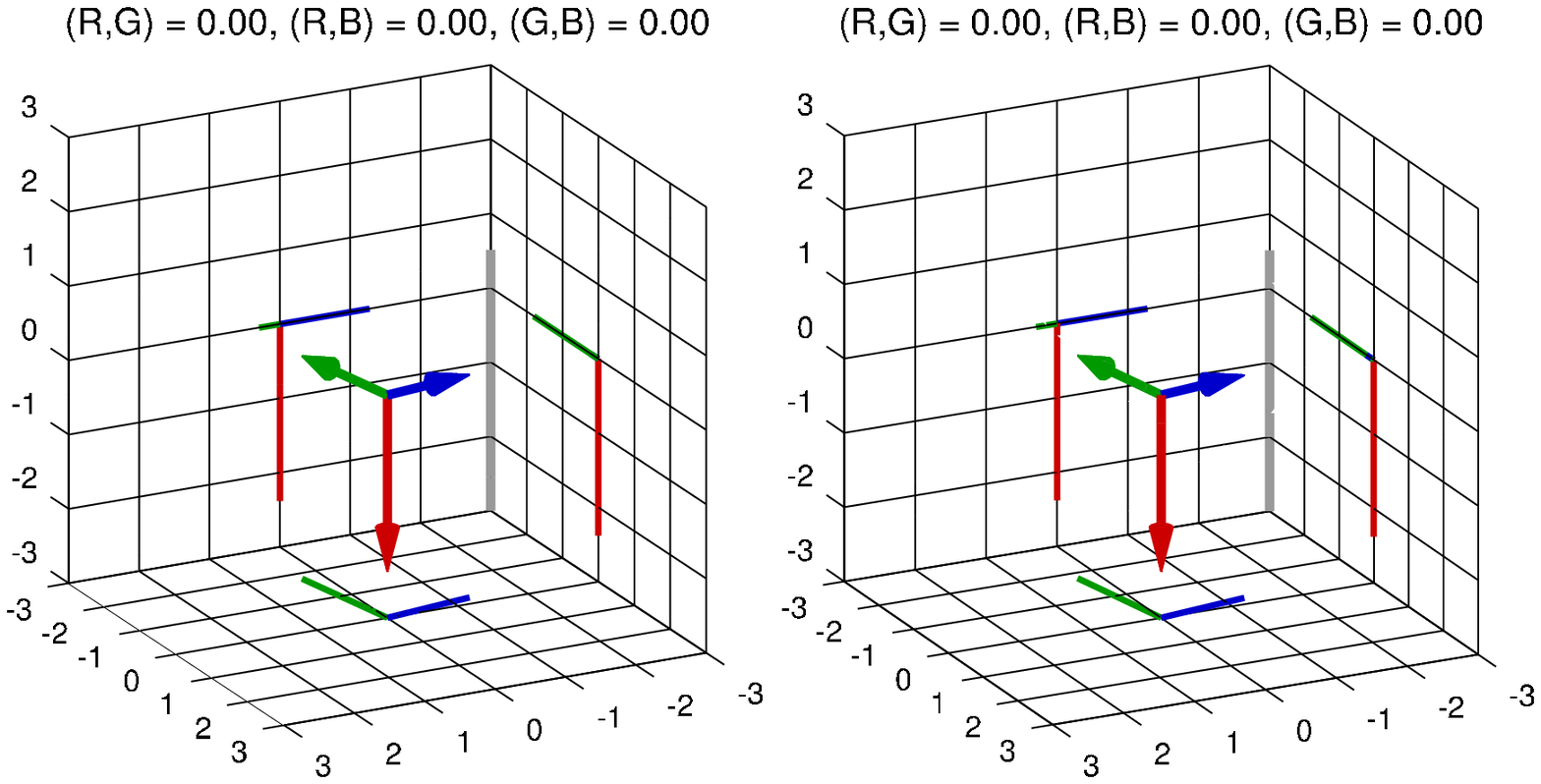}
\end{minipage}
\hspace{0.02 \linewidth}
\begin{minipage}{0.28 \linewidth}
\caption{ }

Diagonalization, step 4:\\ $\color{myred}\vec{\Sigma}^1\color{black}\!\perp\color{mygreen}\vec{\Sigma}^2$,
$\color{myred}\vec{\Sigma}^1\color{black}\!\perp\color{myblue}\vec{\Sigma}^3$, \\ $\color{mygreen}\vec{\Sigma}^2\color{black}\!\perp\color{myblue}\vec{\Sigma}^3$ \\

Coefficient matrix $\Sigma_\kappa^\lambda=$

\small
$\left(\begin{array}{cccc}
3.5 & 0 & 0 & 0\\
0 & \bar{2}.4 & 0 & 0\\
0 & 0 & 0.3 & \bar{1}.3 \\
0 & 0 & \bar{1}.8 & \bar{0}.2
\end{array}\right)$
\normalsize
\end{minipage}
\end{figure}

\vspace{5pt}
\begin{figure}[H]
\begin{minipage}{0.68 \linewidth}
\centering
\includegraphics[trim = 0pt 0pt 0pt 0pt, width=1\textwidth]{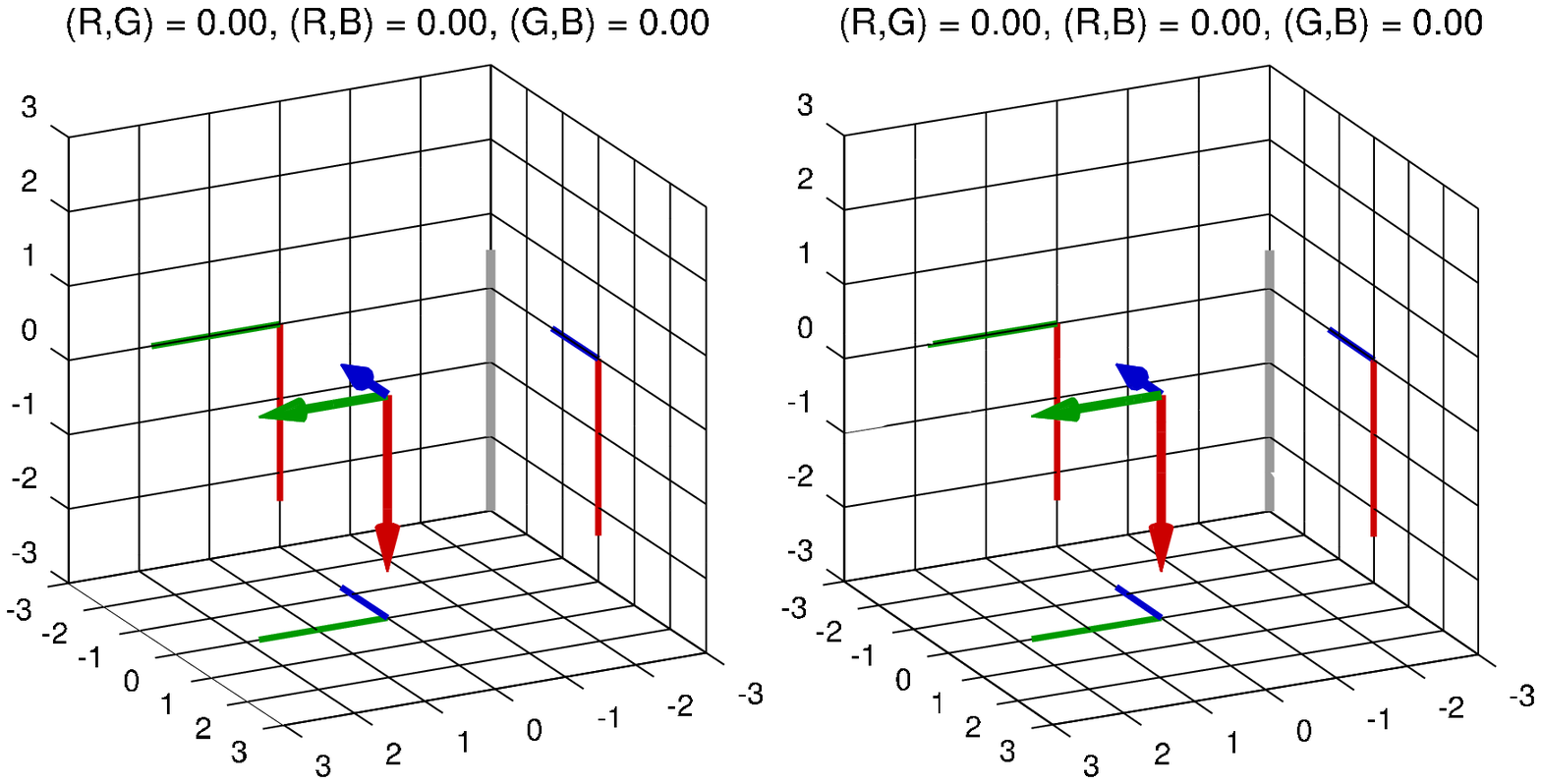}
\end{minipage}
\hspace{0.02 \linewidth}
\begin{minipage}{0.28 \linewidth}
\caption{ }

Diagonalization, step 5:\\
$\color{myred}\vec{\Sigma}^1\color{black}\!\parallel 1$-axis, $\color{mygreen}\vec{\Sigma}^2 \color{black}\!\parallel 2$-axis, \\ $\color{myblue}\vec{\Sigma}^3 \color{black}\!\parallel 3$-axis\\

Coefficient matrix $\Sigma_\kappa^\lambda=$

\small
$\left(\begin{array}{cccc}
3.5 & 0 & 0 & 0\\
0 & \bar{2}.4 & 0 & 0\\
0 & 0 & 1.8 & 0 \\
0 & 0 & 0 & \bar{1}.3
\end{array}\right)$
\normalsize 
\end{minipage}
\end{figure}

\newpage
Here is an alternative diagonalization recipe that doesn't go through block-diagonalization.

\fbox{
\begin{minipage}{0.98\linewidth}
\vspace{5pt}

To diagonalize a beam matrix, proceed as follows.

\vspace{5pt}
\begin{enumerate}
\item 
Decompose $\mathbf{\Sigma}=\mathbf{1}\Sigma^0_0+\vec{\boldsymbol\beta}\mathstrut^m\!\cdot \vec{\Sigma}^m$. \\
Boost $\vec{\Sigma}^2 \!\cdot \vec{\Sigma}^1$ to 0 with $\mathbf{\Sigma'}=\mathbf{R}_{\beta 2}{\mathbf{\Sigma}}\mathbf{R}_{\beta 2}^\intercal$, where 
\begin{equation*}
\mathbf{R}_{\beta 2}=\exp \left( \vec{\boldsymbol\beta}\mathstrut^2 \!\cdot \vec{e} \chi \right) 
\qquad \vec{e}=\frac{\vec{\Sigma}^2\!\wedge\vec{\Sigma}^3-\Sigma^0_0\vec{\Sigma}^1}{|\vec{\Sigma}^2\!\wedge\vec{\Sigma}^3-\Sigma^0_0\vec{\Sigma}^1|} 
\qquad
\tanh (2\chi)=\frac{\vec{\Sigma}^2 \!\cdot \vec{\Sigma}^1} {|\vec{\Sigma}^2\!\wedge\vec{\Sigma}^3-\Sigma^0_0\vec{\Sigma}^1|}
\end{equation*} 
\item 
Rename $\mathbf{\Sigma'} \rightarrow\mathbf{\Sigma}$ and decompose $\mathbf{\Sigma}=\mathbf{1}\Sigma^0_0+\vec{\boldsymbol\beta}\mathstrut^m\!\cdot \vec{\Sigma}^m$. \\
Boost $\vec{\Sigma}^3 \!\cdot \vec{\Sigma}^1$ to 0 (keeping $\vec{\Sigma}^2 \!\cdot \vec{\Sigma}^1 = 0$) with $\mathbf{\Sigma'}=\mathbf{R}_{\beta 3}{\mathbf{\Sigma}}\mathbf{R}_{\beta 3}^\intercal$, where
\begin{equation*}
\mathbf{R}_{\beta 3}=\exp \left( \vec{\boldsymbol\beta}\mathstrut^3 \!\cdot \vec{e} \chi \right) 
\qquad 
\vec{e}=\frac{\vec{\Sigma}^2\!\wedge\vec{\Sigma}^3-\Sigma^0_0\vec{\Sigma}^1}{|\vec{\Sigma}^2\!\wedge\vec{\Sigma}^3-\Sigma^0_0\vec{\Sigma}^1|} 
\qquad
\tanh (2\chi)=\frac{\vec{\Sigma}^3 \!\cdot \vec{\Sigma}^1} {|\vec{\Sigma}^2\!\wedge\vec{\Sigma}^3-\Sigma^0_0\vec{\Sigma}^1|}
\end{equation*} 
\item 
Rename $\mathbf{\Sigma'}\rightarrow\mathbf{\Sigma}$ and decompose $\mathbf{\Sigma}=\mathbf{1}\Sigma^0_0+\vec{\boldsymbol\beta}\mathstrut^m\!\cdot \vec{\Sigma}^m$. \\
Rotate $\vec{\Sigma}^2 \!\cdot \vec{\Sigma}^3$ to 0 (keeping $\vec{\Sigma}^2 \!\cdot \vec{\Sigma}^1 = \vec{\Sigma}^3 \!\cdot \vec{\Sigma}^1 = 0$) with $\mathbf{\Sigma'}= \mathbf{R}_{\gamma}\mathbf{\Sigma}\mathbf{R}_{\gamma}^\intercal$, where
\begin{equation*}
\mathbf{R}_{\gamma}=\exp \left( \boldsymbol\gamma^1 \phi \right) \qquad  \tan (4\phi)=\frac{2 \vec{\Sigma}^2 \!\cdot \vec{\Sigma}^3}{(\vec{\Sigma}^2)^2 - (\vec{\Sigma}^3)^2}
\end{equation*}
\item Rename $\mathbf{\Sigma'} \rightarrow\mathbf{\Sigma}$ and decompose $\mathbf{\Sigma}=\mathbf{1}\Sigma^0_0+\vec{\boldsymbol\beta}\mathstrut^m\!\cdot \vec{\Sigma}^m$. 
The three vectors $\vec{\Sigma}^1$, $\vec{\Sigma}^2$, $\vec{\Sigma}^3$ are now perpendicular to each other. Calculate 
\begin{equation*}
\vec{T}^1=\pm\vec{\Sigma}^1 / |\vec{\Sigma}^1| \qquad 
\vec{T}^2=\pm \vec{\Sigma}^2 / |\vec{\Sigma}^2| \qquad 
\vec{T}^3= \vec{T}^1 \!\wedge \vec{T}^2
\end{equation*}
To make the last rotation angle as small as possible, choose the two (independent) signs such that $T^1_1 + T^2_2 + T^3_3$ becomes maximally positive. Calculate
\begin{equation*}
\cos (2\psi)=\left(T^1_1 + T^2_2 + T^3_3-1\right)/2 \quad \Rightarrow \quad \psi \text{ in } [0,\pi/2] 
\end{equation*}
\begin{equation*}
e_1=\frac{T^3_2 - T^2_3}{2\sin (2\psi)} \qquad 
e_2=\frac{T^1_3 - T^3_1}{2\sin (2\psi)} \qquad 
e_3=\frac{T^2_1 - T^1_2}{2\sin (2\psi)}
\end{equation*}
Rotate $\vec{\Sigma}^1$, $\vec{\Sigma}^2$, $\vec{\Sigma}^3$ parallel to the 1, 2, 3-axis with $\mathbf{\Sigma'}= \mathbf{R}_{\zeta}{\mathbf{\Sigma}}\mathbf{R}_{\zeta}^\intercal$, where
\begin{equation*}
\mathbf{R_\zeta}=\exp \left( \vec{\boldsymbol\zeta} \cdot \vec{e} \psi \right) 
\end{equation*}
\end{enumerate}
\end{minipage}
}

\vspace{5pt}
\textbf{Comment}: The formulas in step 4 for the $\zeta$-rotation parameters $\vec{e}$ and $2\psi$ were obtained as follows. There are 4 possible ways to align $\vec{\Sigma}^1$ to $(1,0,0)^\intercal$, $\vec{\Sigma}^2$ to $(0,1,0)^\intercal$ and 
$\vec{\Sigma}^3$ to $(0,0,1)^\intercal$ (two signs are free). The idea is to choose the alternative which needs the smallest rotation angle. The 3 unit vectors $\vec{T}^1$, $\vec{T}^2$ and $\vec{T}^3$ that are rotated into $(1,0,0)^\intercal$, $(0,1,0)^\intercal$ and $(0,0,1)^\intercal$ and $\vec{e}$  are defined accordingly. 
The matrix $\mathbf{T}=[\vec{T}^1,\vec{T}^2,\vec{T}^3]$ performs the inverse of this rotation in $\mathbb{R}^3$. Eulers rotation formula expresses $\mathbf{T}$ in terms of $\vec{e}$ and $\psi$: $T_{kl}=\delta_{kl}\cos (2\psi) -e_i\epsilon_{ikl}\sin(2\psi) + e_ke_l(1-\cos(2\psi))$. Follows now $T_{kk}=1+2\cos(2\psi)$ and $\epsilon_{ikl}T_{kl}=-2e_i\sin(2\psi)$, and the formulas for $\vec{e}$ \& $\psi$.

\vspace{10pt}
\begin{figure}[H]
\begin{minipage}{0.68 \linewidth}
\centering
\includegraphics[trim = 0pt 0pt 0pt 0pt, width=1\textwidth]{DiracB1.eps}
\end{minipage}
\hspace{0.02 \linewidth}
\begin{minipage}{0.28 \linewidth}
\caption{ }

Diagonalization, step 1:\\
$\color{myred}\vec{\Sigma}^1\color{black}\!\perp\color{mygreen}\vec{\Sigma}^2$\\

Coefficient matrix $\Sigma_\kappa^\lambda=$

\small
$\left(\begin{array}{cccc}
4.3 & 0 & 0 & 0\\
0 & \bar{1}.3 & \bar{1}.9 & 0.7 \\
0 & \bar{1}.6 & 0.6 & \bar{0}.8 \\
0 & 1.8 & \bar{0}.9 & \bar{2}.7
\end{array}\right)$
\normalsize
\end{minipage}
\end{figure}

\begin{figure}[H]
\begin{minipage}{0.68 \linewidth}
\centering
\includegraphics[trim = 0pt 0pt 0pt 0pt, width=1\textwidth]{DiracB2.eps}
\end{minipage}
\hspace{0.02 \linewidth}
\begin{minipage}[Ht]{0.28 \linewidth}
\caption{ }

Diagonalization, step 2:\\
$\color{myred}\vec{\Sigma}^1\color{black}\!\perp\color{mygreen}\vec{\Sigma}^1$, 
$\color{myred}\vec{\Sigma}^1\color{black}\!\perp\color{myblue}\vec{\Sigma}^3$\\

Coefficient matrix $\Sigma_\kappa^\lambda=$

\small
$\left(\begin{array}{cccc}
3.5 & 0 & 0 & 0\\
0 & \bar{1}.1 & \bar{1}.6 & \bar{0}.5 \\
0 & \bar{0}.3 & 0.5 & \bar{1}.2 \\
0 & 2.1 & \bar{0}.7 & \bar{0}.4
\end{array}\right)$
\normalsize
\end{minipage}
\end{figure}

\vspace{5pt}
\begin{figure}[H]
\begin{minipage}{0.68 \linewidth}
\centering
\includegraphics[trim = 0pt 0pt 0pt 0pt, width=1\textwidth]{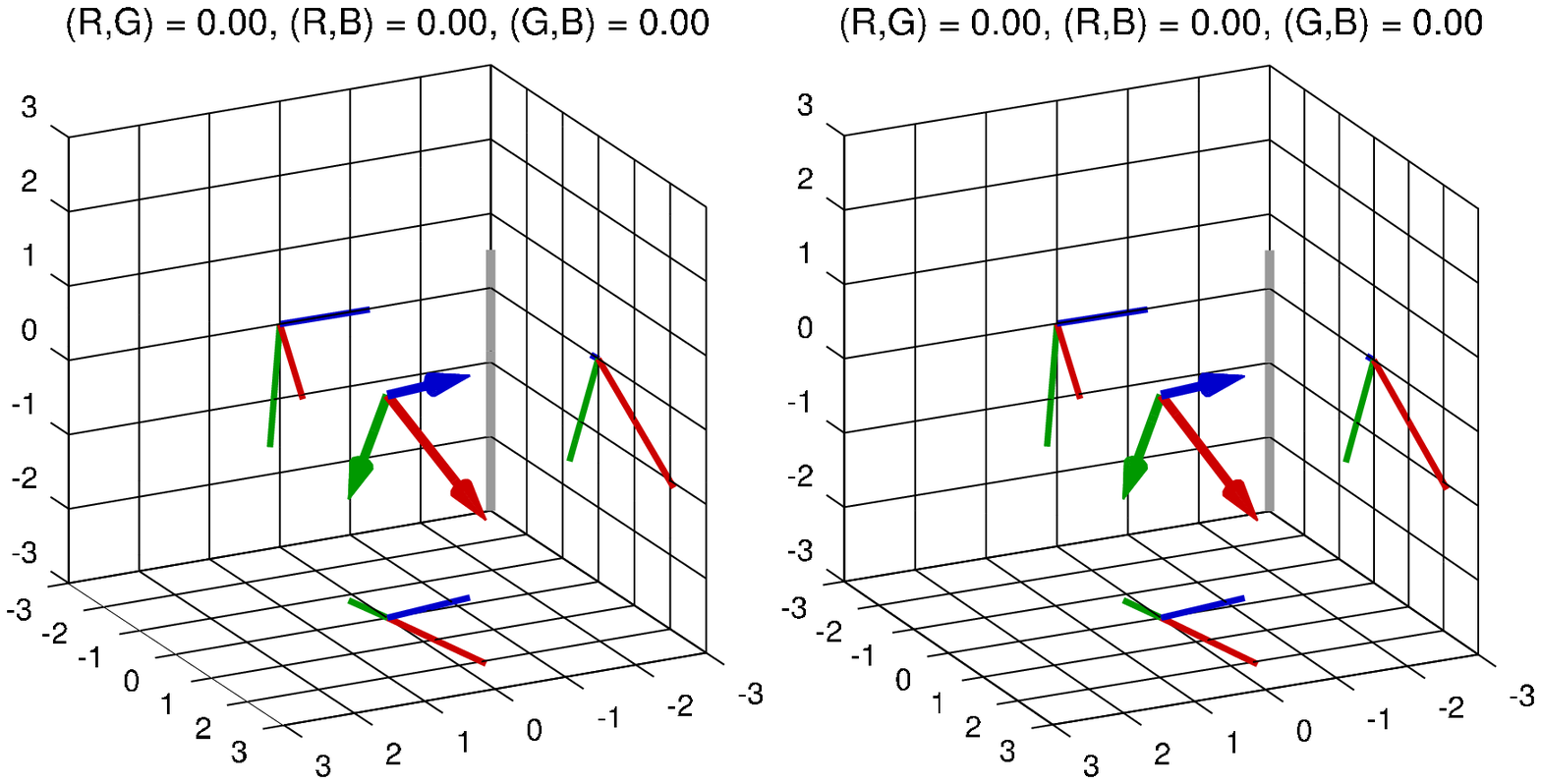}
\end{minipage}
\hspace{0.02 \linewidth}
\begin{minipage}{0.28 \linewidth}
\caption{ }

Diagonalization, step 3:\\
$\color{myred}\vec{\Sigma}^1\color{black}\!\!\perp\color{mygreen}\vec{\Sigma}^2$, $\color{myred}\vec{\Sigma}^1\color{black}\!\!\perp\color{myblue}\vec{\Sigma}^3$\\ $\color{mygreen}\vec{\Sigma}^2\color{black}\!\!\perp\color{myblue}\vec{\Sigma}^3$ \\

Coefficient matrix $\Sigma_\kappa^\lambda=$

\small
$\left(\begin{array}{cccc}
3.5 & 0 & 0 & 0\\
0 & \bar{1}.1 & \bar{1}.6 & 0.0\\
0 & \bar{0}.3 & 0.1 & \bar{1}.3 \\
0 & 2.1 & \bar{0}.8 & \bar{0}.2
\end{array}\right)$
\normalsize
\end{minipage}
\end{figure}

\vspace{5pt}
\begin{figure}[H]
\begin{minipage}{0.68 \linewidth}
\centering
\includegraphics[trim = 0pt 0pt 0pt 0pt, width=1\textwidth]{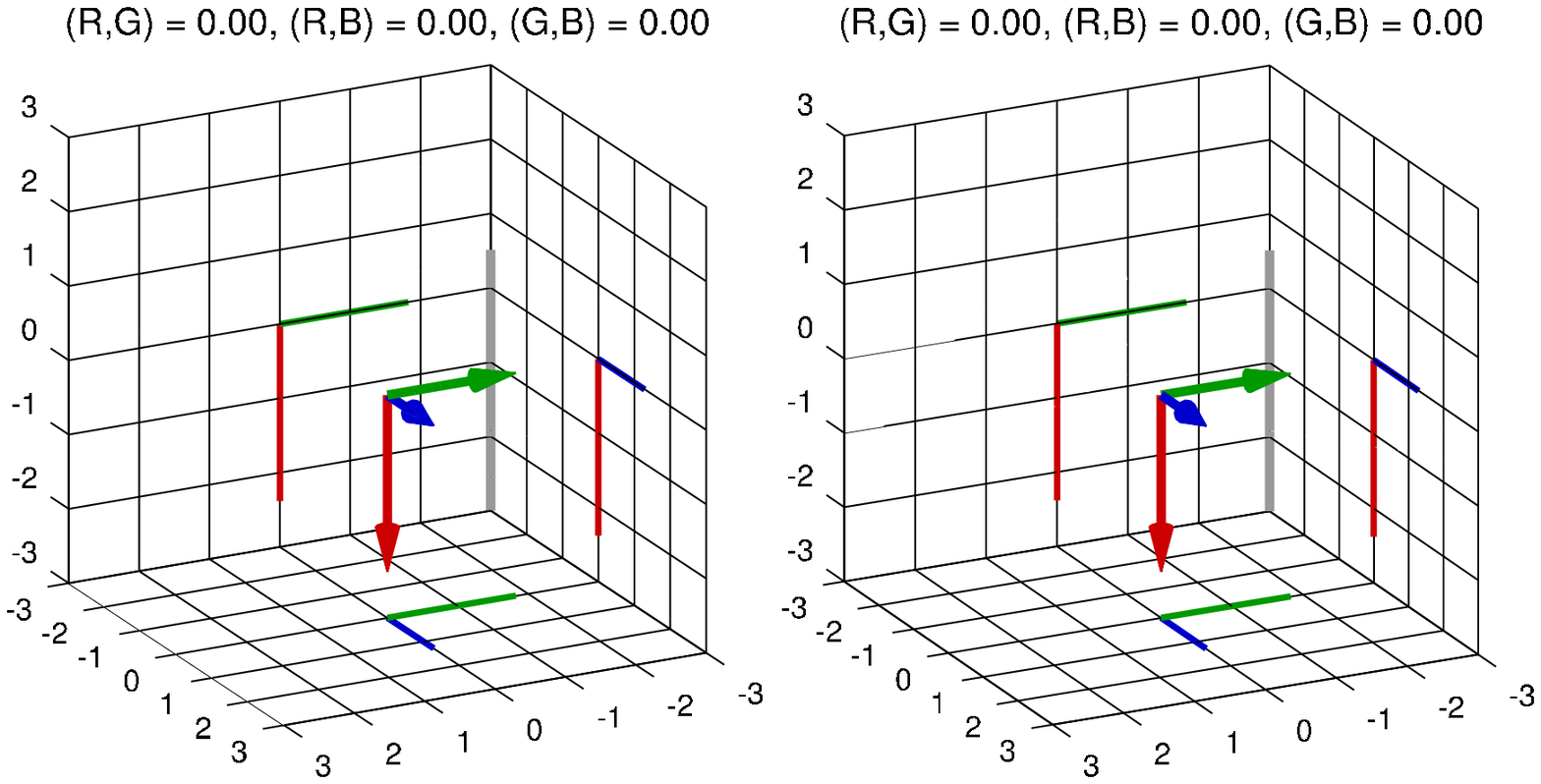}
\end{minipage}
\hspace{0.02 \linewidth}
\begin{minipage}[Ht]{0.28 \linewidth}
\caption{ }

Diagonalization, step 4:\\
$\color{myred}\vec{\Sigma}^1\color{black}\!\parallel\text{1-axis}$,
$\color{mygreen}\vec{\Sigma}^2\color{black}\!\parallel\text{2-axis}$ \\
$\color{myblue}\vec{\Sigma}^3\color{black}\!\parallel\text{3-axis}$\\

Coefficient matrix $\Sigma_\kappa^\lambda=$

\small
$\left(\begin{array}{cccc}
3.5 & 0 & 0 & 0\\
0 & \bar{2}.4 & 0 & 0\\
0 & 0 & \bar{1}.8 & 0 \\
0 & 0 & 0 & \bar{1}.3
\end{array}\right)$
\normalsize
\end{minipage}
\end{figure}

\newpage
\subsection{Normalizing the beam matrix}

A beam matrix in normal form has $\vec{\Sigma}^1$ aligned to the 1-axis, and vanishing $\vec{\Sigma}^2$, $\vec{\Sigma}^3$:

\vspace{5pt}
$\mathbf{\Sigma}=
\left(
\begin{array}{cccc}
\epsilon_{I} & 0 & 0 & 0\\
0 & \epsilon_{I} & 0 & 0\\
0 & 0 & \epsilon_{II} & 0 \\
0 & 0 & 0 & \epsilon_{II}
\end{array}
\right)\qquad \Rightarrow \qquad (\Sigma_\kappa^\lambda)=
\left(
\begin{array}{r r r r}
\Sigma_0^0 & 0 & 0 & 0 \\
0 & \Sigma_1^1 & 0 & 0\\
0 & 0 & 0 & 0 \\
0 & 0 & 0 & 0 \\
\end{array}
\right)$

\vspace{5pt}
\fbox{
\begin{minipage}{0.98\linewidth}
\vspace{5pt}

To normalize a beam matrix, first transform it to diagonal form, see sections 5.6-7. \\Then rename $\mathbf{\Sigma'}\rightarrow\mathbf{\Sigma}$ and decompose $\mathbf{\Sigma}=\mathbf{1} \Sigma^0_0+\boldsymbol\beta^1_1 \Sigma^1_1+\boldsymbol\beta^2_2 \Sigma^2_2+\boldsymbol\beta^3_3 \Sigma^3_3$. \\
Annihilate $\Sigma^2_2$ and $\Sigma^3_3$ with $\tilde{\mathbf{\Sigma}}=\mathbf{R}_{sc}\mathbf{\Sigma}\mathbf{R}_{sc}^\intercal$, where $\mathbf{R}_{sc}$ is the scale transformation
\begin{equation*}
\mathbf{R}_{sc}=\text{diag}
\left(
\sqrt[4]{\frac{\Sigma_{22}}{\Sigma_{11} }},
\sqrt[4]{\frac{\Sigma_{11}}{\Sigma_{22}} },
\sqrt[4]{\frac{\Sigma_{44}}{\Sigma_{33}} },
\sqrt[4]{\frac{\Sigma_{33}}{\Sigma_{44}} }
\right)
\end{equation*}

The normal form is preserved under the transformation group 
$\mathbf{R}=\exp\left( \boldsymbol\zeta_1 \psi +\boldsymbol\gamma^1 \phi\right)$ 
(free parameters $\psi, \phi$). 
\vspace{5pt}
\end{minipage}
}

\vspace{10pt}
\textbf{Comment}: When the beam matrix $\mathbf{\Sigma}$ is diagonal, a symplectic rescaling of the phase space axes achieves normalization. This rescaling could be expressed in terms of a double-boost: $\mathbf{R}_{sc}=\exp \left( \boldsymbol\beta^1_1 \chi_1 + \boldsymbol\beta^2_2 \chi_2 \right)$, but the formulas for the two boost angles are unnecessarily complicated. It is simpler to construct $\mathbf{R}_{sc}$ with the elements of the (diagonal) representative matrix.

\vspace{10pt}
\begin{figure}[H]
\begin{minipage}{0.68 \linewidth}
\centering
\includegraphics[trim = 0pt 0pt 0pt 0pt, width=1\textwidth]{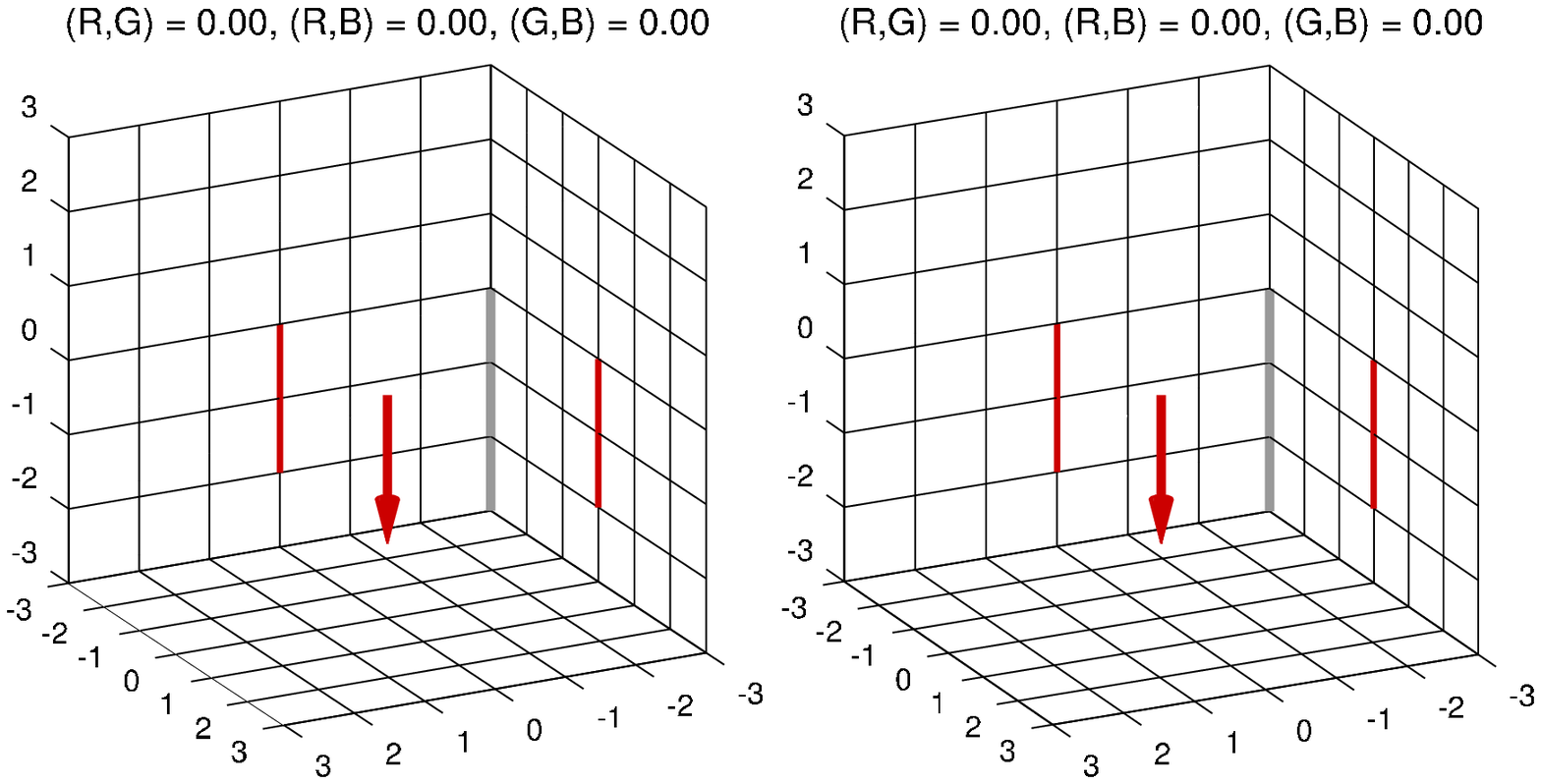}
\end{minipage}
\hspace{0.02 \linewidth}
\begin{minipage}{0.28 \linewidth}
\caption{ }

Normalization:\\
$\color{myred}\vec{\Sigma}^1\color{black}\!\parallel 1, \;
\color{mygreen}\vec{\Sigma}^2\color{black}=\color{myblue}\vec{\Sigma}^3 \color{black}=0$\\

Coefficient matrix $\Sigma_\kappa^\lambda=$

\small
$\left(\begin{array}{cccc}
3 & 0 & 0 & 0\\
0 & \bar{2} & 0 & 0\\
0 & 0 & 0 & 0 \\
0 & 0 & 0 & 0
\end{array}\right)$
\normalsize
\end{minipage}
\end{figure}

\subsection{Invariance group of the beam matrix}

Which transformations $\mathbf{I}$ leave the beam matrix $\mathbf{\Sigma}$ invariant: $\mathbf{I}\mathbf{\Sigma}\mathbf{I}^\intercal= \mathbf{\Sigma}$ ? 

\fbox{
\begin{minipage}{0.98\linewidth}
\vspace{5pt}
Let $\mathbf{N}$ be a normalizing transformation to a beam matrix $\mathbf{\Sigma}$, i. e. \\$\mathbf{\Sigma}=\mathbf{N}\left(\mathbf{1} \Sigma^0_0+\boldsymbol\beta^1_1 \Sigma^1_1\right)\mathbf{N}^\intercal$; the beam matrix is invariant under the group of transformations
$\mathbf{I}\left(\psi, \phi\right)=\mathbf{N}\exp\left( \boldsymbol\zeta_1 \psi +\boldsymbol\gamma^1 \phi\right)\mathbf{N}^{-1}$ (free parameters $\psi, \phi$).
\vspace{5pt}
\end{minipage}
}

\subsection{Alternative transformation formulas}

I haven't addressed yet the problem of decoupling one phase space coordinate from the other three. The reason is, the visual approach used until now didn't suggest to me a way to achieve this kind of decoupling. Instead I had to resort to the following "alternative rotation formulas". (The "alternative boost formulas" are given for completeness; they will not be used.)

The linear transformation formulas for $\mathbf{\Sigma}$ given in section 5.5 can be written  as a $10\times 10$ matrix acting on a 10-component vector $\Sigma_\kappa^\lambda$, or more generally as a $16\times 16$ matrix acting on a 16-component vector $Z_\kappa^\lambda$. Matrix formulas of this size are not very alluring. However, there is a more compact way to express the transformations: by $4\times 4$ matrices acting from the left or right on the component matrix $Z_\kappa^\lambda$. Here is how it is done.

First we express the action of multiplying $\mathbf{Z}$ from left or right with a unit as a matrix multiplication of a unit with the coefficient matrix $(Z_\kappa^\lambda)$:  

$\mathbf{Z}'=\boldsymbol\zeta_k \mathbf{Z}  \enspace \hspace{1pt} \Rightarrow \enspace  (Z_\kappa^\lambda)'=\boldsymbol\zeta_k (Z_\kappa^\lambda) \quad$ \hspace{15pt} and  $\quad \mathbf{Z}'=\mathbf{Z}\boldsymbol\zeta_k  \enspace \hspace{2pt} \Rightarrow \enspace (Z_\kappa^\lambda)'=-\boldsymbol\gamma^k (Z_\kappa^\lambda) \enspace$

$\mathbf{Z}'=\boldsymbol\gamma^l \mathbf{Z}  \enspace \hspace{2pt} \Rightarrow \enspace  (Z_\kappa^\lambda)'=-(Z_\kappa^\lambda)\boldsymbol\zeta^l \quad$\hspace{12pt} and 
$\quad \mathbf{Z}'=\mathbf{Z}\boldsymbol\gamma^l  \enspace \hspace{3pt} \Rightarrow \enspace (Z_\kappa^\lambda)'=(Z_\kappa^\lambda)\boldsymbol\gamma^l \enspace$

$\mathbf{Z}'=\boldsymbol\beta_k^l \mathbf{Z}  \enspace \Rightarrow \enspace  (Z_\kappa^\lambda)'=-\boldsymbol\zeta^k(Z_\kappa^\lambda)\boldsymbol\zeta^l \quad$ and 
$\quad \mathbf{Z}'=\mathbf{Z}\boldsymbol\beta_k^l  \enspace \Rightarrow \enspace (Z_\kappa^\lambda)'=-\boldsymbol\gamma^k(Z_\kappa^\lambda)\boldsymbol\gamma^l \enspace$

From these formulas, the following transformation formulas can be derived.

\vspace{-5pt}
\subsubsection*{$\boldsymbol\zeta$-Rotation}
\vspace{-5pt}
$\mathbf{Z}'=\exp \left( \vec{\boldsymbol\zeta}\cdot\vec{e} \psi \right)\mathbf{Z}\exp \left( -\vec{\boldsymbol\zeta}\cdot\vec{e} \psi \right)  \qquad \Rightarrow \qquad (Z_\kappa^\lambda)'=\exp \left( (\boldsymbol\zeta_k+\boldsymbol\gamma^k) e_k \psi \right)(Z_\kappa^\lambda)$.

\vspace{-5pt}
\subsubsection*{$\boldsymbol\gamma$-Rotation}
\vspace{-5pt}
$\mathbf{Z}'=\exp \left( \boldsymbol\gamma^1 \phi \right)\mathbf{Z}\exp \left( -\boldsymbol\gamma^1 \phi \right) \qquad \qquad \enspace \Rightarrow \qquad (Z_\kappa^\lambda)'=(Z_\kappa^\lambda)\exp \left( -(\boldsymbol\zeta_1+\boldsymbol\gamma^1) \phi \right)$.

\vspace{5pt}
Note that $\boldsymbol\zeta_k+\boldsymbol\gamma^k=2 J_k$, where 

\small
$J_1=\left(
\begin{array}{r r r r}
0 & 0 & 0 & 0 \\
0 & 0 & 0 & 0 \\
0 & 0 & 0 & 1 \\
0 & 0 & -1 & 0 \\
\end{array}\right)\qquad 
J_2=\left(
\begin{array}{r r r r}
0 & 0 & 0 & 0 \\
0 & 0 & 0 & -1 \\
0 & 0 & 0 & 0 \\
0 & 1 & 0 & 0 \\
\end{array}\right)\qquad 
J_3=\left(
\begin{array}{r r r r}
0 & 0 & 0 & 0 \\
0 & 0 & 1 & 0 \\
0 & -1 & 0 & 0 \\
0 & 0 & 0 & 0 \\
\end{array}\right)$
\normalsize

\vspace{-5pt}
\subsubsection*{$\boldsymbol\beta^2$-Boost}
\vspace{-5pt}
$\mathbf{Z}'=\exp \left( \vec{\boldsymbol\beta}\mathstrut^2\!\cdot\vec{e} \chi \right)\mathbf{Z}\exp \left( \vec{\boldsymbol\beta}\mathstrut^2\!\cdot\vec{e} \chi \right) \qquad \: \Rightarrow \qquad (Z_\kappa^\lambda)'=\tfrac{1}{2}\left[(Z_\kappa^\lambda)-\boldsymbol\beta_k^l e_k e_l(Z_\kappa^\lambda)\boldsymbol\beta_2^2\right]+\ldots$ 

$\qquad \qquad \ldots \tfrac{1}{2}\left[(Z_\kappa^\lambda) +\boldsymbol\beta_k^l e_k e_l(Z_\kappa^\lambda) \boldsymbol\beta_2^2\right]\cosh 2\chi +\tfrac{1}{2}\left[-\boldsymbol\zeta_k e_k   (Z_\kappa^\lambda)\boldsymbol\zeta_2-\boldsymbol\gamma^k e_k (Z_\kappa^\lambda)\boldsymbol\gamma^2\right]\sinh 2\chi $

\vspace{-5pt}
\subsubsection*{$\boldsymbol\beta^3$-Boost}
\vspace{-5pt}
$\mathbf{Z}'=\exp \left( \vec{\boldsymbol\beta}\mathstrut^3\!\cdot\vec{e} \chi \right)\mathbf{Z}\exp \left( \vec{\boldsymbol\beta}\mathstrut^3\!\cdot\vec{e} \chi \right) \qquad \: \Rightarrow \qquad (Z_\kappa^\lambda)'=\tfrac{1}{2}\left[(Z_\kappa^\lambda)-\boldsymbol\beta_k^l e_k e_l(Z_\kappa^\lambda)\boldsymbol\beta_3^3\right]+\ldots$ 

$\qquad \qquad \ldots \tfrac{1}{2}\left[(Z_\kappa^\lambda) +\boldsymbol\beta_k^l e_k e_l(Z_\kappa^\lambda) \boldsymbol\beta_3^3\right]\cosh 2\chi +\tfrac{1}{2}\left[-\boldsymbol\zeta_k e_k   (Z_\kappa^\lambda)\boldsymbol\zeta_3-\boldsymbol\gamma^k e_k (Z_\kappa^\lambda)\boldsymbol\gamma^3\right]\sinh 2\chi $

\newpage
\subsection{Decoupling one phase space coordinate}

In the following, $s$ designates a symmetric element $\Sigma_\kappa^\lambda=\Sigma^\kappa_\lambda$ of the component matrix, and $a$ designates a skew-symmetric element $\Sigma_\kappa^\lambda=-\Sigma^\kappa_\lambda$. 

\subsubsection*{Decoupling $x$ from $x', y, y'$}

This means bringing the representative matrix and component matrix into the form

$\mathbf{\Sigma}=
\left(
\begin{array}{c c c c}
\Sigma_{11} & 0 & 0 & 0 \\
0 & \Sigma_{22} & \Sigma_{23} & \Sigma_{24} \\
0 & \Sigma_{32} & \Sigma_{33} & \Sigma_{34} \\
0 & \Sigma_{42} & \Sigma_{43} & \Sigma_{44} \\
\end{array}
\right)\qquad \Rightarrow \qquad (\Sigma_\kappa^\lambda)=
\left(
\begin{array}{r r r r}
\Sigma_0^0 & 0 & 0 & 0 \\
0 & \Sigma_1^1 & s & s\\
0 & s & \Sigma_2^2 & s \\
0 & s & s & \Sigma_3^3 \\
\end{array}
\right)$

\vspace{5pt}
\fbox{
\begin{minipage}{0.98\linewidth}
\vspace{5pt}

To decouple $x$ from $x', y, y'$, calculate the 3D polar decomposition of the 'vector-vector-part' of the component matrix: $(\Sigma_k^l)=\mathbf{O}(\Sigma'^l_{\; k}) \qquad \Leftrightarrow$ 

\vspace{10pt}
$\left(
\begin{array}{r r r r}
\Sigma_1^1 & \Sigma_1^2 & \Sigma_1^3 \\
\Sigma_1^1 & \Sigma_2^2 & \Sigma_2^3 \\
\Sigma_3^1 & \Sigma_3^2 & \Sigma_3^3 \\
\end{array}
\right) = 
\exp \left[2\psi\left(
\begin{array}{c c c}
0 & -e_3 & e_2 \\
e_3 & 0 & -e_1 \\
-e_2 & e_1 & 0 \\
\end{array}
\right)\right]\left(
\begin{array}{r r r r}
\Sigma'^1_{\; 1} & s & s\\
s & \Sigma'^2_{\; 2} & s \\
s & s & \Sigma'^3_{\; 3} \\
\end{array}
\right)$

\vspace{10pt}
Extract $\psi$ and $\vec{e}$ from $\mathbf{O}$:
\begin{equation*}
\cos (2\psi)=\left(O^1_1 + O^2_2 + O^3_3-1\right)/2 \quad \Rightarrow \quad \psi \text{ in } [0,\pi/2] 
\end{equation*}
\begin{equation*}
e_1=\frac{O^3_2 - O^2_3}{2\sin (2\psi)} \qquad 
e_2=\frac{O^1_3 - O^3_1}{2\sin (2\psi)} \qquad 
e_3=\frac{O^2_1 - O^1_2}{2\sin (2\psi)}
\end{equation*}

Then $\mathbf{\Sigma}'=\exp \left( \vec{\boldsymbol\zeta}\cdot\vec{e} \psi \right)\mathbf{\Sigma}\exp \left( -\vec{\boldsymbol\zeta}\cdot\vec{e} \psi \right)$ has $x$ decoupled from $x', y, y'$. 

\vspace{5pt}
The decoupling is preserved under the transformation group 
$\mathbf{R}=\exp\left[\tfrac{1}{2}\left(\boldsymbol\zeta_1+ \boldsymbol\gamma^1\right)\phi \right. $ $\left.  +\tfrac{1}{2}\left(\boldsymbol\beta_3^2 + \boldsymbol\beta_2^3\right)\chi_1 + \boldsymbol\beta_2^2 \chi_2+\boldsymbol\beta_3^3 \chi_3\right]$ (free parameters $\phi, \chi_1, \chi_2, \chi_3$). 
\vspace{5pt}
\end{minipage}
}

\vspace{10pt}
\begin{figure}[H]
\begin{minipage}{0.68 \linewidth}
\centering
\includegraphics[trim = 0pt 0pt 0pt 0pt, width=1\textwidth]{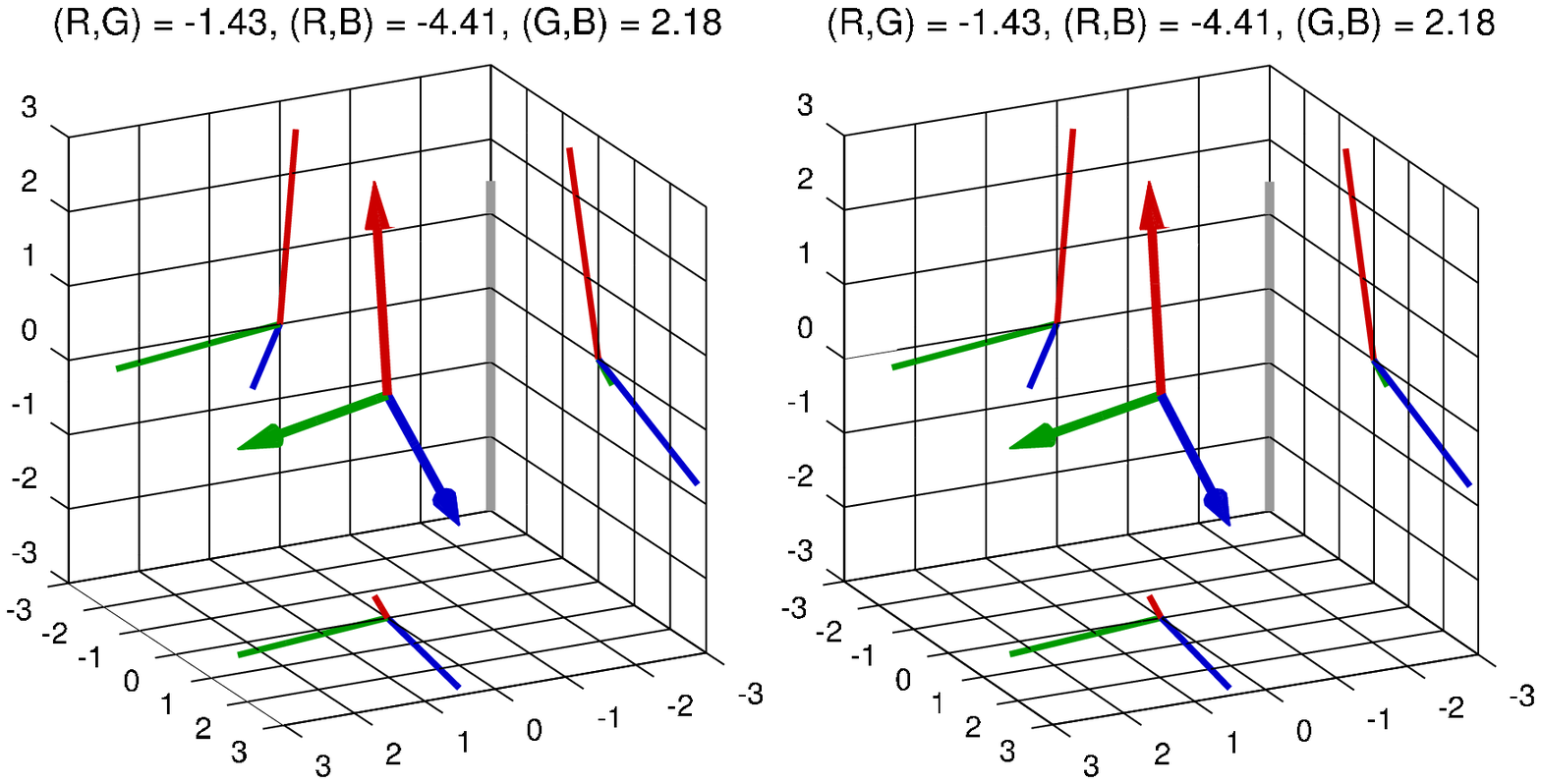}
\end{minipage}
\hspace{0.02 \linewidth}
\begin{minipage}{0.28 \linewidth}
\caption{ }

Beam with $x$ decoupled.\\

Coefficient matrix $\Sigma_\kappa^\lambda=$
\small
$\left(\begin{array}{cccc}
4.4 & 0 & 0 & 0\\
0 & 2.6 & \bar{0}.2 & \bar{0}.8\\
0 & \bar{0}.2 & 2.3 & 0.4 \\
0 & \bar{0}.8 & 0.4 & 2.8
\end{array}\right)$
\normalsize
\end{minipage}
\end{figure}

\newpage
\subsubsection*{Decoupling $x'$ from $x, y, y'$}

This means bringing the representative matrix and component matrix into the form

$\mathbf{\Sigma}=
\left(
\begin{array}{c c c c}
\Sigma_{11} & 0 & \Sigma_{13} & \Sigma_{14}\\
0 & \Sigma_{22} & 0 & 0 \\
\Sigma_{31} & 0 & \Sigma_{33} & \Sigma_{34} \\
\Sigma_{41} & 0 & \Sigma_{43} & \Sigma_{44} \\
\end{array}
\right)\qquad \Rightarrow \qquad (\Sigma_\kappa^\lambda)=
\left(
\begin{array}{c c c c}
\Sigma_0^0 & 0 & 0 & 0 \\
0 & \Sigma_1^1 & a & a\\
0 & a & \Sigma_2^2 & s \\
0 & a & s & \Sigma_3^3 \\
\end{array}
\right)$

\vspace{5pt}
\fbox{
\begin{minipage}{0.98\linewidth}
\vspace{5pt}

To decouple $x'$ from $x, y, y'$, calculate first $\mathbf{\Sigma}'$ with $x$ decoupled from $x', y, y'$ (see above).

Then $\mathbf{\Sigma}''=\boldsymbol\zeta_1\mathbf{\Sigma}' \boldsymbol\zeta_1^\intercal$ has $x'$ decoupled ($\boldsymbol\zeta_1$ is symplectic). 

\vspace{5pt}
The decoupling is preserved under the transformation group 
$\mathbf{R}=\exp\left[\tfrac{1}{2}\left(\boldsymbol\zeta_1+ \boldsymbol\gamma^1\right)\phi \right. $ $\left.+ \tfrac{1}{2}\left(\boldsymbol\beta_3^2 + \boldsymbol\beta_2^3\right)\chi_1 +\boldsymbol\beta_2^2 \chi_2+\boldsymbol\beta_3^3 \chi_3\right]$ (free parameters $\phi, \chi_1, \chi_2, \chi_3$). 
\vspace{5pt}

\end{minipage}
}

\vspace{10pt}
\begin{figure}[H]
\begin{minipage}{0.68 \linewidth}
\centering
\includegraphics[trim = 0pt 0pt 0pt 0pt, width=1\textwidth]{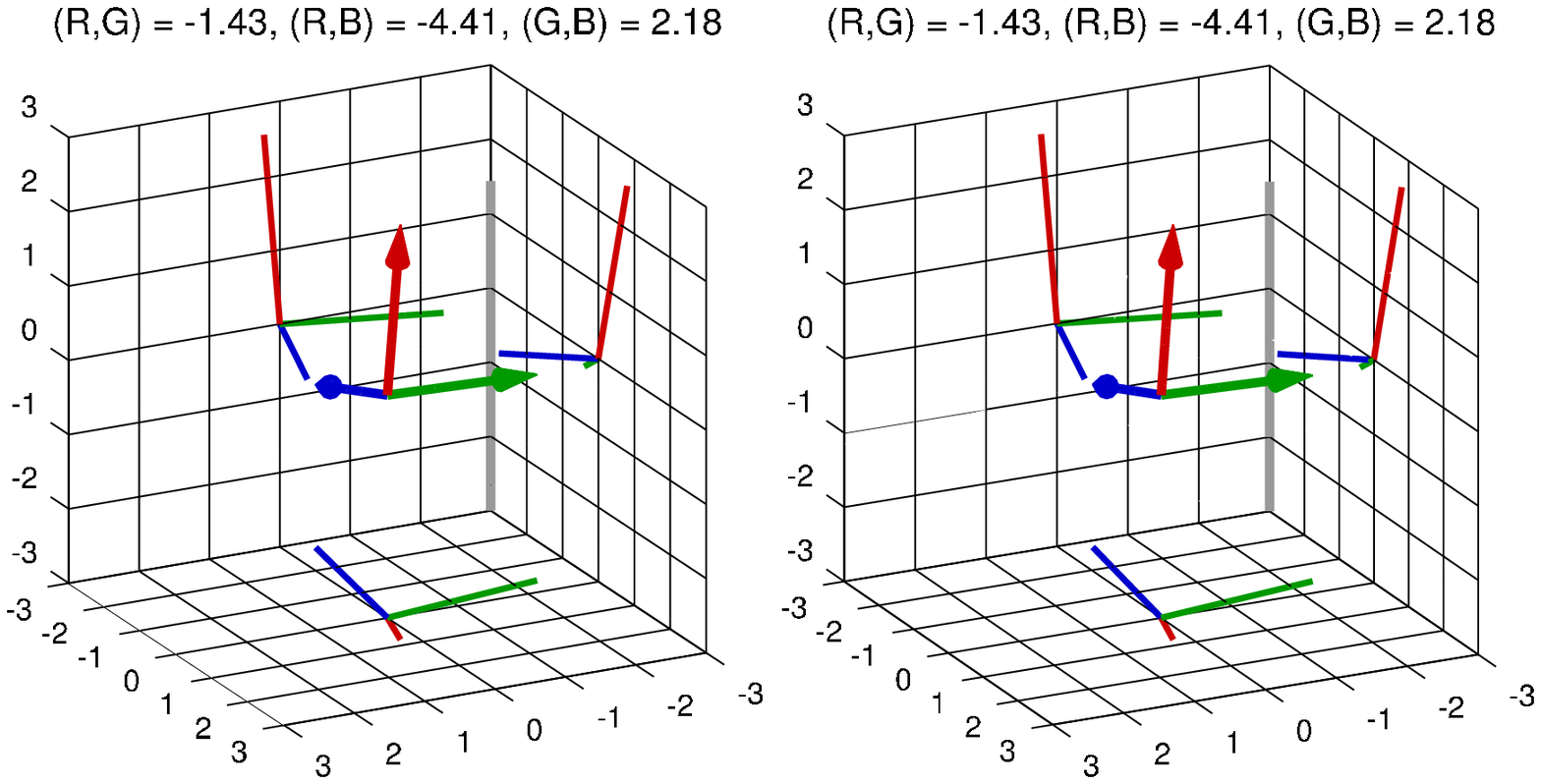}
\end{minipage}
\hspace{0.02 \linewidth}
\begin{minipage}{0.28 \linewidth}
\caption{ }Beam with $x'$ decoupled.\\

Coefficient matrix $\Sigma_\kappa^\lambda=$
\small
$\left(\begin{array}{cccc}
4.4 & 0 & 0 & 0\\
0 & 2.6 & \bar{0}.2 & \bar{0}.8 \\
0 & 0.2 & \bar{2}.3 & \bar{0}.4 \\
0 & 0.8 & \bar{0}.4 & \bar{2}.8
\end{array}\right)$
\normalsize
\end{minipage}
\end{figure}

\subsubsection*{Decoupling $y$ from $x, x', y'$}

This means bringing the representative matrix and component matrix into the form

$\mathbf{\Sigma}=
\left(
\begin{array}{c c c c}
\Sigma_{11} & \Sigma_{12} & 0 & \Sigma_{14}\\
\Sigma_{21} & \Sigma_{22} & 0 & \Sigma_{24} \\
0 & 0 & \Sigma_{33} & 0 \\
\Sigma_{41} & \Sigma_{42}& 0  & \Sigma_{44} \\
\end{array}
\right)\qquad \Rightarrow \qquad (\Sigma_\kappa^\lambda)=
\left(
\begin{array}{r r r r}
\Sigma_0^0 & 0 & 0 & 0 \\
0 & \Sigma_1^1 & a & s\\
0 & a & \Sigma_2^2 & a \\
0 & s & a & \Sigma_3^3 \\
\end{array}
\right)$

\vspace{5pt}
\fbox{
\begin{minipage}{0.98\linewidth}
\vspace{5pt}

To decouple $y$ from $x, x', y'$, calculate first $\mathbf{\Sigma}'$ with $x$ decoupled from $x', y, y'$ (see above).

Then $\mathbf{\Sigma}''= \boldsymbol\zeta_2\mathbf{\Sigma}'\boldsymbol\zeta_2^\intercal$ has $y$ decoupled ($\boldsymbol\zeta_2$ is symplectic). 

\vspace{5pt}
The decoupling is preserved under the transformation group 
$\mathbf{R}=\exp\left[\tfrac{1}{2}\left(\boldsymbol\zeta_1- \boldsymbol\gamma^1\right)\phi \right.$ $\left. + \tfrac{1}{2}\left(\boldsymbol\beta_3^2 - \boldsymbol\beta_2^3\right)\chi_1 +\boldsymbol\beta_2^2 \chi_2+\boldsymbol\beta_3^3 \chi_3\right]$ (free parameters $\phi, \chi_1, \chi_2, \chi_3$). 
\vspace{5pt}
\end{minipage}
}

\vspace{10pt}
\newpage
\begin{figure}[H]
\begin{minipage}{0.68 \linewidth}
\centering
\includegraphics[trim = 0pt 0pt 0pt 0pt, width=1\textwidth]{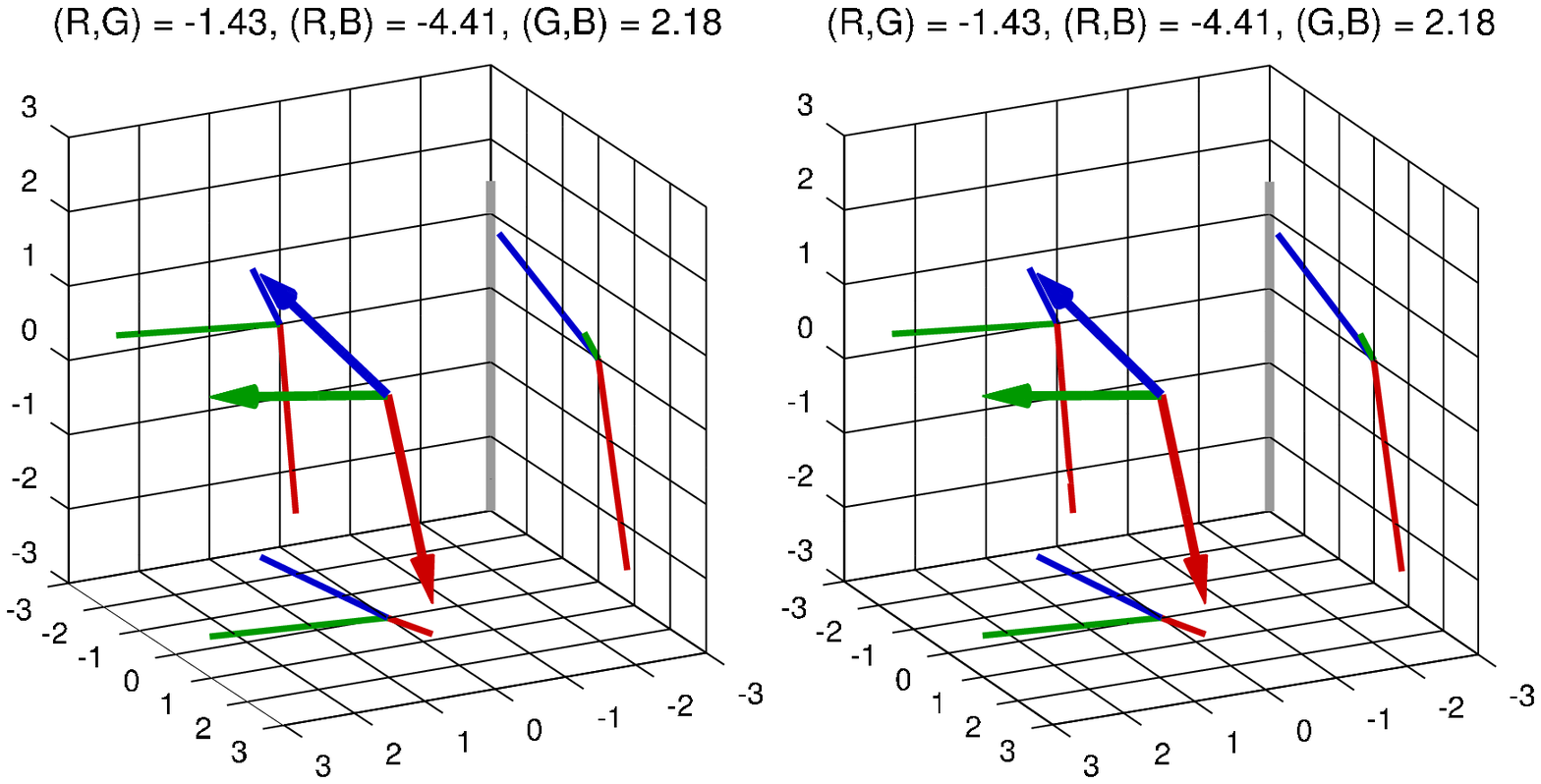}
\end{minipage}
\hspace{0.02 \linewidth}
\begin{minipage}{0.28 \linewidth}
\caption{ }

Beam with $y$ decoupled.\\

Coefficient matrix $\Sigma_\kappa^\lambda=$
\small
$\left(\begin{array}{cccc}
4.4 & 0 & 0 & 0\\
0 & \bar{2}.6 & 0.2 & 0.8 \\
0 & \bar{0}.2 & 2.3 & 0.4 \\
0 & 0.8 & \bar{0}.4 & \bar{2}.8
\end{array}\right)$
\normalsize
\end{minipage}
\end{figure}

\subsubsection*{Decoupling $y'$ from $x, x', y$}

This means bringing the representative matrix and component matrix into the form

$\mathbf{\Sigma}=
\left(
\begin{array}{c c c c}
\Sigma_{11} & \Sigma_{12} & \Sigma_{13} & 0 \\
\Sigma_{21} & \Sigma_{22} & \Sigma_{23} & 0 \\
\Sigma_{31} & \Sigma_{32} & \Sigma_{33} & 0 \\
0 & 0 & 0 & \Sigma_{44} \\
\end{array}
\right)\qquad \Rightarrow \qquad (\Sigma_\kappa^\lambda)=
\left(
\begin{array}{r r r r}
\Sigma_0^0 & 0 & 0 & 0 \\
0 & \Sigma_1^1 & s & a\\
0 & s & \Sigma_2^2 & a \\
0 & a & a & \Sigma_3^3 \\
\end{array}
\right)$

\vspace{5pt}
\fbox{
\begin{minipage}{0.98\linewidth}
\vspace{5pt}

To decouple $y'$ from $x, x', y$, calculate first $\mathbf{\Sigma}'$ with $x$ decoupled from $x', y, y'$ (see above).

Then $\mathbf{\Sigma}''=\boldsymbol\zeta_3\mathbf{\Sigma}' \boldsymbol\zeta_3^\intercal$ has $y'$ decoupled ($\boldsymbol\zeta_3$ is symplectic). 

\vspace{5pt}
The decoupling is preserved under the transformation group 
$\mathbf{R}=\exp\left[\tfrac{1}{2}\left(\boldsymbol\zeta_1- \boldsymbol\gamma^1\right)\phi \right.$ $\left. + \tfrac{1}{2}\left(\boldsymbol\beta_3^2 - \boldsymbol\beta_2^3\right)\chi_1 +\boldsymbol\beta_2^2 \chi_2+\boldsymbol\beta_3^3 \chi_3\right]$ (free parameters $\phi, \chi_1, \chi_2, \chi_3$). 
\vspace{5pt}
\end{minipage}
}

\vspace{10pt}
\begin{figure}[H]
\begin{minipage}{0.68 \linewidth}
\centering
\includegraphics[trim = 0pt 0pt 0pt 0pt, width=1\textwidth]{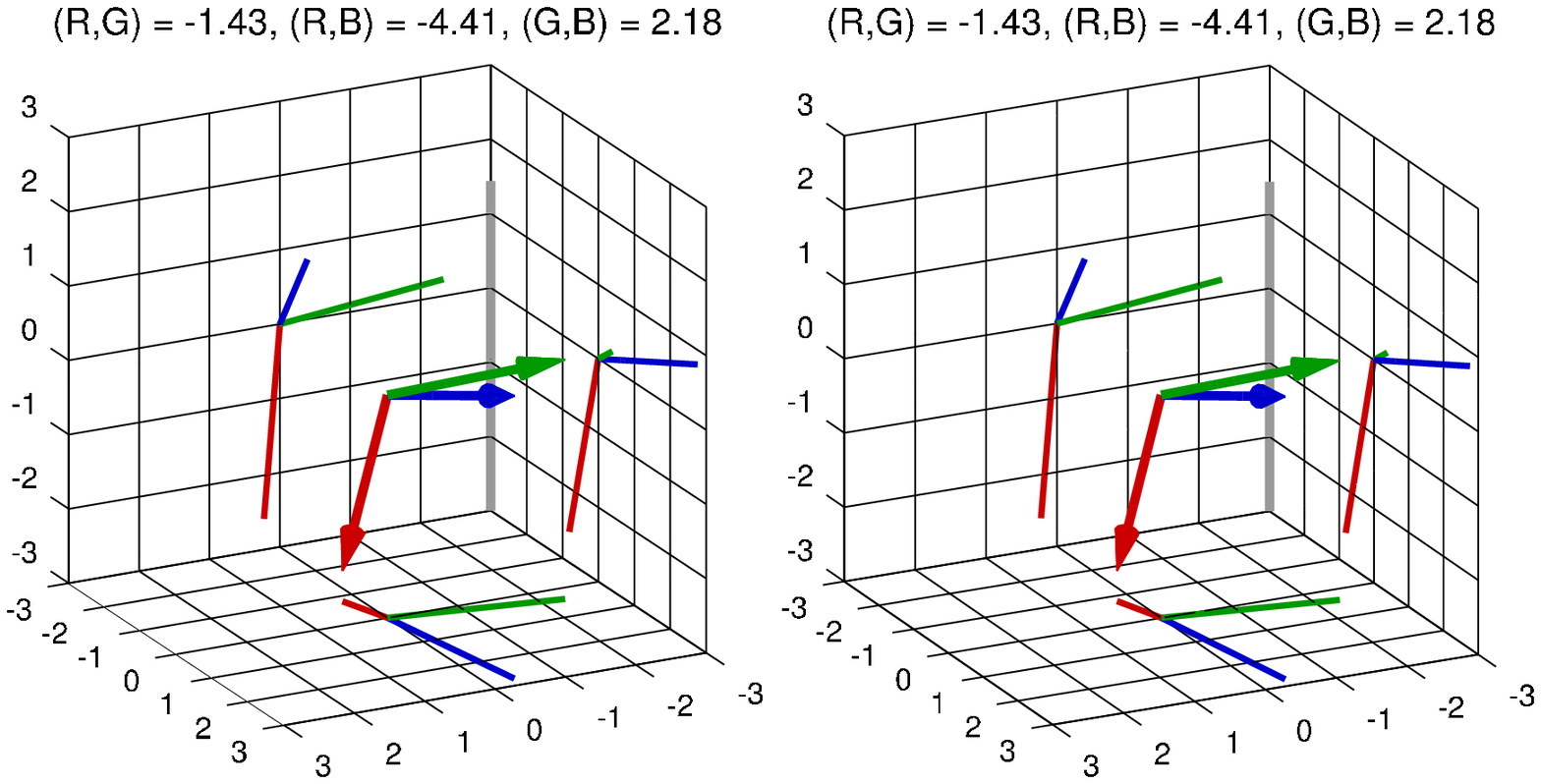}
\end{minipage}
\hspace{0.02 \linewidth}
\begin{minipage}{0.28 \linewidth}
\caption{ }

Beam with $y'$ decoupled.\\

Coefficient matrix $\Sigma_\kappa^\lambda=$
\small
$\left(\begin{array}{cccc}
4.4 & 0 & 0 & 0\\
0 & \bar{2}.6 & 0.2 & 0.8 \\
0 & 0.2 & \bar{2}.3 & \bar{0}.4 \\
0 & \bar{0}.8 & 0.4 & 2.8
\end{array}\right)$
\normalsize
\end{minipage}
\end{figure}

\newpage
\appendix

\section{Calculating the normal decomposition with Octave}

\begin{verbatim}
%% Octave test program (should run on Matlab, too)
%% Create random beam matrix and find normal decomposition

gam=[0 1 0 0 ;-1 0 0 0;0 0 0 1;0 0 -1 0];
bet=abs(gam);Q=(eye(4)+i*bet)/sqrt(2);
sig=randn(4);sig=sig+sig';

% Step 1
bo=(0==0);
while bo
sig=sig+eye(4);
[E,L]=eig(sig*gam);
bo=(norm(real(L))>1E-12);
end
sig
disp('step 1: eigenvalue decomposition'); disp(' ');
diagL=diag(L).'
E=E

% Step 2
disp('step 2: eigenvectors ordered');disp(' ');
L=bet*L*bet;
diagL=diag(L).'
E=E*bet;
% Test: should be 0
testnil=norm(E*L/E-sig*gam)
disp(' ');

% Step 3
disp('step 3: eigenvectors symplectic');disp(' ');
h=E.'*gam*E;
E=E/diag(sqrt([-h(2,1),h(1,2),-h(4,3),h(3,4)]))
%Test: sollte gam sein
testgam=round(100*E.'*gam*E+1E-12)/100

% Step 4
disp('step 4: calculate N'); disp(' ');
N=E*Q
% Test: should be gam
testgam=round(100*N*gam*N.'+1E-12)/100
\end{verbatim}

\section{Supplements to the Clifford algebra $Cl_{3,1}(\mathbb{R})$}
%\section{Supplements to the Clifford algebra Cl(3,1)}

\subsection{How to represent the units of $Cl_{3,1}(\mathbb{R})$}
%\subsection{How to represent the units of Cl(3,1)}

The 16 "real Dirac matrices" $\mathbf{1}$, $\boldsymbol\zeta_k$, $\boldsymbol\gamma^l$, $\boldsymbol\beta_k^l$ representing the units of $Cl_{3,1}(\mathbb{R})$ can be constructed by forming tensor products of the 4 "real Pauli matrices" $1, \boldsymbol\beta_1, \boldsymbol\beta_2, \boldsymbol\gamma$ representing the units of  $Cl_{2,0}(\mathbb{R})$. The tensor product of two $2 \times 2$ matrices is 

\vspace{3pt}
\small
$
\begin{bmatrix} 
a_{11} & a_{12} \\ 
a_{21} & a_{22} \\ 
\end{bmatrix}
\otimes
\begin{bmatrix} 
b_{11} & b_{12} \\ 
b_{21} & b_{22} \\ 
\end{bmatrix}
=
\begin{bmatrix} 
a_{11}  \begin{bmatrix} 
b_{11} & b_{12} \\ 
b_{21} & b_{22} \\ 
\end{bmatrix} & a_{12}  \begin{bmatrix} 
b_{11} & b_{12} \\ 
b_{21} & b_{22} \\ 
\end{bmatrix} \\ 
& \\
a_{21}  \begin{bmatrix} 
b_{11} & b_{12} \\ 
b_{21} & b_{22} \\ 
\end{bmatrix} & a_{22}  \begin{bmatrix} 
b_{11} & b_{12} \\ 
b_{21} & b_{22} \\ 
\end{bmatrix} \\ 
\end{bmatrix}
=
\begin{bmatrix} 
a_{11} b_{11} & a_{11} b_{12} & a_{12} b_{11} & a_{12} b_{12} \\ 
a_{11} b_{21} & a_{11} b_{22} & a_{12} b_{21} & a_{12} b_{22} \\ 
a_{21} b_{11} & a_{21} b_{12} & a_{22} b_{11} & a_{22} b_{12} \\ 
a_{21} b_{21} & a_{21} b_{22} & a_{22} b_{21} & a_{22} b_{22} \\ 
\end{bmatrix}
$
\normalsize

\vspace{3pt}
The matrix multiplication law is $(a \otimes b)(c \otimes d)=(ac \otimes bd)$. 

There are 24 ways (not counting sign change) to represent the generators $\color{myred}\boldsymbol\gamma^1$, $\color{myred}\boldsymbol\beta^2_1$, $\color{myred}\boldsymbol\beta^2_2$, $\color{myred}\boldsymbol\beta^2_3$. The representation of the remaining 12 real Dirac matrices then follows from the multiplication table in section 5.1. A possible (and natural-looking) representation is

\bgroup
\def\arraystretch{1.3}
\begin{tabular}{| l | r r r r |}
\hline
$4\times 4$ & $\begin{bmatrix} \mathbf{1}\\ \vec{\boldsymbol\zeta}\\ \end{bmatrix}$ & $\begin{bmatrix} \color{myred}\boldsymbol\gamma^1 \\ 
\color{black}\vec{\boldsymbol\beta}\mathstrut^1 \\ \end{bmatrix}$
 & $\begin{bmatrix} \boldsymbol\gamma^2 \\ \color{myred}\vec{\boldsymbol\beta}\mathstrut^2 \\ \end{bmatrix}$
  & $\begin{bmatrix} \boldsymbol\gamma^3\color{black} \\  \vec{\boldsymbol\beta} \mathstrut^3 \\ \end{bmatrix}$ \\
\hline
$2\times 2$ & $\mathbf{1} \otimes \mathbf{1}$ & $\color{myred}\mathbf{1} \otimes \boldsymbol\gamma$ & $ \boldsymbol\gamma \otimes \boldsymbol\beta_1$ & $ \boldsymbol\gamma \otimes \boldsymbol\beta_2$ \\
$\phantom{2} \otimes \phantom{2}$ & $- \boldsymbol\beta_1 \otimes \boldsymbol\gamma$ & $ \boldsymbol\beta_1 \otimes \mathbf{1}$ 
& $\color{myred} \boldsymbol\beta_2 \otimes \boldsymbol\beta_2$  & $- \boldsymbol\beta_2 \otimes \boldsymbol\beta_1$ \\
$2\times 2$ & $- \boldsymbol\gamma \otimes \mathbf{1}$ 
& $- \boldsymbol\gamma \otimes \boldsymbol\gamma$& $\color{myred}\mathbf{1} \otimes \boldsymbol\beta_1$ & $\mathbf{1} \otimes \boldsymbol\beta_2$ \\
& $- \boldsymbol\beta_2 \otimes \boldsymbol\gamma$ & $ \boldsymbol\beta_2 \otimes \mathbf{1}$ 
& $\color{myred}- \boldsymbol\beta_1 \otimes \boldsymbol\beta_2$ & $ \boldsymbol\beta_1 \otimes \boldsymbol\beta_1$ \\
\hline
\end{tabular}
\egroup 

\vspace{5pt}
In ref.s \cite{Baumgarten1}, \cite{Baumgarten2}, the units are named $\gamma_n$, $n=0\ldots15$. The correspondence is as follows:

\bgroup
\def\arraystretch{1.3}
\begin{tabular}{| l | r r r r r r r r r r r r r r r r |}
\hline
\cite{Baumgarten1},\cite{Baumgarten2} & $\gamma_0$ & $\gamma_1$ & $\gamma_2$ & $\gamma_3$ & $\gamma_4$ & $\gamma_5$ & $\gamma_6$ & $\gamma_7$ & $\gamma_8$ & $\gamma_9$ & $\!\gamma_{10}$ & $\!\gamma_{11}$ & $\!\gamma_{12}$ & $\!\gamma_{13}$  & $\!\gamma_{14}$ & $\!\!\gamma_{15}$\\ 
\hline
here & $\!\boldsymbol\gamma^1$ & $\!\boldsymbol\beta_3^2$ & $\!\boldsymbol\beta_1^2$ & $\!\!\!-\boldsymbol\beta_2^2$ & $\!\!\!-\boldsymbol\beta_3^3$ & $\!\!\!-\boldsymbol\beta_1^3$ & $\!\boldsymbol\beta_2^3$ & $\!\!\!-\boldsymbol\zeta^3$ & $\!\!\!-\boldsymbol\zeta^1$ & $\!\boldsymbol\zeta^2$ & $\!\boldsymbol\gamma^2$ & $\!\!\!-\boldsymbol\beta_3^1$ & $\!\!\!-\boldsymbol\beta_1^1$ & $\!\boldsymbol\beta_2^1$ & $\!\!\!-\boldsymbol\gamma^3$ & $\!\mathbf{1}$ \\ 
\hline
\end{tabular}
\egroup

\subsection{Calculating the effect of an elementary transformation}

The calculation of the transformed beam matrix $\mathbf{\Sigma}'= \mathbf{R}\mathbf{\Sigma}\mathbf{R}^\intercal$ under an elementary transformation requires the  multiplication table of the algebra units given earlier, but also the following table of "two-sided products" of the form $\mathbf{ABA}$. In this table, $\vec{e}=(e_1,e_2,e_3)^\intercal$ and $(f^1,f^2,f^3)$ are unit-vectors. 

\vspace{5pt}
\bgroup
\def\arraystretch{1.3}
\begin{tabular}{| r | r r r r |}
\hline
$\downarrow \cdot \rightarrow \cdot \downarrow$ & $\mathbf{1}$ & $\boldsymbol\gamma^m$ & $\boldsymbol\zeta_k$ & $\boldsymbol\beta^m_k$\\
\hline
$\mathbf{1} \ldots \mathbf{1}$ & $\mathbf{1}$ & $\boldsymbol\gamma^m$ & $\boldsymbol\zeta_k$ & $\boldsymbol\beta^m_k$\\
$f^o\boldsymbol\gamma^o \ldots \boldsymbol\gamma^nf^n$ & $-\mathbf{1}$ & $\boldsymbol\gamma^m-2f^m(f^n\boldsymbol\gamma^n)$ &  $-\boldsymbol\zeta_k$ & $\boldsymbol\beta^m_k-2f^m(f^n\boldsymbol\beta^n_k)$  \\
$\vec{e}\cdot\vec{\boldsymbol\zeta} \ldots \vec{\boldsymbol\zeta}\cdot\vec{e}$ & $-\mathbf{1}$ & $-\boldsymbol\gamma^m$ & $\boldsymbol\zeta_k-2e_k\vec{e}\cdot\vec{\boldsymbol\zeta}$ & $\boldsymbol\beta^m_k-2e_k(e_l\boldsymbol\beta^m_l)$ \\
$e_if^o\boldsymbol\beta^o_i \ldots \boldsymbol\beta^n_lf^ne_l$ & $\mathbf{1}$ & $-\boldsymbol\gamma^m+2f^m(f^n\boldsymbol\gamma^n)$ & $-\boldsymbol\zeta_k+2e_k\vec{e}\cdot\vec{\boldsymbol\zeta}$ & $\boldsymbol\beta^m_k+4f^me_k(f^ne_l\boldsymbol\beta^n_l)$ \\
& & & & $-2f^m(f^n\boldsymbol\beta^n_k)-2e_k(e_l\boldsymbol\beta^m_l)$ \\
\hline
\end{tabular}
\egroup

\vspace{5pt}
By way of example, these are the calculation steps for the $\gamma^1$-rotation acting on $\mathbf{\Sigma}$:
\begin{align*}
\mathbf{\Sigma}'=&\;\mathbf{R}_{\gamma}\mathbf{\Sigma}\mathbf{R}_{\gamma}^\intercal \\
=&\,\exp \left( \boldsymbol\gamma^1 \phi \right) \left(\mathbf{1} \Sigma^0_0+\vec{\boldsymbol\beta }\mathstrut^m\!\cdot \vec{\Sigma}^m \right) \exp \left( -\boldsymbol\gamma^1 \phi \right) \\
=&\,\mathbf{1} \Sigma^0_0+\left(\cos\phi+\boldsymbol\gamma^1 \sin\phi \right) \left(\vec{\boldsymbol\beta}\mathstrut^m\!\cdot \vec{\Sigma}^m \right) \left(\cos\phi-\boldsymbol\gamma^1 \sin\phi \right)  \\
=&\,\mathbf{1} \Sigma^0_0+\cos^2\phi \vec{\boldsymbol\beta}\mathstrut^m\!\cdot \vec{\Sigma}^m + \cos\phi \sin\phi \left[\boldsymbol\gamma^1, \vec{\boldsymbol\beta}\mathstrut^m\right]\cdot\vec{\Sigma}^m
-\sin^2\phi \left(\boldsymbol\gamma^1\vec{\boldsymbol\beta}\mathstrut^m \boldsymbol\gamma^1\right)  \cdot \vec{\Sigma}^m\\
=&\,\mathbf{1} \Sigma^0_0+\cos^2\phi \left(\vec{\boldsymbol\beta}\mathstrut^1\!\cdot\vec{\Sigma}^1 +\vec{\boldsymbol\beta}\mathstrut^2\!\cdot\vec{\Sigma}^2 +\vec{\boldsymbol\beta}\mathstrut^3\!\cdot\vec{\Sigma}^3 \right) + 2\cos\phi \sin\phi \left( -\vec{\boldsymbol\beta}\mathstrut^3\!\cdot\vec{\Sigma}^2 +\vec{\boldsymbol\beta}\mathstrut^2\!\cdot\vec{\Sigma}^3 \right)  \\
&\,\phantom{1 \Sigma^0_0}+\sin^2\phi \left(\vec{\boldsymbol\beta}\mathstrut^1\!\cdot\vec{\Sigma}^1 -\vec{\boldsymbol\beta}\mathstrut^2\!\cdot\vec{\Sigma}^2 -\vec{\boldsymbol\beta}\mathstrut^3\!\cdot\vec{\Sigma}^3 \right) \\
=&\;\mathbf{1} \Sigma^0_0 + \vec{\boldsymbol\beta}\mathstrut^1 \!\cdot \vec{\Sigma}^1 +\vec{\boldsymbol\beta}\mathstrut^2\!\cdot \left[ \cos \left( 2\phi \right) \vec{\Sigma}^2+\sin \left( 2\phi \right) \vec{\Sigma}^3  \right] +\vec{\boldsymbol\beta}\mathstrut^3\!\cdot \left[ -\sin \left( 2\phi \right) \vec{\Sigma}^2+\cos \left( 2\phi \right) \vec{\Sigma}^3  \right]
\end{align*}

The calculation of the other three elementary transformations runs similar. For the two boosts, $\sin$ and $\cos$ are replaced by $\sinh$ and $\cosh$, and the commutator bracket $[\,,]$ is replaced by the anti-commutator bracket $\left\lbrace\,,\right\rbrace$. 

\subsection{Comparison of four diagonalization recipes}

In ref.s  \cite{Baumgarten1}, \cite{Baumgarten2}, Baumgarten proposes two diagonalization recipes , which are different from each other, and from the two recipes given in section 5.6-7. The four recipes may be summarized as follows:

\vspace{5pt}
\bgroup
\def\arraystretch{1.3}
\begin{tabular}{| c | c c c c|}
\hline
step \# &  ref. \cite{Baumgarten1}, p. 17 & ref. \cite{Baumgarten2}, p. 6 &  section 5.6-7  &  section 5.7 \\
\hline
1 & $\boldsymbol\zeta_2, \boldsymbol\zeta_3 \rightarrow  \color{myblue}\vec{\Sigma}^3 \color{black}\!\parallel 1$  & 
$\boldsymbol\gamma^1\rightarrow \color{myblue}\vec{\Sigma}^3 \color{black}\!\perp\color{myred}\vec{\Sigma}^1$  & 
$\vec{\boldsymbol\beta}\mathstrut^2\rightarrow \color{mygreen}\vec{\Sigma}^2 \color{black}\!\perp\color{myred}\vec{\Sigma}^1$ & 
$\vec{\boldsymbol\beta}\mathstrut^2\rightarrow \color{mygreen}\vec{\Sigma}^2 \color{black}\!\perp\color{myred}\vec{\Sigma}^1$ \\
2 & $\boldsymbol\beta\mathstrut^3_1\rightarrow \color{myblue}\vec{\Sigma}^3 \color{black}\!=\vec{0}$  & 
$\boldsymbol\zeta_2, \boldsymbol\zeta_3 \rightarrow  \vec{A}^0 \color{black}\!\parallel 1$ & 
$\vec{\boldsymbol\beta}\mathstrut^3\rightarrow \color{myblue}\vec{\Sigma}^3 \color{black}\!\perp\color{myred}\vec{\Sigma}^1$ & 
$\vec{\boldsymbol\beta}\mathstrut^3\rightarrow \color{myblue}\vec{\Sigma}^3 \color{black}\!\perp\color{myred}\vec{\Sigma}^1$ \\
3 & $\vec{\boldsymbol\zeta}\rightarrow  \color{mygreen}\Sigma_3^2 \color{black}=0, \color{myred}\vec{\Sigma}^1 \color{black}\!\parallel 1$    & 
$\boldsymbol\beta\mathstrut^2_1\rightarrow \color{mygreen}\vec{\Sigma}^2 \color{black}\!\perp\color{myred}\vec{\Sigma}^1$ & 
$\boldsymbol\zeta_2, \boldsymbol\zeta_3 \rightarrow  \color{myred}\vec{\Sigma}^1 \color{black}\!\parallel 1$ & 
$\boldsymbol\gamma^1\rightarrow \color{mygreen}\vec{\Sigma}^2\color{black}\!\perp\color{myblue}\vec{\Sigma}^3$ \\
4  & 
$\boldsymbol\beta\mathstrut^2_1\rightarrow \vec{\Sigma}^k\!\parallel k$ & 
$\boldsymbol\gamma^1\rightarrow \color{mygreen}\vec{\Sigma}^2 \color{black}\!\perp\color{myblue}\vec{\Sigma}^3$ & 
$\boldsymbol\gamma^1\rightarrow \color{mygreen}\vec{\Sigma}^2 \color{black}\!\perp\color{myblue}\vec{\Sigma}^3$ & 
$\vec{\boldsymbol\zeta}\rightarrow \vec{\Sigma}^k\!\parallel k$ \\
5  & 
  & 
$\boldsymbol\zeta_1\rightarrow \vec{\Sigma}^k\!\parallel k$ & 
$\boldsymbol\zeta_1\rightarrow \vec{\Sigma}^k\!\parallel k$ &  \\
\hline
\end{tabular}
\egroup

Here, $\vec{A}^0=2\left(\vec{\Sigma}^2\!\wedge\vec{\Sigma}^3 -\Sigma^0_0\vec{\Sigma}^1\right)$, and $\vec{\Sigma}^k\!\parallel k$ is shorthand for $\color{myred}\vec{\Sigma}^1 \color{black}\!\parallel 1$-axis, $\color{mygreen}\vec{\Sigma}^2 \color{black}\!\parallel 2$-axis, $\color{myblue}\vec{\Sigma}^3 \color{black}\!\parallel 3$-axis.

\vspace{5pt}
The reader is challenged to try to understand the geometric ideas behind the 4 recipes, and make his own choice - or make up his own recipe. 

\end{flushleft}
\end{document}